\documentclass{_nature2_modified_ori}

\usepackage[utf8]{inputenc}

\usepackage{color}
\usepackage{amsmath,amssymb}
\usepackage{graphicx}
\usepackage{hyperref}
\usepackage{lineno}

\def\tcb{\textcolor{blue}}

\newcommand\arcsec{\mbox{$^{\prime\prime}$}}%
\def\farcs{%
 \mbox{%
  \kern  0.13ex.%
  \kern -0.95ex\arcsec%
  \kern -0.1ex%
 } 
}

\newcommand{\oiii}{[O\,{\sc iii}]}
\newcommand{\oii}{[O\,{\sc ii}]}
\newcommand{\hii}{H\,{\sc ii}}
\newcommand{\cii}{[C\,{\sc ii}]}
\newcommand{\sii}{[S\,{\sc ii}]}
\newcommand{\nii}{[N\,{\sc ii}]}

\newcommand{\hst}{{HST}}
\newcommand{\jwst}{{JWST}}

\newcommand{\cluster}{RXCJ0600-2007}
\newcommand{\targ}{the {\it Cosmic Grapes}}
\newcommand{\Targ}{The {\it Cosmic Grapes}}
\newcommand{\targbf}{\textbf{\textit{the Cosmic Grapes}}}

\newcommand{\targarc}{{$z6\_1,2$ (arc)}}

\newcommand{\targb}{{$z6\_2$}}
\newcommand{\targc}{\targ}
\newcommand{\targd}{{$z6\_4$}}
\newcommand{\targe}{{$z6\_5$}}

\newcommand{\avmu}{{32}}

\newcommand{\method}{{\it Methods}}

\title{\flushleft
Primordial Rotating Disk Composed of $\geq$15 Dense Star-Forming Clumps at Cosmic Dawn
}

\author{
S.~Fujimoto$^{1,2,3*}$,
M.~Ouchi$^{4,5,6,7}$,
K.~Kohno$^{8,9}$,
F.~Valentino$^{10,11}$,
C.~Gim\'{e}nez-Arteaga$^{11,12}$,
G.~B.~Brammer$^{11,12}$,
L.~J.~Furtak$^{13}$,
M.~Kohandel$^{14}$,
M.~Oguri$^{15,16}$,
A.~Pallottini$^{14}$,
J.~Richard$^{17}$,
A.~Zitrin$^{13}$,
F.~E.~Bauer$^{18,19,20}$,
M.~Boylan-Kolchin$^{3}$,
M.~Dessauges-Zavadsky$^{21}$,
E.~Egami$^{22}$,
S.~L.~Finkelstein$^{3}$,
Z.~Ma$^{22}$,
I.~Smail$^{23}$,
D.~Watson$^{11,12}$,
T.~A.~Hutchison$^{24}$,
J.~R.~Rigby$^{24}$,
B.~D.~Welch$^{24,25,26}$,
Y.~Ao$^{27,28}$,
L.~D.~Bradley$^{29}$,
G.~B.~Caminha$^{30}$,
K.~I.~Caputi$^{31}$,
D.~Espada$^{32,33}$,
R.~Endsley$^{3}$,
Y.~Fudamoto$^{15}$,
J.~Gonz\'alez-L\'opez$^{34,35}$,
B.~Hatsukade$^{6,8,36}$,
A.~M.~Koekemoer$^{28}$,
V.~Kokorev$^{30}$,
N.~Laporte$^{37}$,
M.~Lee$^{11,38}$,
G.~E.~Magdis$^{11,12,38}$,
Y.~Ono$^{5}$,
F.~Rizzo$^{11,12}$,
T.~Shibuya$^{39}$,
K.~Shimasaku$^{8,9}$,
F.~Sun$^{17}$,
S.~Toft$^{11,12}$,
H.~Umehata$^{40,41,42}$,
T.~Wang$^{43,44}$,
and H.~Yajima$^{45}$
}

\addaffiliation{$^{1}$ 
David A. Dunlap Department of Astronomy and Astrophysics, University of Toronto, 50 St. George Street, Toronto, Ontario, M5S 3H4, Canada
}
\addaffiliation{$^{2}$ 
Dunlap Institute for Astronomy and Astrophysics, 50 St. George Street, Toronto, Ontario, M5S 3H4, Canada
}
\addaffiliation{$^{3}$ Department of Astronomy, The University of Texas at Austin, Austin, TX 78712, USA}

\addaffiliation{$^{4}$ National Astronomical Observatory of Japan, 2-21-1 Osawa, Mitaka, Tokyo 181-8588, Japan}
\addaffiliation{$^{5}$ Institute for Cosmic Ray Research, The University of Tokyo, 5-1-5 Kashiwanoha, Kashiwa, Chiba 277-8582, Japan}
\addaffiliation{$^{6}$ Department of Astronomical Science, SOKENDAI (The Graduate University for Advanced Studies), 2-21-1 Osawa, Mitaka, Tokyo, 181-8588, Japan}
\addaffiliation{$^{7}$ Kavli Institute for the Physics and Mathematics of the Universe (WPI), University of Tokyo, Kashiwa, Chiba 277-8583, Japan}

\addaffiliation{$^{8}$ Institute of Astronomy, Graduate School of Science, The University of Tokyo, 2-21-1 Osawa, Mitaka, Tokyo 181-0015, Japan}
\addaffiliation{$^{9}$ Research Center for the Early Universe, Graduate School of Science, The University of Tokyo, 7-3-1 Hongo, Bunkyo-ku, Tokyo 113-0033, Japan}

\addaffiliation{$^{10}$ European Southern Observatory, Karl-Schwarzschild-Str. 2, D-85748, Garching, Germany}
\addaffiliation{$^{11}$ Cosmic Dawn Center (DAWN), Denmark}

\addaffiliation{$^{12}$ Niels Bohr Institute, University of Copenhagen, Jagtvej 128, 2200, Copenhagen N, Denmark}

\addaffiliation{$^{13}$ Department of Physics, Ben-Gurion University of the Negev, P.O. Box 653, Beer-Sheva 8410501, Israel}

\addaffiliation{$^{14}$ Scuola Normale Superiore, Piazza dei Cavalieri 7, I-56126 Pisa, Italy}

\addaffiliation{$^{15}$ Center for Frontier Science, Chiba University, 1-33 Yayoi-cho, Inage-ku, Chiba 263-8522, Japan}
\addaffiliation{$^{16}$ Department of Physics, Graduate School of Science, Chiba University, 1-33 Yayoi-Cho, Inage-Ku, Chiba 263-8522, Japan}

\addaffiliation{$^{17}$ Univ Lyon, Univ Lyon1, Ens de Lyon, CNRS, Centre de Recherche Astrophysique de Lyon UMR5574, F-69230, Saint-Genis-Laval, France}

\addaffiliation{$^{18}$ Instituto de Astrofísica and Centro de Astroingeniería, Facultad de Física, Pontificia Universidad Católica de Chile, Campus San Joaquín, Av. Vicuña Mackenna 4860, Macul Santiago, Chile, 7820436} 
\addaffiliation{$^{19}$ Millennium Institute of Astrophysics, Nuncio Monseñor Sótero Sanz 100, Of 104, Providencia, Santiago, Chile}
\addaffiliation{$^{20}$ Space Science Institute, 4750 Walnut Street, Suite 205, Boulder, Colorado 80301, USA} 

\addaffiliation{$^{21}$ Département d'Astronomie, Université de Genève, Chemin Pegasi 51, 1290 Versoix, Switzerland}

\addaffiliation{$^{22}$ Steward Observatory, University of Arizona, 933 N. Cherry Ave, Tucson, AZ 85721, USA}

\addaffiliation{$^{23}$ Centre for Extragalactic Astronomy, Department of Physics, Durham University, South Road, Durham DH1 3LE, UK}  

\addaffiliation{$^{24}$ Astrophysics Science Division, NASA Goddard Space Flight Center, 8800 Greenbelt Rd, Greenbelt, MD 20771, USA}

\addaffiliation{$^{25}$ Department of Astronomy, University of Maryland, College Park, MD 20742, USA}
\addaffiliation{$^{26}$ Center for Research and Exploration in Space Science and Technology, NASA/GSFC, Greenbelt, MD 20771 USA}

\addaffiliation{$^{27}$ Purple Mountain Observatory, Chinese Academy of Sciences, Nanjing, Jiangsu 210023, People’s Republic of China}
\addaffiliation{$^{28}$ School of Astronomy and Space Sciences, University of Science and Technology of China, Hefei, Anhui 230026, People’s Republic of China}

\addaffiliation{$^{29}$ Space Telescope Science Institute, 3700 San Martin Drive, Baltimore, Maryland 21102, USA}

\addaffiliation{$^{30}$ Max-Planck-Institut fur Astrophysik, Karl-Schwarzschild-Str. 1, 85748 Garching, Germany}

\addaffiliation{$^{31}$ Kapteyn Astronomical Institute, University of Groningen, P. O. Box 800, 9700AV Groningen, The Netherlands}

\addaffiliation{$^{32}$ Departamento de Física Teórica y del Cosmos, Campus de Fuentenueva, Edificio Mecenas, Universidad de Granada, E-18071, Granada, Spain}
\addaffiliation{$^{33}$ Instituto Carlos I de Física Teórica y Computacional, Facultad de Ciencias, E-18071, Granada, Spain}

\addaffiliation{$^{34}$ Núcleo de Astronomía de la Facultad de Ingeniería y Ciencias, Universidad Diego Portales, Av. Ejército Libertador 441, Santiago, Chile}
\addaffiliation{$^{35}$ Las Campanas Observatory, Carnegie Institution of Washington, Casilla 601, La Serena, Chile}

\addaffiliation{$^{36}$ National Astronomical Observatory of Japan, 2-21-1 Osawa, Mitaka, Tokyo 181-8588, Japan}

\addaffiliation{$^{37}$ Aix Marseille Université, CNRS, CNES, LAM (Laboratoire d’Astrophysique de Marseille), UMR 7326, 13388 Marseille, France}

\addaffiliation{$^{38}$ DTU-Space, Technical University of Denmark, Elektrovej 327, 2800, Kgs. Lyngby, Denmark}

\addaffiliation{$^{39}$ Kitami Institute of Technology, 165, Koen-cho, Kitami, Hokkaido 090-8507, Japan}

\addaffiliation{$^{40}$ Institute for Advanced Research, Nagoya University, Furocho, Chikusa, Nagoya 464-8602, Japan}
\addaffiliation{$^{41}$ Department of Physics, Graduate School of Science, Nagoya University, Furocho, Chikusa, Nagoya 464-8602, Japan}
\addaffiliation{$^{42}$ Cahill Center for Astronomy and Astrophysics, California Institute of Technology, MS249-17, Pasadena, CA91125, USA}

\addaffiliation{$^{43}$ School of Astronomy and Space Science, Nanjing University, Nanjing, Jiangsu 210093, China}
\addaffiliation{$^{44}$ Key Laboratory of Modern Astronomy and Astrophysics, Nanjing University, Ministry of Education, Nanjing 210093, China}

\addaffiliation{$^{45}$ Center for Computational Sciences, University of Tsukuba, Ten-nodai, 1-1-1 Tsukuba, Ibaraki 305-8577, Japan}

\begin{document}
\maketitle

\vspace{0.6cm}
\begin{abstract} 
{\boldmath
Early galaxy formation, initiated by the dark matter and gas assembly, evolves through frequent mergers and feedback processes into dynamically hot, chaotic structures\cite{hopkins2014}. 
In contrast, dynamically cold, smooth rotating disks have been observed in massive evolved galaxies merely 1.4 billion years after the Big~Bang\cite{rizzo2020}, suggesting rapid morphological and dynamical evolution in the early Universe. 
Probing this evolution mechanism necessitates studies of young galaxies, yet efforts have been hindered by observational limitations in both sensitivity and spatial resolution.
Here we report high-resolution observations of a strongly lensed and quintuply imaged, low-luminosity, young galaxy at $z=6.072$ (dubbed \targbf), 930~million years after the Big Bang. 
Magnified by gravitational lensing, the galaxy is resolved into at least 15 individual star-forming clumps with effective radii of $r_{\rm e}\simeq$~10--60~parsec (pc), which dominate $\simeq$ 70\% of the galaxy's total flux in rest-frame ultraviolet (UV).
The cool gas emission unveils an underlying rotating disk (rotational-to-random motion ratio of $3.58\pm0.74$) characterized by a gravitationally unstable state (Toomre $Q \simeq$ 0.2--0.3), with high surface gas densities comparable to local dusty starbursts with $\simeq10^{3-5}$ solar mass ($M_{\odot}$) per pc$^{2}$. These gas properties suggest that the numerous star-forming clumps are formed through disk instabilities with weak feedback effects. 
The clumpiness of \targbf\ significantly exceeds that of galaxies at later epochs and the predictions from current simulations for early galaxies.
Our findings have unveiled the connection between the host galaxy's internal small substructures and the underlying dynamics along with feedback effects for the first time at cosmic dawn, 
potentially explaining the high abundance of bright galaxies observed in the early Universe\cite{arrabal-halo2023a}}.  
\end{abstract}

Using the James Webb Space Telescope (\jwst) with Near Infrared Camera (NIRCam), we obtained deep near-infrared (NIR) imaging over 1--5$\mu$m wavelengths of a gravitationally-lensed star-forming galaxy, \targ\ at $z=6.072$.  
\Targ\ and the arc were initially discovered as multiply lensed images via the bright \cii\ 158$\mu$m emission lines in the Atacama Large Millimeter Array (ALMA) Lensing Cluster Survey behind the massive galaxy cluster \cluster\cite{laporte2021,fujimoto2021}. 
A total of five multiple images have been spectroscopically confirmed, which provides stringent constraints from their sky positions and morphology, resulting in a well-constrained magnification estimate of $\mu=32^{+0.7}_{-6.8}$ for \targ, even in this high magnification regime (see \method; Extended Data Fig.~\ref{fig:nircam_entire}). 
The previous NIR image taken by the Hubble Space Telescope (\hst) shows a disk-like galaxy with a circularized intrinsic effective radius of $0.68^{+0.09}_{-0.01}$~kpc\cite{fujimoto2021} ($0\farcs12$) and absolute ultra-violet (UV) magnitude of $-19.29^{+0.02}_{-0.26}$ with the above $\mu$ estimate. 
Our spectral energy distribution (SED) analysis, using the new NIRCam filters \cite{clara2024}, infers a star-formation rate of SFR$=2.6^{+1.7}_{-1.5}$\,$M_{\odot}$~yr$^{-1}$ and stellar mass of $M_{\star}=4.5^{+2.7}_{-1.1}\times10^{8}\,M_{\odot}$ after lens correction (see \method; Extended Data Fig.~\ref{fig:nircam_cutout}). 
These SFR and size values are consistent with typical galaxies of similar stellar mass at $z\sim6$\cite{shibuya2015, iyer2018}; based on dark-matter halo and stellar mass evolution tracks predicted in cosmological simulations\cite{behroozi2018}, 
galaxies like \targ\ may represent the progenitor of the Milky Way at $z=6$.

\begin{figure*}[t]
\begin{center}
\includegraphics[angle=0,width=1.0\textwidth]{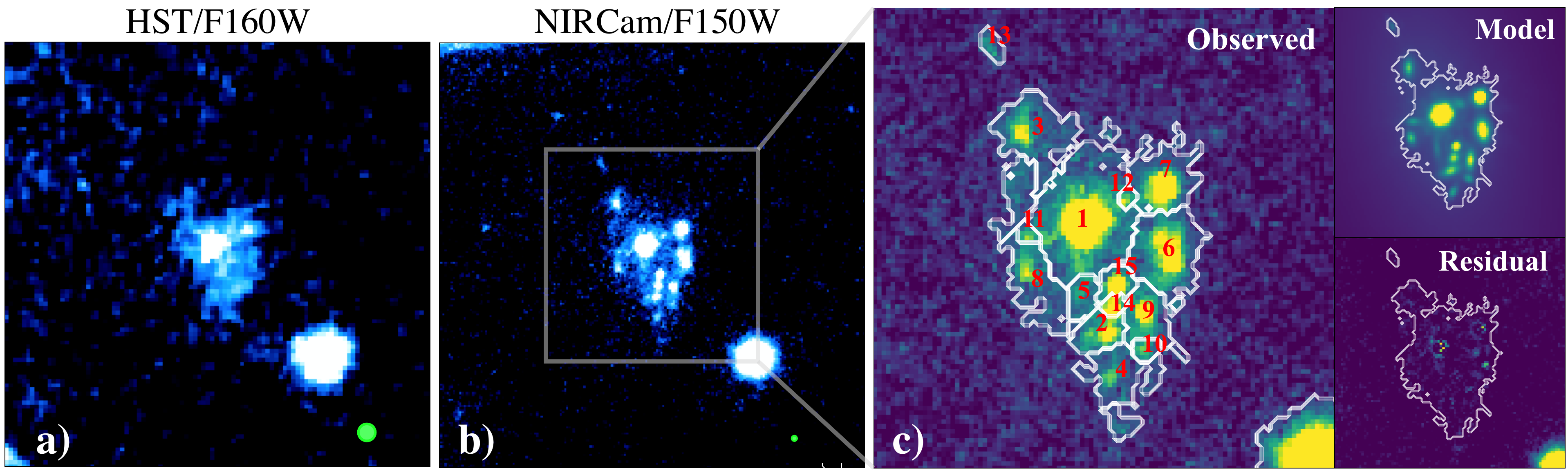}
\end{center}
\vspace{-0.6cm}
\caption{\small 
\textbf{ \boldmath HST and NIRCam near-infrared observed images of 
\targbf} without lens correction. 
\textbf{\emph{(a):}} 
HST/F160W image ($4''\times4''$) around \targ, a strongly lensed ($\mu=$\avmu) low-luminosity, young galaxy at $z=6.072$. 
\textbf{\emph{b:}} 
NIRCam/F150W image ($4''\times4''$) showing the same region as \emph{a}. 
\jwst\ is able to resolve a single disk-like galaxy into 
numerous star-forming clumps due to improved spatial resolution. 
The green circles at bottom right in panels \emph{a} and \emph{b} indicate the PSFs of the HST/F160W and NIRCam/F150W images.  
\textbf{\emph{(c):}} 
A zoomed-in view of the galaxy, highlighting 15 identified star-forming clumps (see \method), as outlined in white.
The right top and bottom panels denote the best-fit S\'ersic models for the clumps obtained using \texttt{GALFIT} (see \method) and the residual map, respectively. 
\label{fig:hst-nircam}}
\end{figure*}

Our deep NIRCam F150W-filter image, equal to rest-frame UV at $\simeq$~2140${\rm \AA}$, resolves the single-disk-like galaxy into numerous individual star-forming clumps (Fig.~\ref{fig:hst-nircam}). 
Most of these clumps remain visible at longer wavelengths (even at lower spatial resolution), and a modest dust reddening of $E(B-V)=0.13\pm0.02$ is measured in spectroscopic follow-up (see \method). 
These results indicate that the distinct clumps are real and not due to highly inhomogeneous dust obscuration effects. 
Our source extraction detects at least 15 star-forming clumps with the current sensitivity and spatial resolution of F150W. We apply simultaneous S\'ersic profile fitting to the 15 individual star-forming clumps, obtaining $r_{\rm e}$ values of $\simeq$10--60~pc after lens correction and find that these clumps dominate $\simeq$70\% of galaxy's total flux in F150W. 
From a pixel-by-pixel based SED analysis\cite{clara2023, clara2024}, we also estimate the SFR and $M_{\star}$ values for the individual star-forming clumps to be $\simeq$0.01--0.2~$M_{\odot}$~yr$^{-1}$ and $\log(M_{\star}/M_{\odot})\simeq$6.2--7.9~$M_{\odot}$, with stellar ages of 2--210 million years (Myr) (see \method). 
These properties suggest that the individual star-forming clumps have similar size and $M_{\star}$ relations as local young massive clusters (YMCs), while exhibiting elevated SFR surface densities compared to clumps identified in lower-redshift galaxies (Extended Data Fig.~\ref{fig:clump_prop})\cite{livermore2015}. 
The elevated surface density of SFR (or $M_{\star}$) among clumps inside galaxies is consistent with recent \jwst/NIRCam results for lensed arc systems at $z\gtrsim6$\cite{vanzella2023, adamo2024}. 
In previous studies, however, effective areas after lens correction are inevitably too small due to their extremely high magnifications in the arcs, making connections from the host galaxy, the small internal substructures, to their underlying dynamics difficult to uncover. 
Uniquely, \targ\ offers the first opportunity to 
investigate such connections, 
owing to its strong $\mu$ yet minimal distortion and differential magnification.

\begin{figure*}[t]
\centering
\includegraphics[angle=0,width=0.85\textwidth]{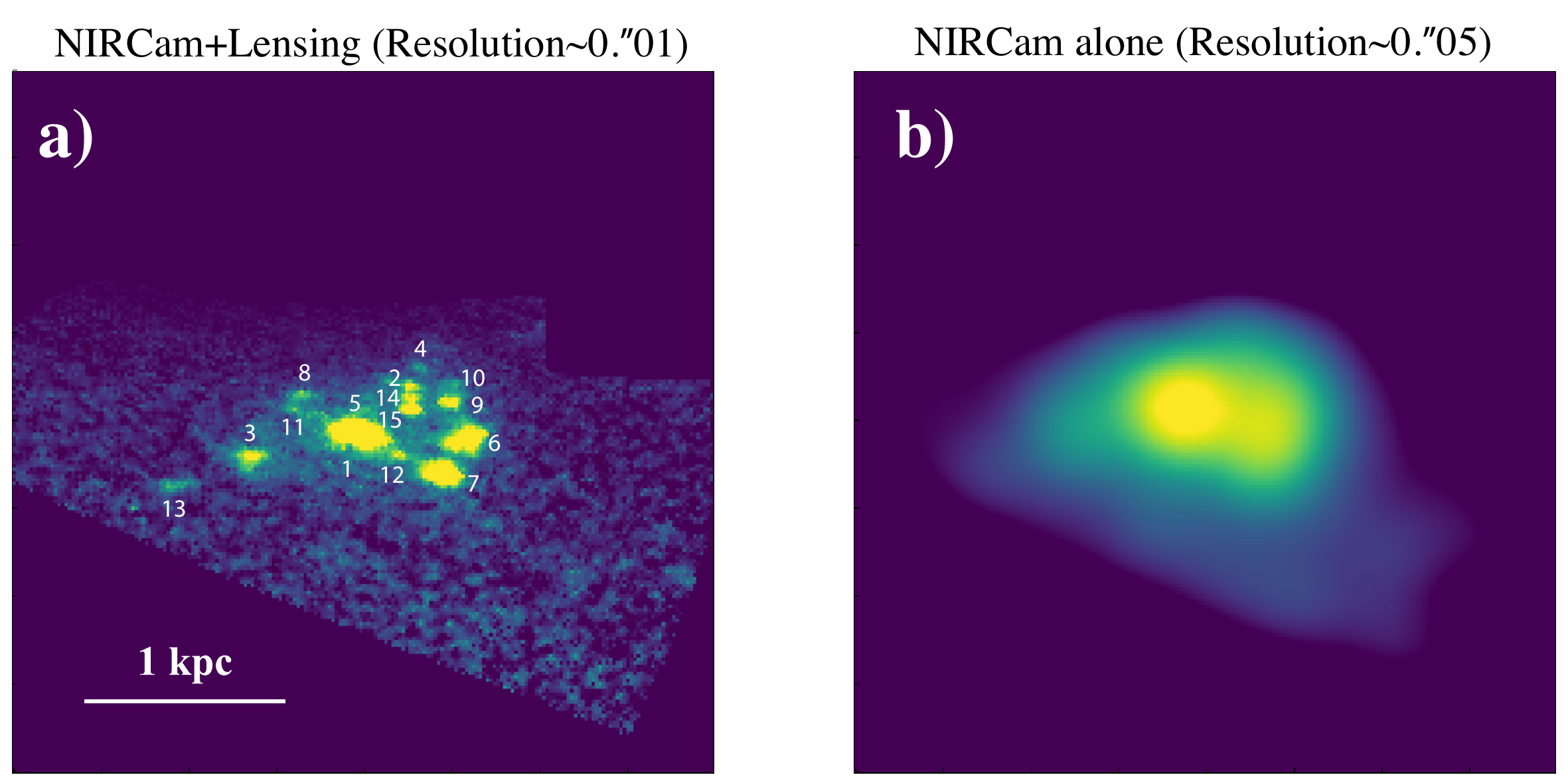}
\caption{\small 
\textbf{ 
\boldmath
Comparison of the views into a typical early galaxy with and without the lensing support.}
\textbf{\emph{(a):}} A source plane reconstructed (i.e., lens-corrected) F150W map of \targ. 
The high magnification with weak distortion uniquely preserves the global-scale intrinsic galaxy morphology with an effective spatial resolution of $\sim0\farcs01$ (58~pc) in the source plane. 
The white labels denote the 15 individual clumps identified and presented in Fig.~\ref{fig:hst-nircam}.
\textbf{\emph{(b):}} The same map as \textbf{\emph{(a)}}, smoothed with a Gaussian kernel to downgrade the spatial resolution to $\sim0\farcs05$  (280~pc), which is comparable to the nominal NIRCam/F150W's resolution without the lensing effect.
The morphology appears to be a single, smooth disk-like galaxy. 
Given \targ\ represents a low-mass main-sequence galaxy at $z=6$ (Extended Data Fig.~\ref{fig:host_prop}),  
many recent HST- and NIRCam-observed early galaxies with similar single-disk-like morphologies\cite{ono2023} may in fact also consist of numerous clumps.  
\label{fig:SP-morph}}
\end{figure*}

In the left panel of Fig.~\ref{fig:SP-morph}, we show the source-plane reconstructed (i.e., lens-corrected) map in F150W. Owing to the strong magnification and the superb spatial resolution of NIRCam in F150W with the Full-width-half-maximum (FWHM) of the point spread function (PSF) being $0\farcs05$ (280~pc) in the image plane, our data effectively achieves the intrinsic spatial resolution of $\simeq0\farcs01$ (58~pc) in the source plane. 
To reveal how this galaxy appears without lensing magnification, the right panel of Fig.~\ref{fig:SP-morph} shows a smoothed F150W map in the source plane, which convolves the existing resolution with a Gaussian kernel to downgrade the source-plane spatial resolution from $0\farcs01$  to $0\farcs05$; the individual clumps disappear in the smoothed map. Since \targ\ represents a typical main-sequence galaxy at $z=6$ (Extended Data Fig.~\ref{fig:host_prop}), this comparison indicates that structures with numerous clumps may be ubiquitous among abundant low-mass galaxies in the early universe, 
and have been overlooked even in recent NIRCam observations of $z{\gtrsim}6$ field galaxies due to its nominal $\sim0\farcs05$ (280~pc) spatial resolution limit. 

\begin{figure*}[t]
\centering
\vspace{-1cm}
\includegraphics[angle=0,width=1.0\textwidth]{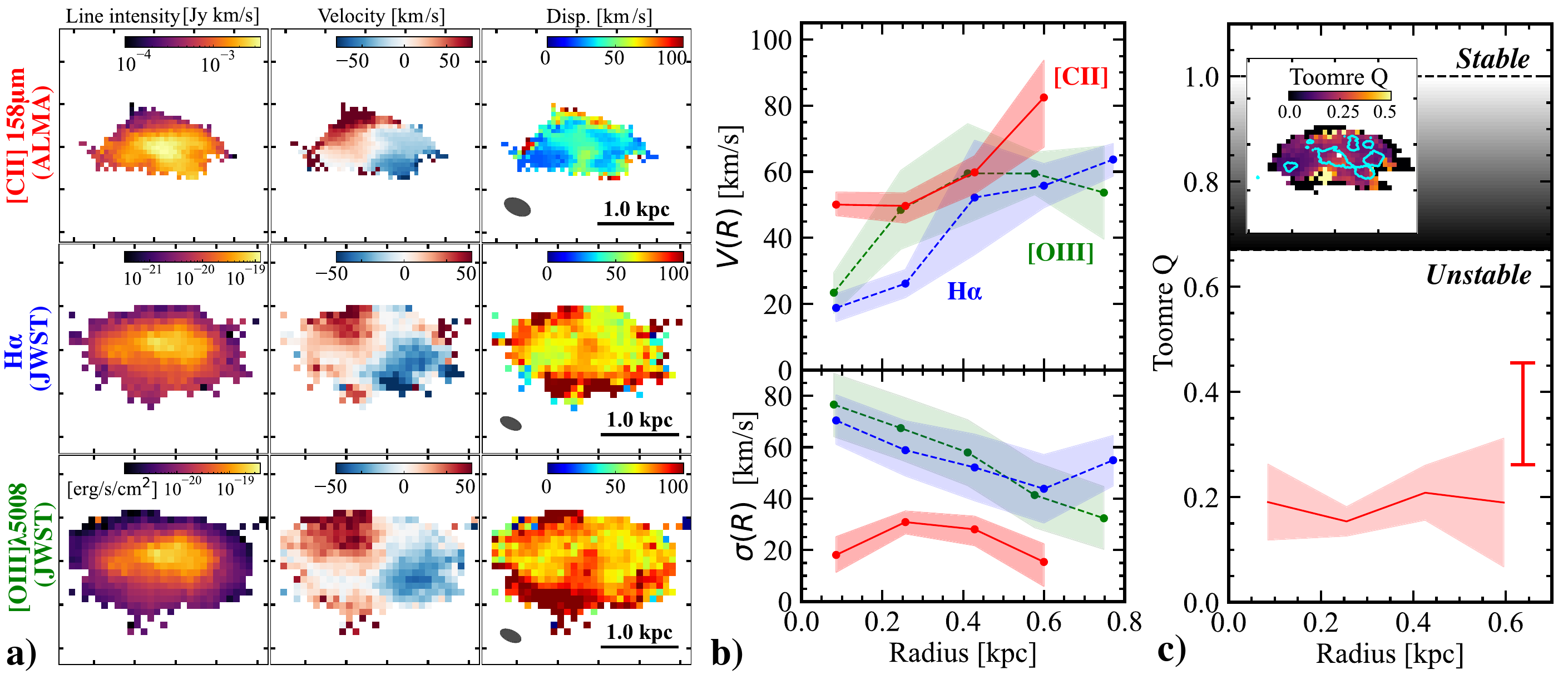}
\vspace{-0.8cm}
\caption{\small 
\textbf{
Kinematic properties of \targbf\ measured by high-resolution IFU observations of ALMA and \jwst/NIRSpec. 
}
\textbf{\emph{(a):}} 
Source-plane reconstructed (i.e., lens-corrected) maps of the line intensity (left), velocity (middle), and dispersion (right) for the 
\cii\ 158$\mu$m, H$\alpha$, and \oiii$\lambda$5008 lines from top to bottom. 
The gray ellipse indicates the effective PSF in the source plane. 
\textbf{\emph{(b):}}
Radial rotation velocity $V(R)$ (top) and velocity dispersion $\sigma(R)$ (bottom) measured using \texttt{3DBarolo} (see \method). 
The red, blue, and green circles and shades indicate the best-fit measurements and 1$\sigma$ uncertainties for \cii, H$\alpha$, and \oiii\ lines, respectively.  
Systematically higher $\sigma(R)$ values observed in the H$\alpha$ and \oiii\ lines, compared to the \cii\ line, could be attributed to ionized gas outflows\cite{kohandel2023} or the limited spectral resolution of NIRSpec ($\simeq$100 km/s). 
We thus use the \cii\ line results to determine an average velocity dispersion $\sigma_{0}$ and maximum rotation velocity ($V_{\rm max}$), and infer $V_{\rm max}/\sigma_{0}=3.58\pm0.74$. 
\textbf{\emph{(c):}}
Radially averaged Toomre $Q$ parameter derived from our kinematic modeling results, while the inset panel displaying a local 2-D $Q$ parameter map obtained from the observed \cii\ data. Cyan contours from the NIRCam/F150W map are overlaid to guide the clump positions.
The vertical bar denotes a possible uncertainty using different conversions between \cii\ luminosity and $M_{\rm gas}$ (see \method). 
The grey-shaded area highlights the critical Toomre Q thresholds: 1.0 for thin disks and 0.67 for thick disks, below which the gas becomes gravitationally unstable.   
The low $Q$ values observed both globally and locally at the clump positions suggest that disk instabilities are the driving force behind numerous clump formations. 
}
\label{fig:kinematics}
\end{figure*}

A unique ensemble of high-resolution IFU observations with 
\jwst\ Near Infrared Spectrograph (NIRSpec) and ALMA have also been carried out on \targ. 
Fig.~\ref{fig:kinematics} summarizes the line intensity, velocity, and dispersion maps for the bright emission lines of \cii158$\mu$m, H$\alpha$, and \oiii$\lambda$5008 reconstructed in the source plane. 
The velocity maps show a smooth velocity gradient from Northeast to Southwest in all emission lines, mostly consistent with the major axis in the galaxy morphology observed in NIRCam. 
Our high-resolution 3D kinematic analysis shows that \targ\ is a rotation-supported system with a symmetric kinematic profile exhibiting a high rotational-to-random motion ratio of $3.58\pm0.74$ (Middle panel of Fig~\ref{fig:kinematics}). 
These kinematic properties indicate that the numerous star-forming clumps in \targ\ are not the result of ongoing merging components but rather are formed through disk-related star-forming activities within the galaxy. 
The oxygen and nitrogen abundances, measured with the NIRSpec IFU data, are in excellent agreement with local \hii\ regions (Extended Data Fig.~\ref{fig:host_prop}), 
which also supports that these clumps are formed through typical star-forming activity within galaxies. 

We compute the local Toomre $Q$-parameter of the cold gas disk pixel-by-pixel, which expresses the balance between the self-gravity of molecular gas and turbulent pressure by stellar radiation and other sources. 
We use the \cii\ line luminosity for the molecular gas estimate\cite{zanella2018, vizgan2022}, which is confirmed to be consistent with the dynamical mass after subtracting the stellar component in this system\cite{fujimoto2021}. 
The $Q$-parameter has critical values of 0.67 and 1.0 for a thick and thin disk, respectively\cite{cacciato2012}, where the higher values correspond to a gravitationally stable disk. 
Our local $Q$ measurements are typically $\lesssim$ 0.3 
over almost the entire disk with potential associations with young (age $<$ 10~Myr) clump positions (Inset in Fig.~\ref{fig:kinematics} right). We also evaluate the radially-averaged $Q$-parameter using the best-fit kinematic parameters that may correct the beam smearing and inclination effects, obtaining similarly low values of $\simeq0.2$--0.3.  
These results imply that gravitational instabilities in the disk indeed induce star formation, leading to the formation of numerous clumps, regardless of the thick or thin disk interpretation.  
Similarly low $Q$ values have been measured in high gas-density star-forming clumps in the disks of star-forming galaxies at later epochs, while they exhibit higher $Q$ values ($\approx$gravitationally stable) toward the galactic center\cite{genzel2011,tadaki2018}. 
The low $Q$ values across the entire galaxy in \targ\ suggest that the galaxy is dominated by high gas-density star-forming clumps without the presence of a massive bulge. This is consistent with the presence of numerous compact star-forming clumps in NIRCam, where the central regions are dominated by young stellar populations (age $<$~10 Myr) (Extended Data Fig.~\ref{fig:nircam_cutout}).

\begin{figure*}[h]
\centering
\includegraphics[angle=0,width=1\textwidth]{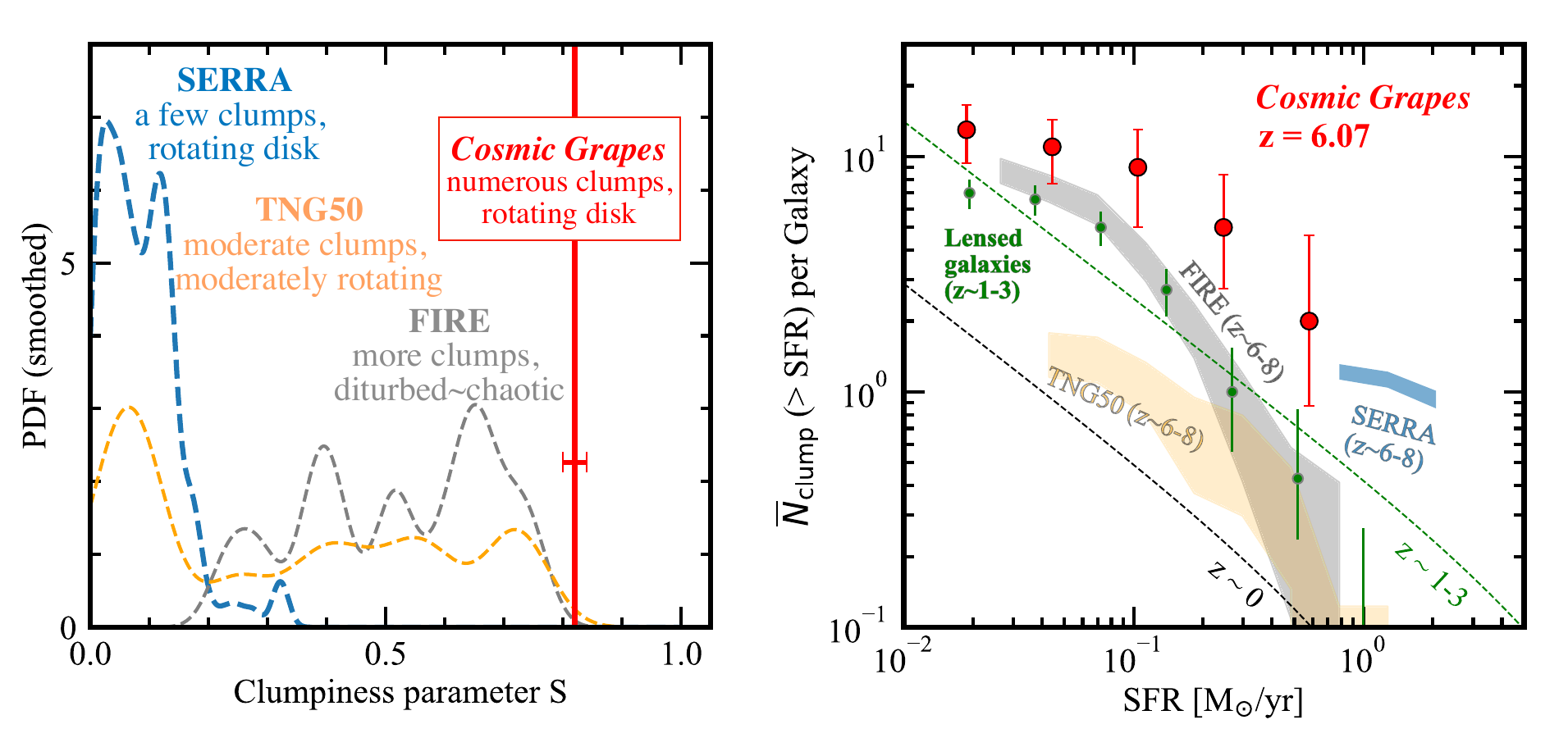}
\vspace{-0.8cm}
\caption{\small \textbf{Comparison of clumpiness in observed and simulated galaxies.} 
\textbf{\emph{(a):}} Clumpiness parameter ($S$)\cite{conselice2003}. 
The red line denotes the observed value in \targ. 
The blue, grey, and orange dashed curves indicate the probability distribution functions (PDFs) drawn from the same measurements for 140 SERRA\cite{pallottini2022}, 13 FIRE\cite{wetzel2023}, and 12 TNG50\cite{pillepich2019} simulated galaxies in their cosmological zoom-in models at $z=$6--8, respectively. 
The PDFs are smoothed with a Gaussian kernel. 
\textbf{\emph{(b):}} Cumulative clump \textit{luminosity function} (LF) as a function of SFR. 
The red circles represent the measurements of \targ.  
The green circles denote lensed galaxies with similar magnifications, SFR~$=1$--10~$M_{\odot}$~yr$^{-1}$ at $z\sim1$--3, and the black and green curves are the best-fit functions for $z\sim0$ and $z\sim$1--3, respectively,  from the literature\cite{livermore2012, livermore2015}.
The blue, grey, and orange shaded regions represent measurements in the same manner for the simulated galaxies of SERRA, FIRE, and TNG50, respectively.  
All error bars are calculated assuming Poisson uncertainties. 
For fair comparisons, we use simulated galaxies whose physical properties (SFR, $M_{\star}$) are similar to those of \targ\ in both $S$ and clump LF measurements (see \method).  
\label{fig:clumpiness}}
\end{figure*}

We quantify the clumpiness of \targ\ using the non-parametric clumpiness parameter $S$ (see \method). We obtain $S=0.81\pm0.02$, which is much higher than average values of $S\simeq$~0.2--0.4 measured in local Spirals and irregular dwarfs in the same rest-frame UV wavelength\cite{mager2018}. 
We also evaluate the clumpiness of \targ\ via the clump \textit{luminosity function} (LF) in a cumulative format as a function of SFR ($\bar{N}_{\rm clump}$) in Fig.~\ref{fig:clumpiness}, together with previous measurements for lensed galaxies at $z\sim$1-3 that have the intrinsic spatial resolution similar to our measurement (see \method)\cite{livermore2012, livermore2015}.  
While an increasing trend of clumpiness has been reported in previous studies from $z=0$ to $z\sim3$ \cite{livermore2012, livermore2015}, our results show that this trend continues out to $z=6$ at least, with \targ\ being consistently a factor of $\gtrsim$ 2 higher at all SFRs compared to lower-redshift results. 

To further investigate the observed clumpiness, we also evaluate the $S$ parameter and the clump LF for galaxies in the state-of-the-art cosmological zoom-in simulations of FIRE\cite{wetzel2023} and SERRA\cite{pallottini2022}. 
We select simulated galaxies at $z=$6--8 with physical properties (SFR, $M_{\rm star}$) similar to \targ, generate image cutouts with similar spatial resolution and sensitivity to our rest-frame UV data, and evaluate the $S$ parameter and the clump LF in the same manner (see \method). 
We find that these simulated galaxies have lower $S$ values and lower clump LFs compared to \targ\ (Fig.~\ref{fig:clumpiness}). SERRA galaxies reproduce smooth-rotating disks but generally only with a few young stellar clumps; FIRE galaxies generally have higher clumpiness owing to more frequent starbursts implemented in the model but show chaotic gas motions with no rotating gas disk due to the subsequent feedback effects from the frequent starbursts; TNG50 galaxies generally fall between the two, while the number of clumps is still much less than \targ.
These discrepancies highlight the challenges of reproducing the observed clumpy disk without resulting in a system dominated by random motions. 
Note that the formation of small star-forming clumps is also regulated by the spatial and time resolutions in the simulations\cite{ma2020}. 
This gap between observations and simulations might suggest that we are reaching a critical resolution limit to compare the substructures inside early galaxies and the zoom-in simulations. 

Our finding may shed new light on several open questions. 
Recent \jwst\ studies show the high abundance of bright galaxies at $z\gtrsim9$, exceeding the theoretical predictions made pre-\jwst\cite{finkelstein2022b, harikane2023}. 
It has been suggested that these early galaxies exhibit a high surface gas density ($\Sigma_{\rm gas}$), leading to ineffective feedback mechanisms and, consequently, a high efficiency of star formation ($\epsilon_{\star}$) \cite{finkelstein2022b, harikane2023}. 
This weak feedback scenario naturally explains the presence of clumps in a rotating gas disk, as high $\Sigma_{\rm gas}$ leads to gravitational fragmentation, producing self-bound clumps with high $\epsilon_{\star}$\cite{grudic2021}.
The bright clumps in \targ\ indeed show high surface densities reaching 
$\Sigma_{\rm gas}\simeq10^{3-5}M_{\odot}$~pc$^{-2}$ (Extended Data Fig.~\ref{fig:clump_prop}), where radiation hydrodynamical simulations suggest $\epsilon_{\star}$ may increase out to $>0.6$--$0.9$.\cite{fukushima2021}  
The bright galaxies observed at $z\gtrsim9$ may also consist of clumps with similarly high values of $\Sigma_{\rm gas}$ and $\epsilon_{\star}$ with weak feedback.
The numerous star-forming clumps could also be key to the dark matter cusp-core problem. Dynamical friction from the numerous star-forming clumps may provide sufficient energy to heat the central dark matter component, potentially creating a density core in the dark matter distribution\cite{elzant2001}. While this depends on how long-lived the clumps are, it is unlikely that the simultaneous formation of nearly 15 short-lived clumps was captured by another unique event of the strong lensing phenomenon by chance. Instead, we might be witnessing an average presence of $\simeq$15 clumps that continuously form (and some may dissipate) via the disk instability. 
Our results provide the first insights into relating the host galaxy's internal small substructures and the underlying dynamics in a primordial galaxy at cosmic dawn. This stresses the importance of high-resolution experiments, which will be further accelerated in the next generation of large telescopes (e.g., Extremely Large Telescope, Thirty Meter Telescope, Giant Magellan Telescope, next-generation VLA).

\vspace{1.0cm}

\vspace{-0.6cm}
\subsection{Acknowledgements}

We thank Jiyai Sun, Filippo Fraternali, and Amina Helmi for discussions on our target properties based on the comparison from systems in the local Universe; Takashi Kojima for helpful inputs for the optical emission line analyses; John Silverman for engaging discussions regarding the potential impacts of clumps on the dark matter distribution; and researchers at Cosmic Frontier Center for valuable feedback on interpretations for the numerous clumps and the gas dynamics observed in our target. 
This work is based on observations made with the NASA/ESA/CSA James Webb Space Telescope (program ID: 1567), ALMA (program ID: 2021.1.00055.S; 2021.1.00247.S; 2021.1.00181.S; 2022.1.00195.S), and MUSE (program ID: 0103.A-0871(B) and 198.A-2008(E)). 
For JWST, the data were obtained from the Mikulski Archive for Space Telescopes at the Space Telescope Science Institute, which is operated by the Association of Universities for Research in Astronomy, Inc., under NASA contract NAS 5-03127 for JWST. 
Support for program \#1567 was provided by NASA through a grant from the Space Telescope Science Institute, which is operated by the Association of Universities for Research in Astronomy, Inc., under NASA contract NAS 5-03127.
ALMA is a partnership of ESO (representing its member states), NSF (USA) and NINS (Japan), together with NRC (Canada), MOST and ASIAA (Taiwan), and KASI (Republic of Korea), in cooperation with the Republic of Chile. The Joint ALMA Observatory is operated by ESO, AUI/NRAO and NAOJ.
We acknowledge support from: the Danish National Research Foundation under grant DNRF140; the NASA Hubble Fellowship grant HST-HF2-51505.001-A awarded by the Space Telescope Science Institute (STScI), which is operated by the Association of Universities for Research in Astronomy, Incorporated, under NASA contract NAS5-26555; JSPS KAKENHI Grant Numbers JP22K21349, JP20H05856, JP22H01260, JP17H06130, JP22H04939, JP23K20035; 
the NAOJ ALMA Scientific Research Grant Number 2017-06B; 
ANID grants for the Millennium Science Initiative Program \#ICN12\_009 (FEB), CATA-BASAL \#FB210003 (FEB), and FONDECYT Regular \#1200495 (FEB);
STFC (ST/X001075/1);
Grant No. 2020750 from the United States-Israel Binational Science Foundation (BSF) and Grant No. 2109066 from the United States National Science Foundation (NSF); by the Ministry of Science \& Technology, Israel; and by the Israel Science Foundation Grant No. 864/23;
the National Natural Science Foundation of China (12173089); 
NSF CAREER award AST-1752913, NSF grants AST-1910346 and AST-2108962, NASA grant 80NSSC22K0827, and HST-GO-15658, HST-GO-15901, HST-AR-16159, HST-GO-16686, HST-AR-17028, and HST-AR-17043 from STScI; 
the European Union (ERC, HEAVYMETAL, 101071865);
the Netherlands Research School for Astronomy (NOVA) and the Dutch Research Council (NWO) through the award of the Vici Grant VI.C.212.036. 
Views and opinions expressed are, however, those of the authors only and do not necessarily reflect those of the European Union or the European Research Council. Neither the European Union nor the granting authority can be held responsible for them.

\subsection{Author contributions}

Y.A., F.E.B., L.D.B., G.B.B, G.B.C., K.I.C., M.D., D.E., J.G., B.H., E.E., A.M.K, K.K, M.O., N.L. M.L., G.E.M., M.O., M.O., J.R., K.S., I.S., F.S., H.U., F.V., A.Z., and S.F. discussed and planned the follow-up observing strategy of \targ\ and its multiple images, writing the telescope proposals. 
G.B. reduced the \jwst/NIRCam images. 
S.F. reduced the \jwst/NIRSpec IFU and ALMA data, and Z.M. and E.E. also reduced the \jwst/NIRSpec IFU data independently to cross-check our results. 
T.H., B.W., and G.R. supported reducing the \jwst/NIRSpec data. 
J.R. reduced the MUSE data and identified the emission lines, generating the spectroscopic sample in the target field.  
C.G.A. performed the pixel-by-pixel based SED fitting using the NIRCam data and produced the SED output maps. 
F.V. conducted the far-infrared SED analysis using the constraints from ALMA and Hershel/SPIRE. 
S.F. and F.R. carried out the kinematic modeling using \texttt{3DBarolo}. 
L.J.F. and A.Z. constructed the updated lens model, and M.O., J.R., and G.B.C. supported it.  
M.K. and A.P. produced image cutouts for the simulated galaxies in the cosmological zoom-in simulation SERRA.  
S.F. and M.B. producing image cutouts for the simulated galaxies in the cosmological zoom-in simulation FIRE using public resources. 
All authors discussed the results and commented on the manuscript. 
S.F. led the team, being the Principal Investigator of the follow-up \jwst, ALMA, and MUSE programs, wrote the main text and the Methods section, and produced all figures and tables.

\subsection{Competing interests}
The authors declare no competing interests. 

\subsection{Corresponding Author} 
Seiji Fujimoto (fujimoto@utexas.edu)

\printaffiliations

\newpage

\clearpage
\newpage

\begin{methods}
\vspace{-0.4cm}

\newcounter{extfigure}
\renewcommand{\thefigure}{\arabic{extfigure}}

\renewcommand{\figurename}{Extended Data Figure}

\renewcommand{\theHfigure}{ext.\arabic{extfigure}}

\setcounter{figure}{0} 


In this paper, error values represent the $1\sigma$ uncertainty, where $\sigma$ denotes the root-mean-square or standard deviation; upper limits are indicated at the 3$\sigma$ level; error values for intrinsic physical properties after the lens correction are obtained by propagating the 1$\sigma$ uncertainties of the magnification estimate and the measurement of the physical properties; red symbols in figures denote \targ, unless otherwise specified. 
We adopt cosmological parameters measured by the Planck mission\cite{planck2014}, i.e.\ a $\Lambda$ cold dark matter ($\Lambda$CDM) model with total matter, 
vacuum and baryonic densities in units of the critical density,
$\Omega_{\Lambda}=$ 0.692,
$\Omega_{\rm m}$ =  0.308, 
$\Omega_{\rm b} =$ 0.0481, 
and Hubble constant, $H_{0}=100$ $h$\,km\,$s^{-1}$\,Mpc$^{-1}$, 
with $h= 0.678$. 
Based on these parameters, we adopt the angular size distance of 5.80 kpc/arcsec at the source redshift of $z=6.072$ in this paper. 

\stepcounter{extfigure} 
\begin{figure*}[h]
\begin{center}
\includegraphics[angle=0,width=1.0\textwidth]{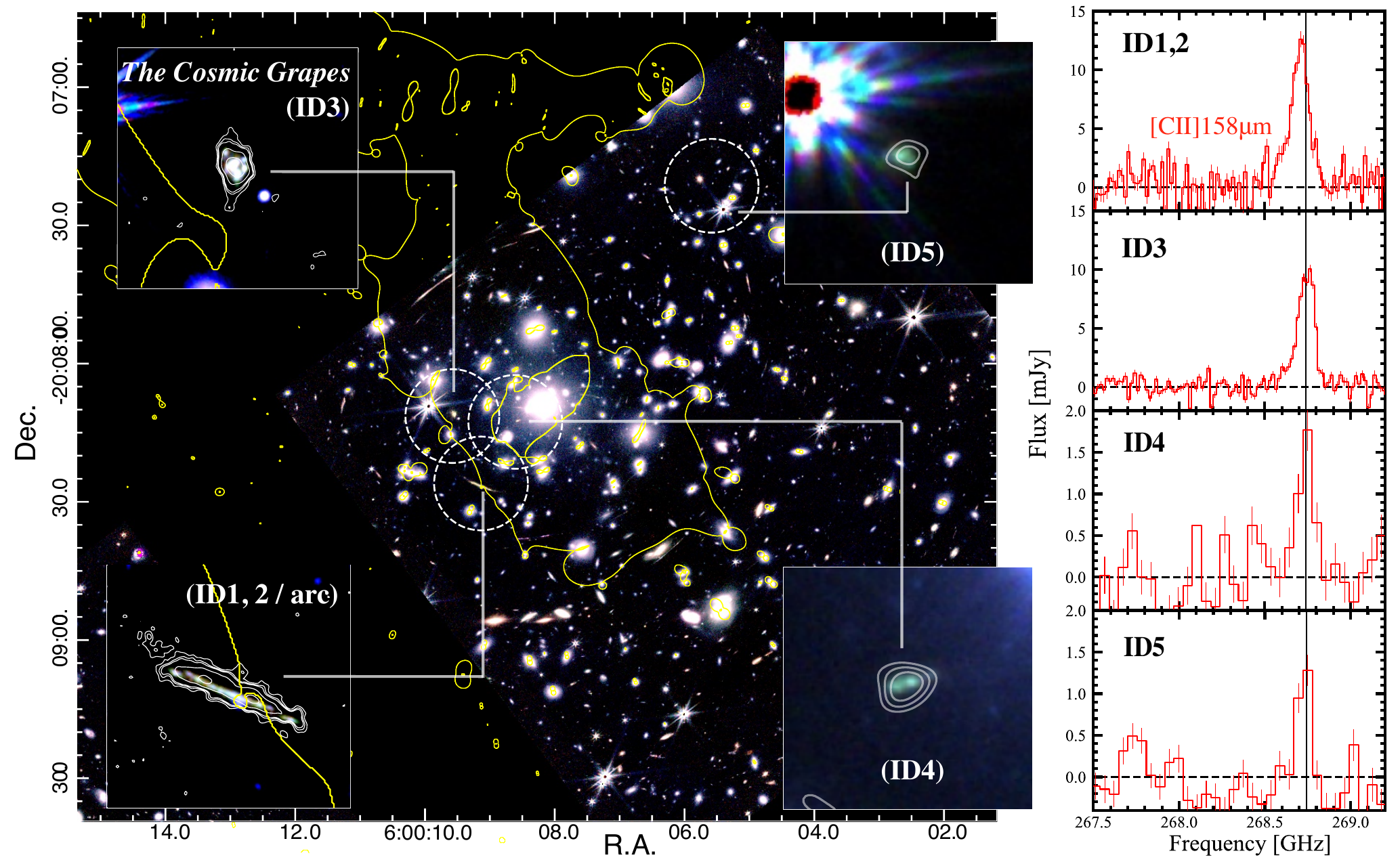}
\end{center}
\vspace{-0.6cm}
\caption{\small 
\textbf{\boldmath NIRCam and ALMA overviews of the cluster RXCJ0600-2007 and the quintuply lensed images of \targbf\ at $z=6.072$.}
\textbf{\emph{(a):}} NIRCam false-color (R: F444W, G: F356W, B: F277W) image. The yellow curves represent the critical lines at $z=6.072$. 
The white dashed circles denote the FoVs of the ALMA follow-up observations. 
The inset panels show the zoom-in NIRCam false-color images for the multiple images of ID1,2 and the rest in $8''\times8''$ and $4''\times4''$ scales, respectively. 
The white contours indicate $3\sigma$, $4\sigma$, $5\sigma$, $10\sigma$, and $20\sigma$ levels of the \cii\ line intensity. 
\textbf{\emph{(b):}} Follow-up ALMA Band~6 spectra of \cii\ for all multiple images. 
Black vertical lines indicate the observed \cii\ frequency at $z=6.072$. Although the \cii\ line detected in \targarc\ is slightly redshifted ($\sim$90~km~s$^{-1}$), this is consistent with the interpretation that \targarc\ traces the outskirt region of the galaxy where it crosses the caustic line in the source plane \cite{fujimoto2021}. 
}
\label{fig:nircam_entire}
\end{figure*}

\subsection{1. Target -- A quintuply lensed galaxy \targbf\ at \boldmath $z=6.072$:}  

\Targ\ was initially discovered in the RXCJ0600-2007 field\cite{ebeling2001} as a part of the ALMA Lensing Cluster Survey (ALCS), an ALMA Large Program in Cycle~6 (\#2018.1.00035.L, PI: K.~Kohno) studying 33 massive lensing clusters observed in the HST treasury programs of RELICS\cite{coe2013}, CLASH\cite{postman2012}, and HFF\cite{lotz2017}. 
In the RXCJ0600-2007 field, two bright emission lines were identified at the same frequency of $\sim$268.7~GHz, exactly at the positions of two Lyman-break galaxies (LBGs) at $z\sim6$, whose redshifts were confirmed in follow-up Gemini GMOS spectroscopy\cite{laporte2021}. This determined the bright lines to be \cii\ 158~$\mu$m at $z=6.072$.\cite{fujimoto2021} 
Three independent lens models (Glafic\cite{oguri2010}, Lenstool\cite{jullo2007}, and Light-Traces-Mass\cite{zitrin2015}) consistently suggest that these two \cii\ emitters are multiple images of an intrinsically faint, sub-$L^{\star}$ LBG. 
These models also predict the presence of two additional multiple images, and the spectroscopic confirmation for the other two multiple images has been achieved in recent ALMA follow-up observations (Section~3). 
The extended arc structure consists of a pair of two multiple images, and thus a total of five multiple images have been spectroscopically confirmed, referred to as $z6\_1$, $z6\_2$, $z6\_3$, $z6\_4$, and $z6\_5$, respectively\cite{fujimoto2021}.   
In this paper, we focus on the results obtained from $z6\_3$, dubbed ``\targ'', while results from the other multiple images will be presented in a separate paper\cite{fujimoto2024prep}.
The sky positions, morphology, and flux ratios among the five multiple images provide stringent constraints on the lens model, which makes the magnification estimate for these sources robust even in the high magnification regime\cite{bouwens2017} (Section~4).  
Extended Data Fig.~\ref{fig:nircam_entire} shows the sky positions of all these multiple images. 

In the source plane, the outskirt of the background galaxy crosses the caustic and is lensed into the long ($\sim6''$) arc, corresponding to \targarc, with local magnifications of $>$100, while \targc\ represents a less sheared image of the galaxy on the global scale with a still high magnification of $32^{+0.7}_{-6.8}$ (see Figure 4 in Fujimoto et al. 2021\cite{fujimoto2021}). 
In contrast to \targarc\ showing significant distortion tracing a local region, the entire galaxy is magnified with minimal distortion in \targc, which is an optimal target for this paper to study the galaxy morphology and internal structure.

\subsection{2. NIRCam Observations: }

The massive galaxy cluster RXCJ0600-2007 was observed with \jwst/NIRCam as a part of \jwst\ GO cycle~1 program (\#1567, PI: S. Fujimoto) in January 2023. 
The center and position angle of the NIRCam field-of-view (FoV) were optimized to include all five multiple images of the background LBG at $z=6.072$ and several unique high-redshift sources that were identified in previous \hst\ and ALMA observations \cite{fujimoto2023b}. 
The NIRCam images were taken in five bands: F115W, F150W, F277W, F356W, and F444W with the exposure times of 1890~sec, 4982~sec, 1890~sec, 2491~sec, and 2491~sec, respectively.  
We adopted the standard subpixel ($N=3$) dithering. 

We reduced and calibrated the NIRCam data, following the procedure described in the DAWN \jwst\ Archive (DJA)\footnote{https://dawn-cph.github.io/dja/} (see also Valentino et al.\cite{valentino2023} for the photometry procedure), and here we briefly explain the reduction and calibration procedure.
The \jwst\ pipeline calibrated level-2 NIRCam imaging products were retrieved and processed with the \texttt{grizli} pipeline \cite{brammer2021, brammer2023}. 
The NIRCam photometric zero-point correction was applied with the Calibration Reference Data System (CRDS) context \texttt{jwst\_1039.pmap}, including detector variations\footnote{
https://github.com/gbrammer/grizli/pull/107
}. 
The fully-calibrated images in each filter were aligned with the GAIA DR3 catalog \cite{gaia2021}, co-added, and drizzled at 20~mas and 40~mas pixel scales for the short-wavelength (SW: F150W, F200W) and long-wavelength (LW: F277W, F356W, F444W) NIRCam bands, respectively. 
In Extended Data Fig.~\ref{fig:nircam_entire}, we show a NIRCam RGB color image for the RXCJ0600-2007 field, and zoom-in cutout RGB images for all the multiple images.
We also present the NIRCam cutouts around \targ\ in Extended Data Fig.~\ref{fig:nircam_cutout}. 

\stepcounter{extfigure} 
\begin{figure*}[t]
\begin{center}
\includegraphics[angle=0,width=1.0\textwidth]{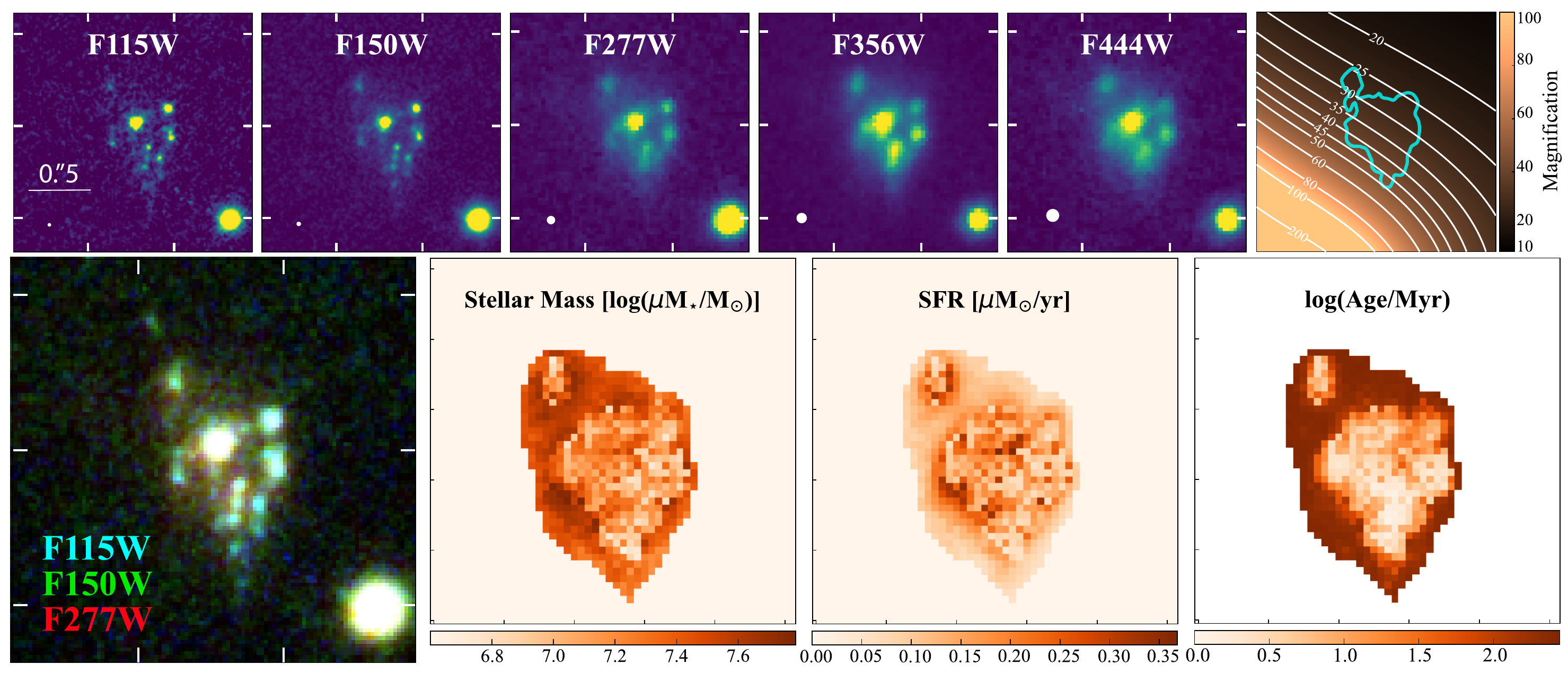}
\end{center}
\vspace{-0.6cm}
\caption{\small 
{\small 
\textbf{\boldmath NIRCam cutouts and spatially-resolved SED fitting outputs for \targbf}. 
\textbf{\emph{Top:}}
NIRCam cutouts ($2\farcs6\times2\farcs6$) in the observed frame (i.e., no lens correction). The cyan circles at the bottom right denote the PSF in each filter. 
The structure remains clumpy in the LW filters (F277W, F356W, F444W), suggesting it is not caused just by patchy dust obscuration (see also Extended Data Fig.~\ref{fig:ebv}). 
The right panel displays the magnification distribution around \targ\ at the same $2\farcs6 \times 2\farcs6$ scale. A cyan contour derived from a smoothed F150W map is overlaid to indicate the galaxy's general position and structure. The map demonstrates that the differential magnification effects across \targ\ are minimal, with variations of $\lesssim30\%$.
}
\textbf{\emph{Bottom:}}
NIRCam false-color image and the pixel-by-pixel SED fitting\cite{clara2023} (see \method) outputs from left to right, showing the centrally located clumps of young stellar populations ($<$10 Myr), surrounded by an extended region of older stellar populations. 
}
\label{fig:nircam_cutout}
\end{figure*}

\subsection{3. ALMA Observations: }

Follow-up ALMA Band~6 observations were carried out on the five multiple images in \cluster\ from January to August 2022 as a part of Cycle~8 program (\#2021.1.00055.S, PI: S. Fujimoto), targeting their \cii\ 158-$\mu$m emission line.
Because the \cii\ line was detected from the brightest multiple images of \targarc\ and \targc\ in the previous ALCS data cube with relatively short exposure, 
these follow-up observations aimed to perform deep, high-resolution \cii\ line spectroscopy to probe the internal structures and kinematics. 
The configurations of C43-2 and C43-5 were used, with the on-source integration times of 49.4~mins and 107.8~mins, respectively, for \targarc\ and \targb. 
Note that the \cii\ line was not clearly detected in the other two multiple images of \targd\ and \targe\ in the previous ALCS data cube, and thus, additional follow-up observations for \targd\ and \targe\ were also designed to detect the \cii\ line to obtain the spectroscopic confirmations of these multiple images. 
The C43-2 configuration was used, with the on-source integration time of 12.6~mins for each. 

We reduced and calibrated the ALMA data with the Common Astronomy Software Applications package CASA version 6.4.1.12\cite{casa2022}, using the pipeline script in the standard manner. 
We imaged the calibrated visibilities with a pixel scale of $0\farcs05$. 
For \targarc\ and \targc, we adopt Briggs weighting with a robust parameter of $0.5$ to maximize the balance between sensitivity and spatial resolution. 
For \targd\ and \targe, we adopt natural weighting to maximize the sensitivity for a detection experiment. 
We adopted a common spectral channel bin of 20~km~s$^{-1}$ for the deep data of \targarc\ and \targc\ and 60~km~s$^{-1}$ for \targd\ and \targe\ and applied the \texttt{tclean} routines down to the 2$\sigma$ level with a maximum iteration number of 100,000 in the automask mode.
The continuum subtraction was applied to the visibility by performing a power-law fit with the line-detected channels masked.  
The generated cubes achieved full-width-half-maximum (FWHM) sizes of the synthesized beam of $0\farcs28\times0\farcs25$ ($0\farcs80\times0\farcs65$) with $1\sigma$ line sensitivities in a 20-km~s$^{-1}$ (60-km s$^{-1}$) width channel of 0.17 mJy~beam$^{-1}$ (0.18 mJy~beam$^{-1}$) for the \targarc\ and \targc\ (\targd\ and \targe) data. 
For a high-resolution experiment to study the potential differential distribution of dust, we also generate a high-resolution dust continuum map with Briggs weighting and a robust parameter of $0.5$, only using the C43-5 configuration data. This map shows the beam size of $0\farcs25\times0.21$ with $1\sigma$ sensitivity of 12~$\mu$Jy~beam$^{-1}$.

\subsection{3. NIRSpec IFU Observations: }

Follow-up \jwst/NIRSpec IFU observations were also performed for the brightest multiple images of \targarc\ and \targc\ from September to December 2022, under the same GO program (\#1567, PI: S. Fujimoto) as the NIRCam observations (Section~1). 
We adopted a 4-point dither with the off-scene nod for both \targarc\ and \targc\ in order to probe for potential extended emission. We used the grating/filter setup G395H/F290LP, whose wavelengths cover the rest-frame wavelength from $\sim$4000${\rm \AA}$ to $\sim7300{\rm \AA}$ for our targets, and thus include key optical emission lines such as H$\alpha$ and \oiii$\lambda$5008. The on-source integration is 3.2 hrs for each target. 
Because of the scope of this paper, here we present observations on \targc, while the other observations are presented elsewhere\cite{fujimoto2024prep}. 

\stepcounter{extfigure} 
\begin{figure*}[t]
\begin{center}
\includegraphics[angle=0,width=1.0\textwidth]{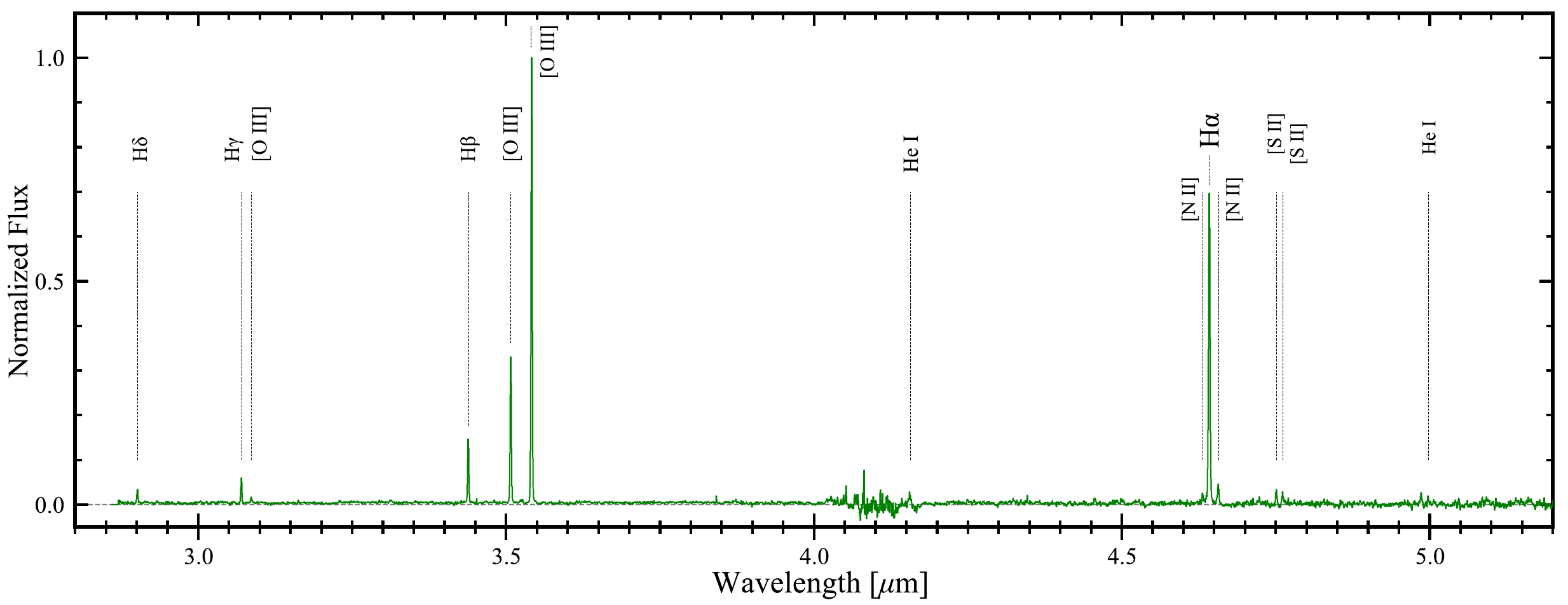}
\end{center}
\vspace{-0.6cm}
\caption{\small 
{\small 
\textbf{NIRSpec G395H/F290LP 1D spectrum of \targ.}
The spectrum is extracted from the IFU data cube with a $0\farcs7$-radius aperture. 
The emission lines detected at $>5\sigma$ and consistent with the \cii\ line redshift are highlighted with the label. 
}
}
\label{fig:ifu_spectrum}
\end{figure*}

We reduced the NIRSpec/IFU raw data using the STScI pipeline (version 1.12.5) with the CRDS context \texttt{jwst\_1039.pmap}, following the procedure developed by the ERS TEMPLATES team (PI: Rigby, Co-PI: Vieira; PID 01355; Rigby et al.\cite{rigby2023}). A detailed description is provided in Rigby et al.\cite{rigby2023} (also described in \cite{welch2023, birkin2023}) with associated data reduction code publicly available\footnote{
https://github.com/JWST-Templates/Notebooks/blob/main/nirspec\_ifu\_cookbook.ipynb}. 
Here, we briefly explain the reduction and calibration procedure.
We processed the raw data through the general three stages with default parameters and added the following steps. 
After Stage~1, we also applied the \texttt{NSClean} package\cite{rauscher2023} to the Level1 products, which mitigates the systematic vertical pattern noise and snowballs. 
We confirmed that the remaining vertical pattern noise and snowballs are less in the products processed with \texttt{NSClean}. 
In Stage~2, we found that the pipeline background subtraction did not work properly due to the different pixel units between the on-source and background data, and thus, we did not apply it inside the pipeline. Instead, we generated the reduced on-source and background data cubes based on the same procedure, calculated the median count in each channel in the background cube, and performed the subtraction in all channels after all steps. 
We also confirmed that our background estimate in our target field from the background data is consistent with the \jwst\ background tool prediction within $\sim$5--20\% over the 2.8--5.3$\mu$m range. 
In Stage~3, although several previous NIRSpec IFU studies\cite{cresci2023, marshall2023, perna2023} have reported that this step can lead to false outlier detections corresponding to bright sources in dithered exposures, we find that spurious outlier detections do not occur in our data, likely because our targets are much fainter than those in previous studies such as the luminous quasars\cite{vanzella2023b}.
We thus ran the outlier detection step with the default parameters. 
After completing these procedures, we obtain the final calibrated data cube with the default spatial scale of $0\farcs1$ for each spaxel. 
Note that we find in the data cube that the structures of the bright emission lines are always shifted by $\sim0\farcs1$ systematically in the same direction. We speculate that this is due to an uncertainty of the astrometry and correct it by using the Southwest bright object as a reference, whose continuum is also detected within the IFU FoV. 
We also determine the PSF size of the NIRSpec IFU by analyzing a nearby bright object. By fitting a 2D Gaussian profile using \texttt{GALFIT}\cite{peng2010} to the NIRCam/F444W filter map, we measure an intrinsic source size of FWHM$=0\farcs06$. From a continuum map generated by collapsing the NIRSpec IFU data cube channels between 3.5--4.5~$\mu$m, we observe an FWHM size of $0\farcs23$ for this object. Consequently, we estimate the PSF size to be $0\farcs22$, based on Gaussian convolution assumptions.
We note that another NIRSpec IFU study reports a slightly smaller PSF size ($\simeq0\farcs18$) at similar wavelengths, based on a serendipitous star within the IFU field of view\cite{deugenio2024}. We confirm that this potential PSF uncertainty has a negligible impact on our dynamical measurements with the NIRSpec IFU, with changes in the results remaining within 10\%.
Regarding the line spread function (LSF), we assume spectral resolutions of $R=2400$ at $\sim3.5\mu$m (\oiii) and $R=3100$ at $\sim4.6\mu$m (H$\alpha$), following the \jwst\ Users Guide
\footnote{\url{https://jwst-docs.stsci.edu/jwst-near-infrared-spectrograph/nirspec-instrumentation/nirspec-dispersers-and-filters\#gsc.tab=0}}. These nominal values have been validated through analyses of the JWST calibration program, showing consistency within $\sim10$--20\% \cite{isobe2023}. We further confirm that potential uncertainties in the LSF have a minimal impact on our NIRSpec IFU dynamical results, with changes in the derived quantities in the following analyses being $<5\%$.

In Extended Data Fig.~\ref{fig:ifu_spectrum}, we show the 1D spectrum of NIRSpec IFU G395H/F290LP extracted with a $0\farcs7$-radius aperture for \targc. In addition to the bright optical emission lines of H$\alpha$, \oiii$\lambda$5008, and H$\beta$, further faint emission lines of H$\delta$, H$\gamma$, \oiii$\lambda$4363, HeI$\lambda$5877, \nii, and \sii\ are successfully detected at $\gtrsim5\sigma$ that are all consistent with the redshift determined by the \cii\ emission line. 
We measured the line fluxes by fitting a Gaussian profile with a fixed line center based on the source redshift. Several neighboring emission lines were fitted simultaneously by assuming their line widths were the same. 
We summarize our line flux measurements in Extended Data Table~2. 
While we show spatially-resolved H$\alpha$/H$\beta$ line measurements in Section~8, full spatially-resolved line analyses and results will be presented in a separate paper\cite{fujimoto2024prep},

\subsection{4. Lens model: } 

Several lens models were previously constructed in \cluster\ using the \hst\ maps taken in the RELICS program \cite{coe2019} and an 0.8-hour VLT/MUSE IFU dataset (\#0100.A-0792, P.I.; A.~Edge)\cite{fujimoto2021, laporte2021}. 
Since then, in addition to the \jwst/NIRCam observations described above, a new 8.9 hour MUSE observation was obtained (\#109.22VV.001, P.I.: S.~Fujimoto). 
Furthermore, new Keck/DEIMOS and Gemini/GMOS observations have also been carried out to confirm the presence of additional massive cluster structures in the northeast, which was hinted at in the previous \hst\ data. These data improve the identification of cluster member galaxies and the multiply lensed images in \cluster, and thus, we use an updated lens model in this study. The details of the new MUSE, DEIMOS, and GMOS data and the construction of the updated lens model are presented in separate papers\cite{furtak2024}, while here we briefly describe the basic procedure of the lens model. 

The model used in this work was constructed using a revised version of the parametric strong lensing (SL) pipeline by Zitrin et al.\cite{zitrin2015}, which was successfully used for modeling several clusters imaged with JWST\cite{pascale2022,furtak2023a}. The smooth dark matter (DM) component of the cluster was modeled by four pseudo-isothermal elliptical mass distributions\cite{jaffe1983} (PIEMDs). This approach effectively captures the elliptical symmetry and core structures of the DM halos. The individual cluster member galaxies were modeled using dual pseudo-isothermal ellipsoids\cite{eliasdottir2007} (dPIEs), accounting for the galaxy profiles' core and truncation radii. The SL model comprises a total of 504 cluster member galaxies, 67 of which have spectroscopic redshifts from the MUSE, DEIMOS or GMOS observations. The final SL model was constrained using a total of 91 multiple images and candidates, belonging to 35 distinct sources. Among these, 31 sources with a total of 83 images have spectroscopic redshifts, providing a robust basis for the lensing analysis. The model was then optimized through a series of Monte-Carlo Markov Chains (MCMCs) of several $10^4$ steps in total. The final SL model of the cluster has an average lens plane image reproduction error of $\Delta_{\mathrm{RMS}}=0.72\arcsec$ which corresponds to reduced $\chi^2\simeq4.2$. 
In particular, the five images of the object studied in this work were exceptionally well reproduced with an average image reproduction error of $\Delta_{\mathrm{RMS}}=0.12\arcsec$ and the flux ratios consistent within $\sim$10\% of the best-fit model. 
Although we do not explicitly include the systematic uncertainty from different lens models in this paper, a previous study of \targ\cite{fujimoto2021} constructed three independent models, allowing us to comment on the potential systematic uncertainty. Combining the magnification estimate from our fiducial model with the three previous models, the standard deviation among these four estimates is calculated to be $\sim$25\%, reflecting the potential systematic uncertainty.
This relatively modest uncertainty, despite the large magnification factor (cf. typically $\gtrsim100\%$ for systems with $\mu>$10--30\cite{zitrin2015,bouwens2017,fujimoto2023b}), is attributed to the spectroscopic confirmation of the multiple images. The positions, distortions, and flux ratios of these images provide exceptional constraints on the lens model, regardless of the specific methodology used for its construction.
Moreover, the lens model in this study significantly benefits from newly available data, including the spectroscopic confirmation of the fifth multiple image and deeper MUSE and NIRCam observations. If these additional constraints were incorporated into the previous three models, their magnification estimates would likely converge closer to the current model, further reducing the systematic uncertainty for this specific system. Therefore, we conclude that the magnification uncertainty in this study does not exceed $\sim$25\%, even when accounting for potential systematic uncertainties across different models. 
The details of the lens model and the discovery of the extended structures of MACS0600 are presented in Furtak et al. (2024)\cite{furtak2024}. 

Throughout this paper, we mainly employ this latest model, while we use the previous model constructed with \texttt{GLAFIC}\cite{fujimoto2021} for the source-plane reconstruction analysis, which incorporates calculations on a higher-resolution grid, including perturbations from the object nearby \targ, making it better suited for the purposes. It is noteworthy that the \texttt{GLAFIC} model magnification and distortion estimates align with those of the latest one within a 10\% accuracy, and thus the impact on our results is minimal. 

\subsection{5. SED analysis: }
\label{sec:sed}

Taking advantage of the superb sensitivity and spatial resolution of \jwst/NIRCam, we conduct the SED fitting in a spatially-resolved manner, 
which reduces the so-called ``outshining'' effect\cite{papovich2001, maraston2010} due to UV-luminous young stellar populations formed in recent bursts\cite{clara2023}.
Note that the PSF size is much worse in NIRSpec/IFU ($\sim0\farcs22$; Section~3) than those in NIRCam ($\sim0\farcs04$--$0\farcs14$\footnote{\url{https://jwst-docs.stsci.edu/jwst-near-infrared-camera/nircam-performance/nircam-point-spread-functions}}), and thus we adopt the outputs from the spatially-resolved SED analyses for characterizing the basic physical properties (SFR, $M_{\star}$, age) of \targ\ and clumps inside in this paper. 
We basically follow the pixel-by-pixel method presented in Gim\'enez-Arteaga et al. (2023)\cite{clara2023}. 
The details of the method and the results are presented in Gim\'enez-Arteaga et al. (2024)\cite{clara2024}, while here we briefly provide the analysis procedure. 
Again, we focus on ID3 in this paper to study the SED properties of \targ\ from the global scale of the galaxy and the individual star-forming clumps inside, and the SED fitting results in the other multiple images are further presented in a separate paper\cite{fujimoto2024prep}. 
Note that we do not include the ALMA measurements in the SED fitting in this study because of the lower-spatial resolution of that data (beam FWHM$=0\farcs28\times0\farcs25$) compared to NIRCam. 

We first matched PSFs in all NIRCam images to the one in F444W using the PSF models generated with \texttt{WebbPSF}\cite{perrin2012, perrin2014} and Gaussian kernel. 
We then resized the PSF-matched images to a common pixel scale of 40~mas per pixel. 
We use Agglomerative Clustering (\texttt{sklearn.cluster}) with a `single' linkage method and a 1.5 distance threshold in the pixel selection for the SED fitting. 
After initial S/N~$>$~1 thresholding in all bands and subsequent masking, a secondary threshold was applied based on the mean S/N per band. The final pixel selection was determined by combining these thresholds, resulting in 625 pixels satisfying the S/N criteria across all bands, ensuring a $\sim2\sigma$ detection minimum. The following analysis was performed only for these selected pixels. 

We ran the SED fitting code of \textsc{Bagpipes}\cite{carnall2018} on the photometry obtained in each selected pixel, fixing the redshift to $z=6.072$. The ionisation parameter was varied ($-3 < \log_{10} U < -1$), incorporating nebular emission with \textsc{Cloudy} \cite{ferland2017} and stellar population synthesis (SPS) models by Bruzual \& Charlot\cite{bruzual2003}. We assumed a Calzetti\cite{calzetti2000} attenuation curve and a constant star-formation history (SFH). We adhered to a \cite{kroupa2001} IMF, setting $t_{\rm bc}=10$~Myr, and established uniform priors for $A_V\in[0,3]$ and $\log_{10}(M_*/M_\odot) \in [5,11]$. A Gaussian prior centered at 0.1 solar metallicity $Z_{\odot}$ (with $\sigma = 0.2~Z_{\odot}$) was used for metallicity.  

In Extended Data Fig.~\ref{fig:nircam_cutout}, we show the SED outputs of stellar mass $M_{\star}$, star-formation rate (SFR) averaged over the past 100~Myr, and age of the stellar populations weighted by mass ($t_{\rm age}$). 
We find that young ($t_{\rm age}<$10~Myr) stellar populations are formed in the central regions surrounded by older stellar populations ($t_{\rm age}>$100~Myr). 
We confirm that predicted line fluxes and spatial distributions of H$\alpha$ and \oiii$\lambda$5008 from our NIRCam-based SED fitting remain in excellent agreement with the observed values obtained from our NIRSpec IFU, assuring our SED results\cite{clara2024}. 
Adopting a pixel-by-pixel lens correction, we obtain the total $M_{\star}$ and SFR values of $2.6^{+1.7}_{-1.5}$\,$M_{\odot}$~yr$^{-1}$ and $M_{\star}=4.5^{+2.7}_{-1.1}\times10^{8}\,M_{\odot}$, respectively. 

\subsection{6. Global-scale galaxy properties of \targbf: }
\label{sec:host_prop}

\stepcounter{extfigure} 
\begin{figure*}[t!]
\begin{center}
\includegraphics[angle=0,width=1.0\textwidth]{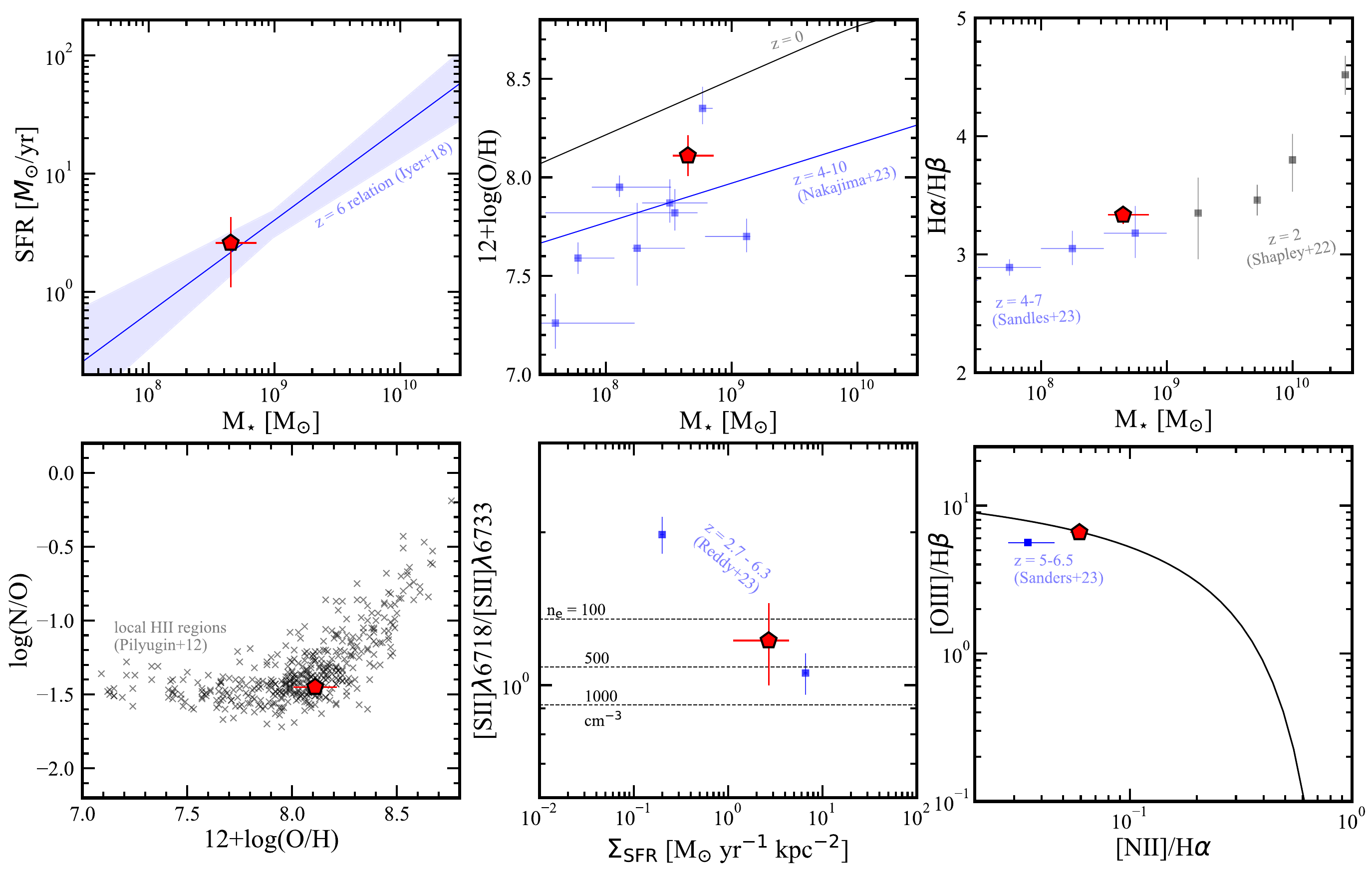}
\end{center}
\vspace{-0.8cm}
\caption{\small 
{\small 
\textbf{Global-scale physical properties of \targbf}.
\textbf{\emph{Top:}} Relations of SFR, metallicity, and Balmer decrement as a function of $M_{\star}$, from left to right. 
The blue and black symbols, lines, and shades present the typical relations estimated in the literature\cite{iyer2018, nakajima2023, sandles2023, shapley2022}. 
\targ\ falls well within the scatter or $1\sigma$ errors of these typical relations at similar redshifts. 
\textbf{\emph{Bottom:}}
Relations of N/O and O/H abundances, line ratio of the [S{\sc ii}] doublet ($\approx$electron density $n_{\rm e}$) and $\Sigma_{\rm SFR}$, and the BPT diagram from left to right. 
The black and blue symbols denote the typical relations estimated in the literature\cite{pilyugin2012, reddy2023, sanders2023}. 
The black dashed horizontal lines in the middle panel are obtained from the \texttt{PyNeb} package
The black curve in the right panel is the diagnostical line between star-forming and AGN populations, taken from the literature\cite{kauffmann2003}
\targ\ falls in the regimes generally consistent within the scatter or the $\sim$1-2$\sigma$ errors of the previous measurements. 
These results suggest that \targ\ represents a typical star-forming galaxy at $z=6$. 
}
}
\label{fig:host_prop}
\end{figure*}

In Extended Data Fig.~\ref{fig:host_prop}, we show the global galaxy scale SFR-$M_{\star}$ relation of \targ. We find that \targ\ falls on the main sequence at $z=6$\cite{iyer2018}, indicating that \targ\ represents a typical star-forming galaxy.  

Using the optical line fluxes measured with NIRSpec IFU, we further evaluate the Balmer decrement, electron density $n_{\rm e}$, oxygen abundance of 12+$\log$(O/H), nitrogen abundance of $\log$(N/O), and BPT diagnostic for \targ\ on the global galaxy scale. 
For the Balmer decrement, we use H$\alpha$/H$\beta$ and obtain the line ratio of H$\alpha$/H$\beta = 3.33\pm0.07$. Assuming an intrinsic ratio of 2.86 under the case B recombination\cite{osterbrock1989}, we estimate a dust reddening of $E(B-V)=0.03\pm0.01$. We apply the dust correction based on this $E(B-V)$ value in the following analyses. 
For $n_{\rm e}$, we use the \sii$\lambda\lambda$6718,6733 doublet. 
Based on the line ratio of \sii$\lambda$6718/\sii$\lambda6733=1.22\pm0.20$, 
we obtain $n_{\rm e}=260^{+400}_{-230}$ using the nebular emission code of \texttt{PyNeb}\footnote{http://research.iac.es/proyecto/PyNeb/}. 
For 12+$\log$(O/H), we use the direct electron temperature $T_{\rm e}$ method with the auroral \oiii$\lambda$4364 line. Based on the line ratio of \oiii$\lambda$4363/\oiii$\lambda$5008, we obtain $T_{\rm e}$ of $14,100\pm600$~K using \texttt{PyNeb}. 
We then estimate the oxygen abundance, following the equations in Izotov et al. (2006)\cite{izotov2006}, where we ignore oxygen ions of O$^{3+}$ due to their exceptionally high ionization potential (54.9~eV). Because our NIRSpec G395H/F290LP data does not cover the wavelength of the \oii$\lambda$3727 line from \targ, we infer the \oii\ line flux from the tight correlation between $R$ ($\equiv I_{\rm [OIII]\lambda4363}$/$I_{\rm [OIII]\lambda5008}$) and $R_{23}$ ($\equiv (I_{\rm [OII]\lambda\lambda3727,3729}+I_{\rm [OIII]\lambda4960,5008})$/$I_{\rm H\beta}$), calibrated in Pilyugin \& Thuan (2005)\cite{pilyugin2005} as 
$\log R = -4.264 + 3.087\log R_{23}$. 
Because auroral lines of \oii$\lambda\lambda$7320,7330 are not observed in our data, we also assume
$t_{2} = 0.7t_{3} + 0.3$,\cite{campbell1986, garnett1992} 
where $t_{3}=10^{-4}T_{\rm e}$(\oiii) and $t_{2}=10^{-4}T_{\rm e}$(\oii). 
We obtain 12+$\log$(O/H) = $8.11\pm0.03$. 
We confirm the general consistency between our $T_{\rm e}$-based measurement and those from other strong line methods with R3 ($\equiv I_{\rm [OIII]\lambda5008}$/$I_{\rm H\beta}$), N2 ($\equiv I_{\rm [NII]\lambda6585}$/$I_{\rm H\alpha}$), O3N2 ($\equiv$ R3/N2), S2 ($\equiv I_{\rm [SII]\lambda\lambda6718,6733}$/$I_{\rm H\alpha}$), and O3S2 ($\equiv$ R3/S2) calibrated in the literature\cite{curti2020b, nakajima2022} (Extended Data Fig.~\ref{fig:comp_metallicity}).
For $\log$(N/O), we follow the equations in Izotov et al. (2006)\cite{izotov2006}, assuming N/O $\simeq$ N$^{+}$/O$^{+}$, as their ionization potentials are almost the same. Here, we use the \oii\ line flux inferred above. We obtain $\log$(N/O)$=-1.46\pm0.04$. 
We summarize these estimates in Extended Data Table~3. 

\stepcounter{extfigure} 
\begin{figure*}[t]
\begin{center}
\includegraphics[angle=0,width=0.5\textwidth]{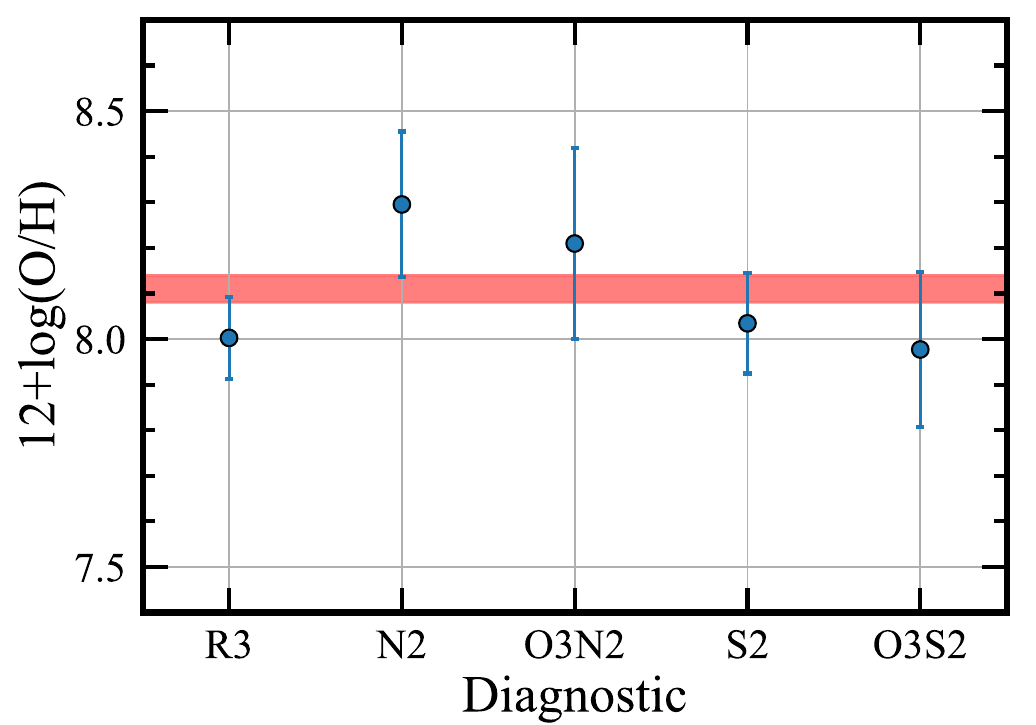}
\end{center}
\vspace{-0.8cm}
\caption{\small 
{\small 
\textbf{Comparison of metallicity measurements for \targbf\ with different calibrations}. 
The red horizontal line indicates our fiducial estimate based on the $T_{\rm e}$ method (see \method). 
The blue circles are estimated from the strong line methods based on calibrations in the literature\cite{curti2020b}, showing the general consistency among the calibrations. 
}
}
\label{fig:comp_metallicity}
\end{figure*}

In Extended Data Fig.~\ref{fig:host_prop}, we summarize the physical properties of \targ\ evaluated on the global galaxy scale described above. For comparison, we also show previous measurements among high-redshift galaxies in the literature, including recent \jwst\ results when available. We find that the chemical enrichment, dust attenuation, and nebular conditions (density, ionization state) in \targ\ agree well with the typical relations measured in similarly high-redshift galaxies, indicating again that \targ\ represents a typical star-forming galaxy at $z=6$. 
In the $\log$(N/O)--12+$\log$(O/H) relation, \targ\ is in excellent agreement with the measurements obtained in local \hii\ regions. Although recent \jwst\ studies report the presence of significantly nitrogen-abundant systems with $\log$(N/O)$\gtrsim-0.5$, hinting at their possible connection to Globular cluster and intermediate blackhole formations\cite{bunker2023,marques-chaves2024, isobe2023b, topping2024}, our results indicate that it is not the case in \targ. 
In the BPT diagram, \targ\ falls on the border between star-forming and AGN activities\cite{kauffmann2003}. However, the location remains consistent within the 1-2$\sigma$ range with the recent \jwst\ measurements for galaxies at similar redshifts\cite{sanders2023}. Besides, the difficulty in using the BPT diagram to distinguish the star formation and AGN in metal-poor systems with the gas-phase metallicity of $Z_{\rm gas}\lesssim0.2Z_{\odot}$ has been argued and become clear at $z>5$,\cite{ubler2023} which is likely the case in \targ. In the NIRSpec IFU spectrum, we find that the line profiles of the Balmer emission lines and \oiii$\lambda$5008 are consistent. 
We thus conclude that the current data does not show any clear evidence of the presence of AGN.

\stepcounter{extfigure} 
\begin{figure*}
\begin{center}
\vspace{-1.0cm}
\includegraphics[angle=0,width=0.95\textwidth]{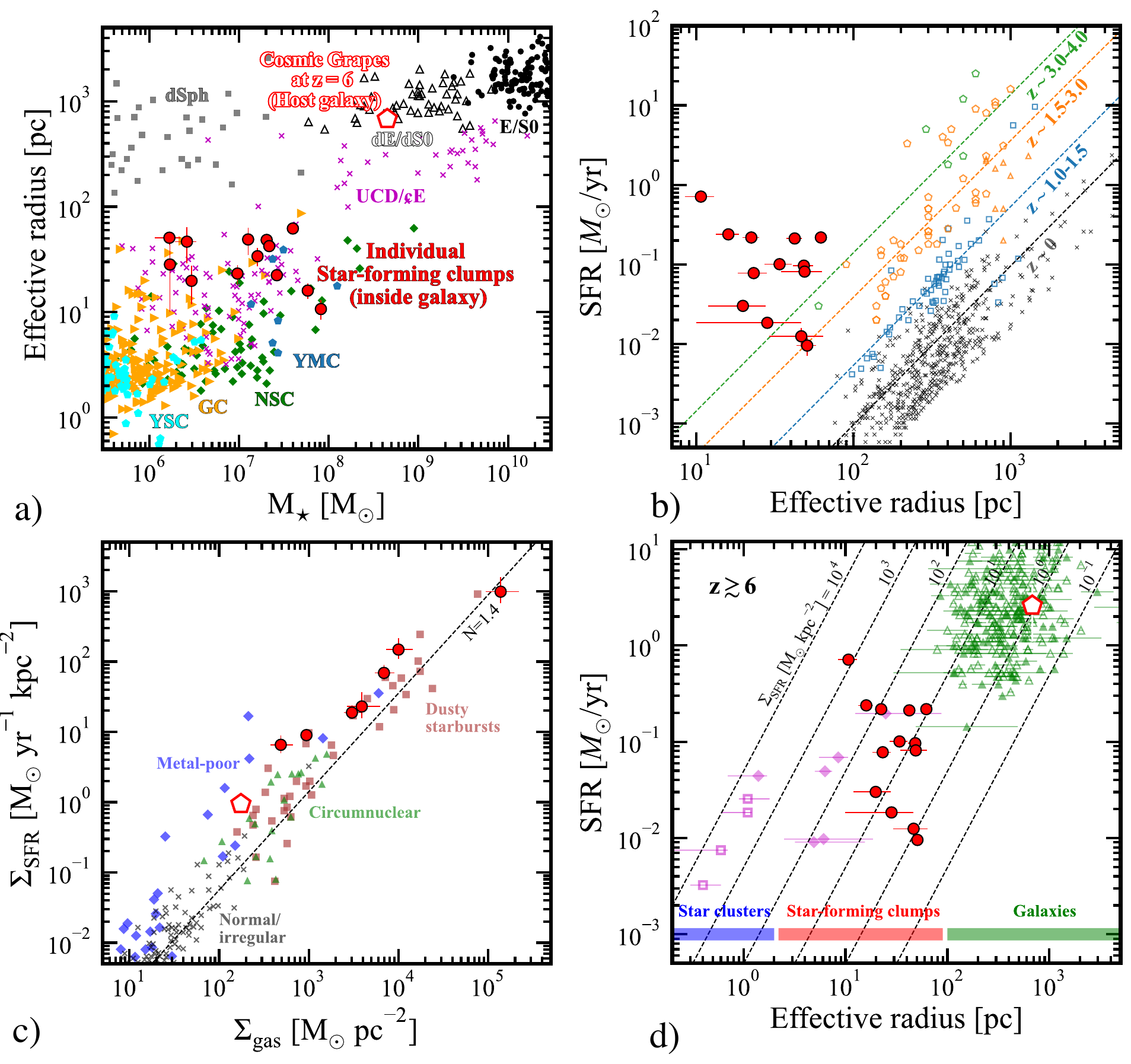}
\end{center}
\vspace{-0.8cm}
\caption{\small 
{\small 
\textbf{Physical properties of the individual star-forming clumps identified in \targ}. 
\textbf{\emph{(a):}} Size--$M_{\star}$ relation. 
The red circles represent the star-forming clumps, while the open red pentagon shows the measurement as the entire galaxy.
For comparison, the other symbols indicate local objects of elliptical/S0 galaxies (E/S0; black circles), dwarf elliptical/S0 galaxies (dEs/dS0; open black triangles), dwarf spheroids (dSph; grey squares), ultra-compact dwarfs/compact elliptical galaxies (UCD/cE; magenta crosses), young massive star clusters (YMC; blue pentagon), nuclear star clusters (NSCs; green diamonds), globular clusters (GCs; orange triangles), young star clusters (YSC; cyan pentagon) taken from the compilation in the literature\cite{norris2014}. 
\textbf{\emph{(b):}} Size--SFR relation. 
For comparison, the other symbols represent lower-redshift star-forming clumps at $z\sim0$ (black crosses), $z\sim1.0-1.5$ (blue open squares), $z\sim1.5-3.0$ (orange open pentagons), and $z\sim3.0-4.0$ (green open pentagons), with the best-fit power-law functions (dashed lines) estimated in the literature\cite{livermore2015}. 
\textbf{\emph{(c):}} $\Sigma_{\rm SFR}$--$\Sigma_{\rm gas}$ relation. The color symbols indicate different classes of local star-forming galaxies compiled in the literature\cite{kennicutt2012}. 
The dashed line presents the Kennicutt-Schmidt (KS) law with the power-law slope of 1.4\cite{kennicutt2012}. 
\textbf{\emph{d:}} Same as $b$, but comparing with other $z\gtrsim6$ objects recently observed with NIRCam. The green triangles represent the field galaxies at $z\sim6$--10\cite{morishita2023}, and the magenta diamonds\cite{vanzella2023} and squares\cite{adamo2024} show the individual clumps identified in strongly lensed arc systems. The filled and open symbols indicate spectroscopic and photometric samples, respectively. The horizontal labels denote the typical physical scales of star clusters, star-forming clumps, and galaxies, where \targ\ and its star-forming clumps uniquely bridge the relation between the galaxy and the star-forming clumps inside.  
}
}
\label{fig:clump_prop}
\end{figure*}

\subsection{7. Properties of the individual star-forming clumps inside \targbf: }
\label{sec:clump_prop}

We use \texttt{SExtractor} version 2.25.0\cite{bertin1996} to quantify the number of individual star-forming clumps that appear in the \jwst/NIRCam maps (Extended Data Fig.~\ref{fig:nircam_cutout}). Here, we adopt the NIRCam/F150W map, which provides one of the highest spatial resolutions (PSF FWHM $\simeq 0\farcs05$) and the best sensitivity owing to the longest exposure among our data. 
We run \texttt{SExtractor} using the default parameter values\footnote{https://github.com/astromatic/sextractor/blob/master/tests/default.sex}, except for DETECT\_MINAREA = 9, DETECT\_THRESH = 2.0, DEBLEND\_NTHRESH = 64, and DEBLEND\_MINCONT = 0.0001 that are optimized to identify the small individual star-forming clumps separately. 
In the right panel of Fig.~\ref{fig:hst-nircam}, we show the segmentation map obtained from the above parameter set. 
The above parameter set identifies 20 segmented areas, 
although visual inspection shows that some of these clumps are likely too faint or diffuse to be considered secure detections. 
Therefore, we have refined our selection to 15 clumps that visually exhibited clear luminosity concentrations among the 20 segmented areas to ensure robust results.
These clumps are labeled from ID1 to ID15 in the right panel of Fig.~\ref{fig:hst-nircam}. 
For these 15 clumps, we calculate the integrated flux and the aperture-based photometry uncertainty according to each segmented area size. The significance of the detections ranges from S/N = 17 (ID5) to 355 (ID1), confirming the robust identification of individual star-forming clumps.
Despite a different methodology adopted for clump identification, visual inspections already tell a stark difference from the number of clumps identified in lower redshift galaxies in recent JWST/NIRCam studies ($N<5$)\cite{kalita2024}, which is likely because of a possible redshift evolution of the morphology\cite{shibuya2016} and the higher spatial resolution achieved in this study aided by the gravitational lensing effect. 
Numerous clump identifications ($N=10$) have also been reported in a strongly lensed arc system at $z=8.3$\cite{mowla2024}. 
Given the remarkably low likelihood of capturing extraordinary phenomena through strong lensing twice, these numerous clumpy structures observed by the JWST, in conjunction with lensing effects, might be a common feature in early galaxies. In this paper, we regard $N=15$ as the count of the individual star-forming clumps in \targ, while this number could be increased with further high-resolution observations in the future. 

Using our pixel-by-pixel SED outputs and the \cii\ line intensity map, we characterize the physical properties of $M_{\star}$, SFR, and gas mass $M_{\rm gas}$ for these individual star-forming clumps. 
We sum the pixel-based output values for $M_{\star}$ and SFR, following the segmentation presented in Fig.~\ref{fig:hst-nircam}. 
We note that the pixels of star-forming clump-13 did not pass the pixel mask procedure in our SED fitting, so characterizations are performed for all of the clumps except that one. 
For $M_{\rm gas}$, we first convert the \cii\ luminosity $L_{\rm [CII]}$ to $M_{\rm gas}$ pixel-by-pixel using the conversion factor calibrated in the literature\cite{zanella2018}, and sum the values in the same manner as above. We discuss potential uncertainties in this conversion in Section~10. Note that because the ALMA beam size in the \cii\ map ($0\farcs28\times0\farcs25$) is larger than those in NIRCam, the $M_{\rm gas}$ estimate is only performed for star-forming clump-1, 3, 4, 6, 7, 8, and 9 whose segmentation areas are larger than the ALMA beam size.
We summarize the $M_{\star}$, SFR, and $M_{\rm gas}$ values for each star-forming clump in Extended Data Table~4. 
The posterior distributions of these SED outputs for each clump are also summarized in Extended Data Fig.~\ref{fig:clump_pdf}. 
We also derive $M_{\rm gas}$ in the global galaxy scale in the same manner, resulting in $1.1^{+0.4}_{-0.1}\times10^{9}\,M_{\odot}$, which is listed in Extended Data Table~3. 

The spatial sizes of the individual star-forming clumps are also constrained by S\'ersic profile fitting to the NIRCam/F150W map using the \texttt{GALFIT} software\cite{peng2010}. 
In the fitting, we adopt 16 S\'ersic profiles for the 15 individual star-forming clumps and an underlying diffuse stellar disk component. 
While the surface density profile of star clusters has traditionally been fitted with the King model\cite{king1966}, we find that a S\'ersic profile with $n\sim0.8$ closely approximates the King model out to several core radii. For simplicity and to ensure stable results, we fix the S\'ersic index to $n=1.0$ for all components except the brightest star-forming clump (clump-1).
Other parameters are used as free parameters with initial values of $r_{\rm e}=1$~pix (pix $=0\farcs02=116$~pc), axis ratio $q = 0.9$, position angle PA = 0~deg, spatial position $(x, y)$ taken from the center of the segmentation area for each clump, and magnitude calculated by summing the pixel fluxes within the segmentation area for each clump. 
For the underlying disk component, we adopt the luminosity-weighted center and 50\% of the total flux of \targ\ for the initial values of the spatial position and the magnitude in the fitting. 
We also set the boundary limits of $x = [-3:3]$~pix, $y = [-3:3]$~pix (from the initial values), $r_{\rm e} = [0.1:10.0]$~pix, and $q=[0.1:1.0]$ for the 15 star-forming clumps, and $n = [0.5:5]$, $r_{\rm e} = [10:60]$~pix, and $q = [0.3:0.8]$ for the underlying disk component. 
To ensure a balanced fit of both compact and diffuse components, the area defined by the segmentation map is expanded by applying smoothing to the outer edge, extending the fitting mask by approximately $0\farcs2$. All 16 components are then fitted simultaneously within this adjusted region.
In Fig.~\ref{fig:hst-nircam}, we show the best-fit model and the residual maps. 
We confirm that the individual star-forming clumps are reasonably modeled. 
Based on the best-fit models, the sum of the fluxes from the 15 clumps accounts for 69.9\% of the total flux of \targ\ in F150W, indicating that these clumps are the dominant luminosity component of the galaxy. 
We summarize the best-fit S\'ersic profiles for each star-forming clump in Extended Data Table~4. 
Note that we obtain $n=2.0\pm1.2$ from the brightest clump (clump-1). If we adopt $n=2.0$ for all S\`ersic components in the fitting, the resulting output sizes are consistent with our original results with the $n=1$ assumption within $\sim50\%$. This comparison acknowledges the potential uncertainty introduced by our choice of S\'ersic index $n$, while it would not change our size measurements beyond a factor of 2. 
Another note is that although the emission from \targ\ observed in F150W corresponds to the rest-frame UV continuum, recent \jwst/NIRCam studies show that spatial sizes in rest-UV and optical are comparable\cite{morishita2023}. We thus refer to the rest-frame UV-based size measurement as the size throughout this paper.

In Extended Data Fig.~\ref{fig:clump_prop}, we summarize the size--$M_{\star}$, size--SFR, and surface gas density $\Sigma_{\rm gas}$ -- surface SFR density $\Sigma_{\rm SFR}$ relations. In the calculation of $\Sigma_{\rm gas}$, we assume that the size of the gas clouds is two times larger than that of the rest-UV emitting region based on previous studies suggesting $r_{\rm e}$ (\cii) 
 $\simeq2\times r_{\rm e}$ (UV) in high-redshift galaxies\cite{fujimoto2020b}. 
In panels of $a$ and $b$, we find that our numerous star-forming clumps identified in \targ\ fall in the same parameter space as young massive clusters (YMCs) in the local galaxies on the size--$M_{\star}$ plane, while their $\Sigma_{\rm SFR}$ is much higher than that of the star-forming clumps at lower redshifts. Such an increasing trend of $\Sigma_{\rm SFR}$ in star-forming clumps towards high redshifts is consistent with recent studies for high-redshift lensed systems\cite{livermore2012, livermore2015,welch2023a,vanzella2023, adamo2024}. 
In panel $c$, we find that our star-forming clumps have very high $\Sigma_{\rm gas}$ and $\Sigma_{\rm SFR}$ values comparable to dusty starburst systems, following the local Kennicutt-Schmidt law with the power-law slope of 1.4\cite{kennicutt2012}. 
The panel $d$ is the same as the panel $b$, but comparing the size--SFR relation with those measured in recent NIRCam-observed objects at $z\gtrsim6$\cite{morishita2023, vanzella2023, adamo2024}. 
This comparison shows that our star-forming clumps are $\sim$1--2 orders magnitude more compact than galaxies (green triangles\cite{morishita2023}), while $\sim$1--2 orders magnitude larger than the compact star clusters (SCs) recently identified in the strongly lensed arc system at $z\simeq10.2$\cite{adamo2024}. 
Although spectroscopic confirmation has not yet been made for the lensed arc system at $z\simeq10.2$, this indicates that we are witnessing the three important regimes in early galaxies from our and recent NIRCam+lensing studies: global galaxy scales, star-forming clumps, and SCs. In this context, \targ\ and its numerous star-forming clumps uniquely bridge the galaxy and SC layers, enabling us to study their roles in early galaxy formation and evolution through the multi-layered components. 
Examining the details of the internal structures while maintaining a comprehensive picture of the entire galaxy has been challenging in strongly distorted arc systems because of the difficulty in reconstructing the entire galaxy from the significant local magnification in a specific narrow region in the source plane. Therefore, \targ\ is offering us an invaluable laboratory at the epoch of reionization, not only because of its strong magnification ($\mu=\avmu$), but also because of its weak distortion and differential magnification.

\subsection{8. Dust reddening in \targbf: }
\label{sec:ebv}

\stepcounter{extfigure} 
\begin{figure*}
\begin{center}
\includegraphics[angle=0,width=0.95\textwidth]{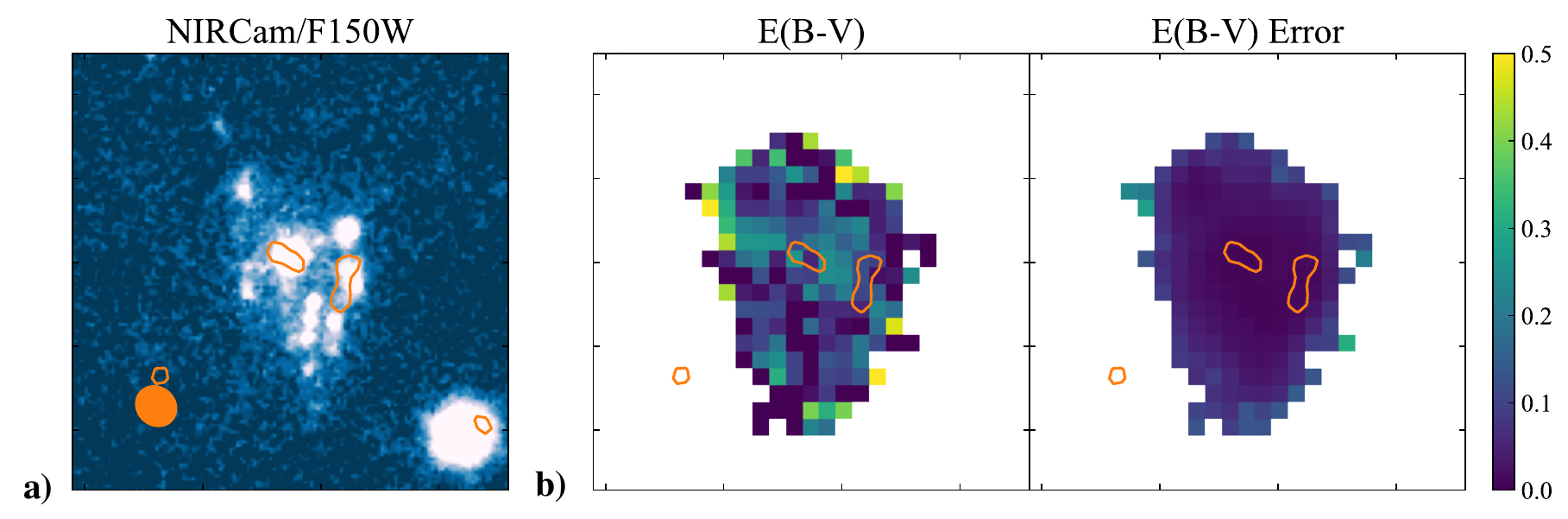}
\end{center}
\vspace{-0.8cm}
\caption{\small 
{\small
\textbf{Spatially-resolved views of modest dust reddening in \targbf.}
\textbf{\textit{(a):}} Overlaid dust continuum on the NIRCam/F150W image cutout ($2\farcs2 \times 2\farcs2$) in the image plane (i.e., before lens correction). The orange contour represents the 3$\sigma$ level dust continuum detected in a high-resolution ALMA map, utilizing only the C43-5 configuration. 
The orange ellipse at bottom left denotes the ALMA beam.  
A single component, recently detected in lower-resolution deep ALMA follow-up observations \cite{valentino2024} is resolved into two peaks that align with the bright clumps seen in F150W.
\textbf{\textit{(b):}} Pixel-by-pixel $E(B-V)$ and error estimates derived from H$\alpha$/H$\beta$ ratios observed with NIRSpec IFU, covering the same field as \textbf{\textit{(a)}}. 
 \Targ\ exhibits modest dust reddening, with $E(B-V) \lesssim 0.2$. The alignment of dust emission with the clump positions, along with the modest dust reddening, suggests that the distinctively clumpy morphology of \targ\ is not a result of dust obscuration.
}}
\label{fig:ebv}
\end{figure*}

While \targ\ represents a sub-$L^{\star}$ galaxy ($0.025L^{\star}$ of UV LF at $z=6$\cite{ono2018}), characterized by its youth ($0.26\,Z_{\odot}$), lower mass ($4.5\times10^{8}\,M_{\odot}$), and typical dust reddening ($E(B-V)=0.13$) for its mass range at $z=6$, the modest dust emission is successfully detected in deep ALMA follow-up observations aided by the strong lensing effect\cite{valentino2024}. To explore the influence of dust reddening on the observed clumpy morphology of \targ, we examine the spatial distributions of dust continuum and dust reddening via the Balmer decrement of H$\alpha$/H$\beta$.

In Extended Data Fig.\ref{fig:ebv}a, we overlay the high-resolution ALMA dust continuum map, produced using only the C43-5 configuration (Section 3), on the NIRCam/F150W map to pinpoint dust-emitting peak positions, relative to the rest-frame UV continuum seen in F150W. The dust continuum is resolved into two peaks above the $3\sigma$ level, coinciding with bright clumps in F150W. The absence of dust emission peaks at the gaps between the F150W clumps suggests that the clumpy morphology of \targ\ is unlikely due to the presence of heavily dust-obscured star-forming regions. 
Additionally, we analyze spatially-resolved H$\alpha$/H$\beta$ line ratios on a pixel-by-pixel basis using NIRSpec IFU data, considering only pixels where both lines are detected at $\geq3\sigma$.
In Extended Data Fig.\ref{fig:ebv}b, we present $E(B-V)$ and its error distributions, assuming a Calzetti\cite{calzetti2000} attenuation curve.
Apart from several pixels at the edges, \targ\ predominantly features modestly dust-reddened regions with $E(B-V)\lesssim0.2$, aligning with the $E(B-V)$$=0.13\pm0.02$ measured in the global galaxy scale (Section~6). The minimal variation within $E(B-V)\lesssim0.06$ further suggests that the gaps between the clumps do not always show slightly high $E(B-V)$ values across \targ.  
We, therefore, conclude that the remarkably clumpy morphology of \targ\ is not driven by dust reddening.

\subsection{9. Kinematic modeling: }
\label{sec:kin}

\stepcounter{extfigure} 
\begin{figure*}[t]
\begin{center}
\includegraphics[angle=0,width=0.9\textwidth]{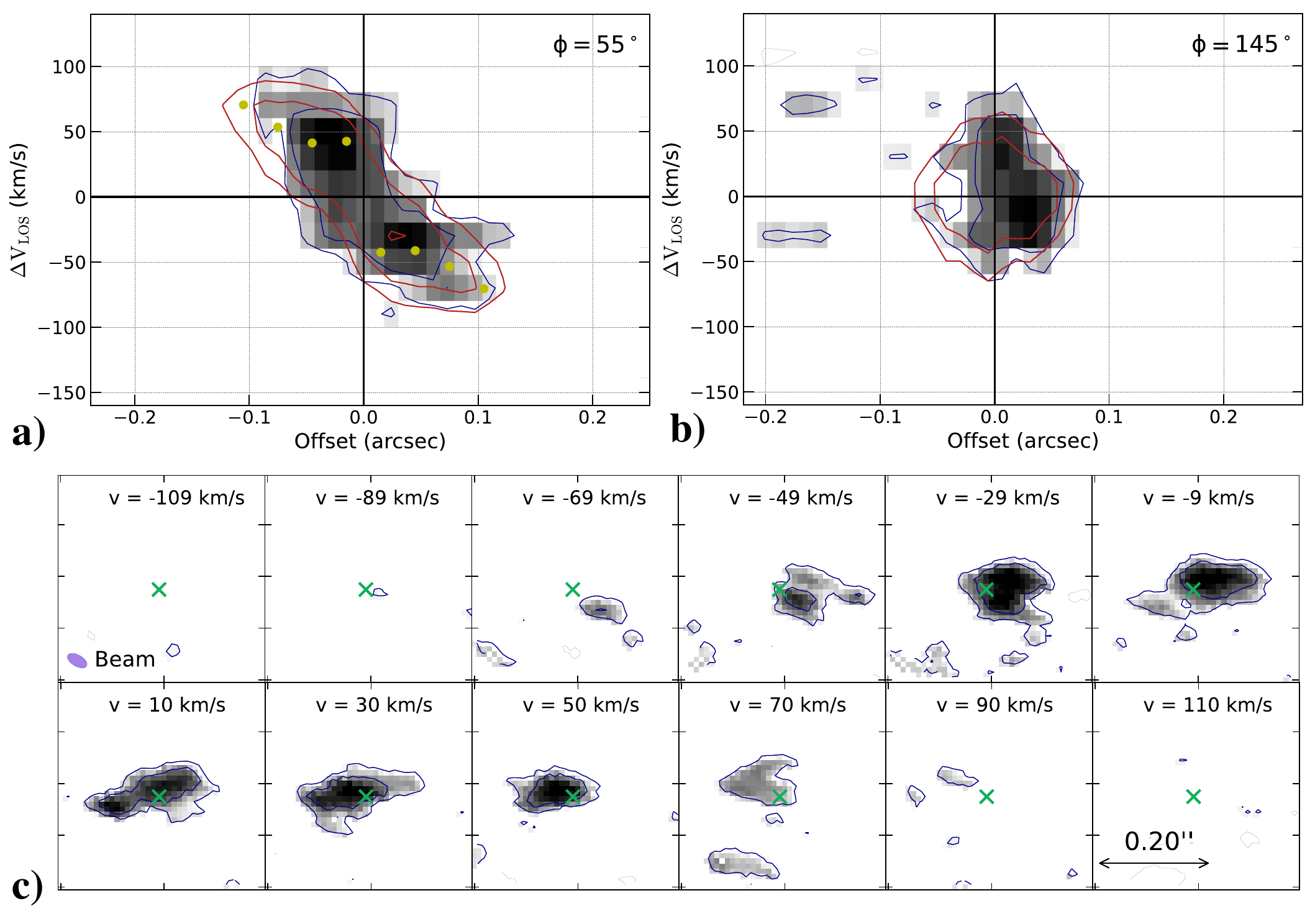}
\end{center}
\vspace{-0.8cm}
\caption{\small 
{\small 
\textbf{
Cool gas dynamics observed in \targbf\ via \cii\ 158$\mu$m line with ALMA. 
} 
These maps are obtained in the source plane (i.e., after the lens correction).
\textbf{\textit{(a):}} Position-velocity (PV) diagram along the major axis for the disk extracted using \texttt{3DBarolo}. \textbf{\textit{(b):}} Same as \textbf{\textit{a}}, but along with the minor axis for the disk. 
\textbf{\textit{(c):}} Channel maps for the \cii\ line. The green cross denotes the kinematic center, and the magenta ellipse at left bottom indicates the effective beam in the source plane (i.e., after lens correction). 
These maps show a symmetric 3D structure, indicating that \targ\ is a rotation-supported system\cite{rizzo2022}, rather than an ongoing merging system.     
}}
\label{fig:pvd}
\end{figure*}

We analyze the kinematic properties of \cii\ 158$\mu$m, H$\alpha$, and \oiii$\lambda$5008 observed with ALMA and \jwst/NIRSpec IFU. We perform the following analyses using 3D data cubes in the source plane, reconstructed in a pixel-by-pixel approach.
First, we extracted radial rotation velocity (\( V(R) \)) and velocity dispersion (\( \sigma(R) \)) profiles, using the 3D-based analysis tool \texttt{3DBarolo}\cite{diteodoro2015}. 
A 3D tilted-ring model is employed in \texttt{3DBarolo}, allowing us to derive rotation curves and gas dynamics in a non-parametric fashion, under the assumption of circular rotation. 
The smearing effects by the spectral and spatial resolutions are taken into account in \texttt{3DBarolo}. 
Each tilted ring has flexibility with parameters such as rotation velocity $V$, velocity dispersion $\sigma$, inclination ($i$), position angle (PA), and scale height ($Z_{0}$). 
With a 2D Gaussian fitting for the target morphology in NIRCam/F150W in the source plane, we obtain the axis ratio of $q\simeq0.53$ and calculate $i=\arccos(q)\simeq57.9$~deg.
Here we use the convolved image (right panel of Fig.~\ref{fig:SP-morph}) to obtain the global-scale morphology better.
We adopt the AZIM normalization to study the global scale kinematics. 
We confirm that the PA and inclination estimates do not change much when we derive them with the emission line morphology. 
The dynamical center is fixed based on an iterative fitting output that initiates at the peak of the line intensity and velocity dispersion. 
The $Z_{0}$ value is also fixed around the effective spatial resolution of each instrument in the source plane. 
We use $V$, $\sigma$, PA, and inclination as free parameters. 
Owing to the weak distortion and differential magnification of our target, variations of the PSF shape in the source plane are minimal, and we adopt an average PSF shape over the target area. 
We set the width of the tilted ring between (0.5--1.0) $\times$ FWHM of the average PSF. 

\stepcounter{extfigure} 
\begin{figure*}[t]
\begin{center}
\includegraphics[angle=0,width=1\textwidth]{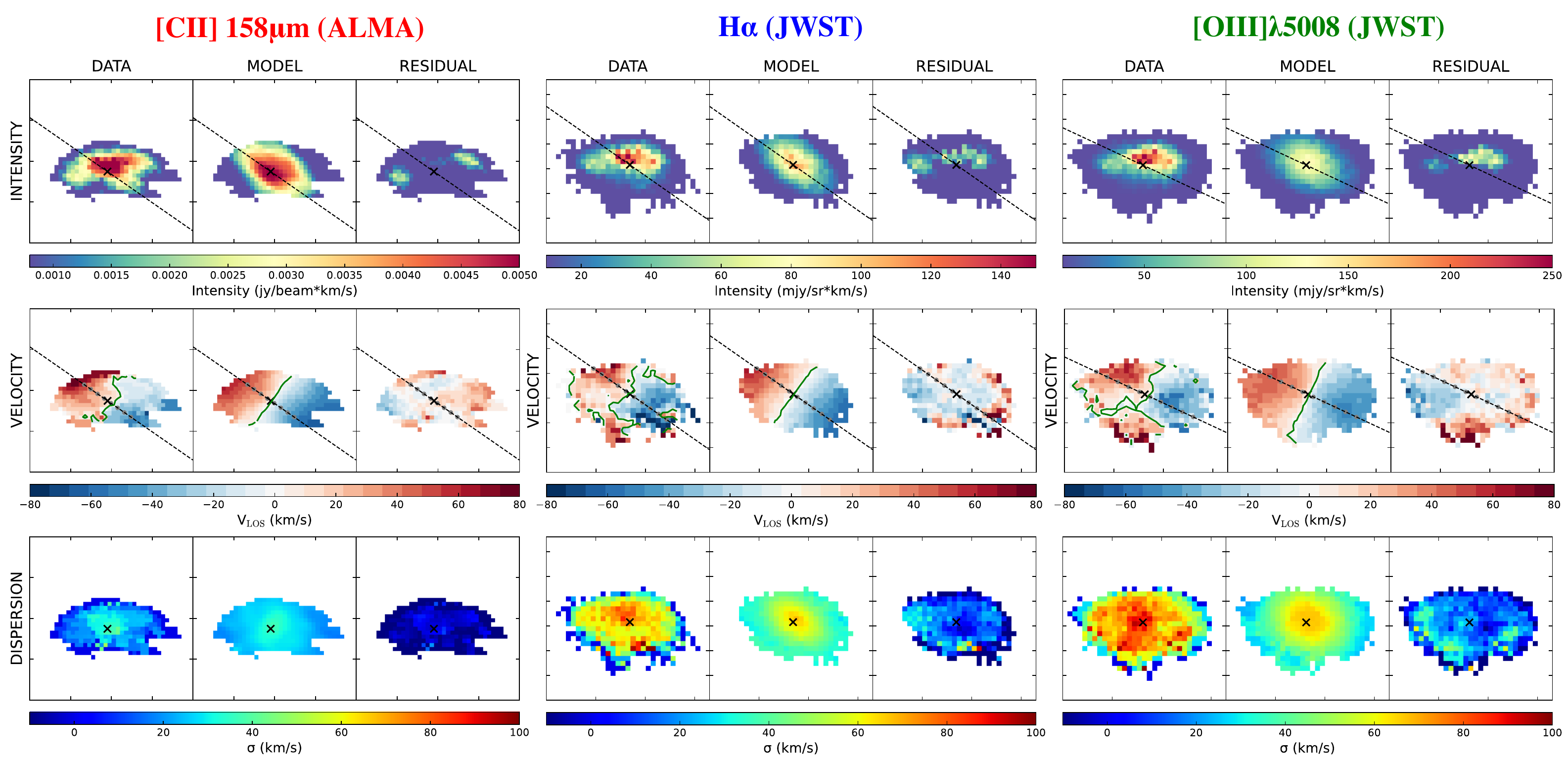}
\end{center}
\vspace{-0.8cm}
\caption{\small 
{\small 
\textbf{
Kinematic modeling of the \cii\ 158~$\mu$m, H$\alpha$, and \oiii$\lambda$5008 lines across the galaxy}, using \texttt{3DBarolo}. 
All maps are reconstructed in the source plane (i.e., lens-corrected). 
\textbf{\textit{Top row:}} Line intensity maps for each emission line, with the observed data, model fit, and residuals displayed from left to right. 
\textbf{\textit{Middle row:}} Line of sight velocity ($V_{\rm LOS}$), probing the rotational dynamics and any potential perturbations. 
\textbf{\textit{Bottom row:}} Velocity dispersion maps, highlighting the kinematic thermal and non-thermal broadening effects. 
Contours overlaying the  $V_{\rm LOS}$ maps denote the $V_{\rm LOS}=0$~km~s$^{-1}$, and crosses mark the galaxy's dynamical center. The dashed line is the positional angle, which is fixed during the fitting. 
}
}
\label{fig:kin_residual}
\end{figure*}

In Extended Data Fig.~\ref{fig:pvd}, we highlight the position and velocity (PV) diagram and velocity channel maps for the \cii\ line observed with the high spatial and spectral resolution ALMA data. The PV diagram shows a symmetric distribution with the two peaks, providing strong evidence that \targ\ is a rotation-supported system\cite{rizzo2022},\footnote{
We apply the \texttt{PVsplit} classification\cite{rizzo2022} and obtain ($P_{\rm major}$, $P_{\rm v}$, $P_{\rm r}$) $= (-2.10, 0.01, 0.48)$ which falls in the parameter space of the single disk galaxy, well separated from mergers.
}
rather than the result of ongoing merging components\cite{spilker2022, nakazato2024}. 
In Extended Data Fig.~\ref{fig:kin_residual}, we also summarize the observed, best-fit model, and residual maps for the \cii\ (left), H$\alpha$ (middle), and \oiii\ (right) lines. 
While some residual components stand in the line intensity map, this is a natural consequence of the AZIM normalization in the fitting, and they are interpreted as the elevated line emissivity due to local-scale substructures (e.g., clumps). 
The residuals in the velocity and dispersion maps are generally within $\simeq \pm20$~km~s$^{-1}$ along with the dynamical major axis, and the global-scale 3D structures of these emission lines are well modeled by \texttt{3DBarolo}. 
Despite the marginal S/N ($\sim$3), a redshifted velocity component along both the dynamical major and minor axes is identified in the H$\alpha$ and \oiii\ lines. Neither stellar continuum nor dust continuum emission is detected in these regions. Based on the current upper limits from NIRCam and assuming the typical mass-to-light ratio of individual clumps in \targ\ (Section~5), the absence of rest-frame UV to FIR counterparts suggests that these features may result from perturbations in the disk structure caused by the accretion of very low-mass star clusters ($\lesssim10^{5}M_{\odot}$) or from an outflowing ionized gas component, if real. This scenario could also explain the residual components observed in the velocity dispersion maps for H$\alpha$ and \oiii, particularly at larger radii. This will be further addressed in a separate paper with further deep follow-up in the future.
In Fig.~\ref{fig:kinematics}b, the red, blue, and green points indicate the measurements obtained for \cii, H$\alpha$, and \oiii\ lines, respectively. While the $V(R)$ measurements are generally consistent among the lines, we find that H$\alpha$ and \oiii\ lines show systematically higher $\sigma(R)$ measurements than \cii. This is consistent with predictions from cosmological zoom-in simulations and is interpreted that ionized gas is likely more influenced by feedback in galaxies\cite{kohandel2023}. We caution that the limited spectral resolution of NIRSpec ($\simeq100$~km~s$^{-1}$ in FWHM of the line spread function) might also contribute to the elevated $\sigma(R)$ measurements in the H$\alpha$ and \oiii\ lines. 
We thus determine an average velocity dispersion $\sigma_{0}$ and the maximum rotation velocity $V_{\rm max}$ (inclination corrected) from the \cii\ results.  
We obtain $V_{\rm max}/\sigma_{0}=3.58\pm0.74$.
This $V/\sigma$ value is broadly consistent with recent measurements for star-forming galaxies at similar redshifts and simulation predictions (see e.g., Fig.2 in Kohandel et al.\cite{kohandel2023}), while these measurements can be underestimated given the increased stability parameter when considering the 3D galaxy structure\cite{bacchini2024}.
In Extended Data Table~5, we list the best-fit $V$ and $\sigma$ measurements and the derived $V_{\rm max}/\sigma_{0}$ values.

Note that we have tested the correlated noise effects through Monte Carlo simulations by creating mock galaxy cubes and running the same kinematic fitting procedure using \texttt{3DBarolo}. We find that the $V_{\rm max}/\sigma_{0}$ value could be overestimated by $4\pm2\%$ for our observation setup (PSF/pixel size ratio $\approx$ 4). This effect, while present, is small ($<5\%$) and does not significantly impact our kinematic results or conclusions. Similarly negligible impacts on the outputs from the correlated noise have been obtained in previous studies using \texttt{3DBarolo}\cite{diteodoro2015, rizzo2022}. 

\subsection{10. Toomre $Q$ parameter: }
\label{sec:qparam}

We estimate the Toomre $Q$ parameter, which gauges the stability of the disk against gravitational fragmentation. 
We use the moderately high-resolution ($0\farcs28\times0\farcs25$) ALMA \cii\ data for this estimate. 
Notably, the latest higher-resolution ($\simeq0\farcs05$) \cii\ follow-up observations with ALMA reveal that $\gtrsim90\%$ of the \cii\ flux detected in the current data originates from the diffuse ISM, rather than compact clumps, different from what is observed in the rest-frame UV with NIRCam \cite{fujimoto2024prep}. This confirms that the current ALMA \cii\ data is well-suited for measuring the Toomre $Q$ parameter. 
The critical threshold of $Q_{\rm crit}$ is 1.0 (0.67) for a thin (thick) gas disk galaxy\cite{genzel2011, cacciato2012}, where the self-gravity of gas overcomes the repelling forces by pressure and differential rotation at $Q<Q_{\rm crit}$. 
The $Q$ parameter is given by 
\begin{equation}
Q = \frac{\kappa \sigma}{\pi G \Sigma_{\text{gas}}},
\end{equation}
where \(G\) is the gravitational constant, $\sigma$ is the velocity dispersion of the gas, and $\Sigma_{\rm gas}$ is the surface density of $M_{\rm gas}$, and $\kappa$ is the epicyclic frequency defined by 
\begin{equation}
\kappa^2 = 2\Omega R\left(\frac{2\Omega}{R} + \frac{d\Omega}{dR}\right),
\end{equation}
where \(\Omega = V/R\) is the angular velocity. 

To compute $\Sigma_{\rm gas}$, we first convert $L_{\rm [CII]}$ to $M_{\rm gas}$ using a conversion factor 
of $\alpha_{\rm [CII]}=31\,M_{\odot}/L_{\odot}$ calibrated with star-forming galaxies at $z\sim2$\cite{zanella2018}. 
While a systematic uncertainty may remain in this conversion, 
a consistent $M_{\rm gas}$ value is obtained from the dynamical mass $M_{\rm dyn}$ estimate, after subtracting $M_{\star}$ in \targ\ in the global galaxy scale\cite{fujimoto2021}. 
Based on our latest \cii\ data and the kinematic modeling results (Section~9), we obtain $M_{\rm dyn}=1.7^{+0.7}_{-0.8}\times10^{9}\,M_{\odot}$ within 2 $\times r_{\rm e}$ of the rest-UV emitting region, and infer $M_{\rm gas} \approx M_{\rm dyn}-M_{\star}=(1.3\pm0.8)\times10^{9}\,M_{\odot}$, assuming that the dark matter contribution is negligible. This is in excellent agreement with the $M_{\rm gas}$ estimate of $1.1^{+0.4}_{-0.1}\times10^{9}\,M_{\odot}$ in the global galaxy scale using the above conversion. With an analytical model proposed based on the KS law, empirical relations, and a star-formation burstiness parameter (i.e., deviation from the KS relation)\cite{sommovigo2021}, we infer $\alpha_{\rm [CII]]}\simeq15\,M_{\odot}/L_{\odot}$, which is consistent with $\alpha_{\rm [CII]}=18\,M_{\odot}/L_{\odot}$ predicated from cosmological simulations for $z\sim6$ galaxies\cite{vizgan2022}. These models imply that using $\alpha_{\rm [CII]}=31\,M_{\odot}/L_{\odot}$ could result in overestimating $M_{\rm gas}$ by a factor of $\sim$2. On the other hand, the dust mass is also estimated to be $M_{\rm dust}=1.2^{+1.2}_{-0.7}\times10^{6}\,M_{\odot}$ for \targ\cite{valentino2024}, based on a dedicated far-infrared (FIR) SED analysis for \targ\ with Hershel and ALMA multiple band constraints that cover observed wavelengths of 0.35--4~mm. Using the gas-to-dust mass ratio of $\sim10^{3}$ calibrated at a low-metallicity regime\cite{remy-ruyer2014} like \targ, we obtain $M_{\rm gas}=1.2^{+1.2}_{-0.7}\times10^{9}\,M_{\odot}$, which is again well consistent with the estimate based on $\alpha_{\rm [CII]}=31\,M_{\odot}/L_{\odot}$. In this paper, we adopt the $M_{\rm gas}$ estimate using $\alpha_{\rm [CII]}=31,M_{\odot}/L_{\odot}$ as our fiducial estimate, while we acknowledge a potential uncertainty up to a factor of 2 in Fig.~\ref{fig:kinematics}c. 
In Extended Data Table~3, we also list the $M_{\rm gas}$ estimates from $M_{\rm dyn}$ and $M_{\rm dust}$.

We applied the following approaches to assess the $Q$ parameter. 
First, we derived the radial profile of the Toomre $Q$ parameter by employing non-parametric radial profiles of the best-fit $V(R)$ and $\sigma(R)$ and the radial-averaged profile of the \(L_{\text{[CII]}}\) luminosity obtained from \texttt{3DBarolo}. This approach allowed us to correct for the beam-smearing effect, with the results presented in Fig.~\ref{fig:kinematics}c. 
Second, as a complementary check, while retaining the beam-smearing effect, we directly computed the 2D distribution of the $Q$ parameter by computing the $V$ and $\sigma$ pixel-by-pixel across the observed \cii\ cube. 
In both approaches, we used the best-fit non-parametric $V(R)$ profile obtained from \texttt{3DBarolo} to compute $\kappa$ via the numerical fashion. 
The results of this direct computation are displayed in the inset panel of Fig.~\ref{fig:kinematics}c. 
Both approaches yielded consistent values of the Toomre $Q$ parameter in a range of $\simeq$ 0.2--0.3, well below $Q_{\rm crit}=$0.67--1.0, suggesting that the gas is unstable over the entire galaxy. This instability likely facilitates star formation throughout the disk and forms numerous star-forming clumps (Fig.~\ref{fig:hst-nircam}).

\subsection{11. Clumpiness parameter $S$: }

To quantify the clumpy structure observed in \targ, we evaluate the clumpiness parameter $S$, defined as
\begin{equation}
    S = 10\times \Sigma^{N,N}_{{\rm x,y}=1,1}\frac{(I_{\rm x,y}-I_{\rm x,y}^{\sigma})-B_{\rm x,y}}{I_{\rm x,y}}, 
\end{equation}
where $I_{\rm x,y}$ is the sky-subtracted flux values of the galaxy at position $(x, y)$, 
$I_{\rm x,y}^{\sigma}$ is the value of the galaxy's flux at $(x,y)$ once it has been reduced in resolution by a smoothing filter of width $\sigma$, $N$ is the size of the galaxy in pixels, and $B_{\rm x,y}$ is the background pixel values in an area of the sky which is equal to the galaxy's area\cite{conselice2003}. 
For this measurement, we adopt $\sigma=0\farcs2$ and use the pixels in NIRCam/F150W map within a circular area with $r=0\farcs7$ ($\approx$ 0.7~kpc in the source plane) from the peak pixel in \targ, given that circularized $r_{\rm e}$ of \targ\ on the global galaxy scale is estimated to be 0.7~kpc\cite{fujimoto2021}. 
We obtain $S=0.81\pm0.02$, which is much higher than the average values observed among local Spirals ($0.08\pm0.08$), dwarf irregulars ($0.40\pm0.20$), and dwarf ellipticals ($0.00\pm0.06$) observed in the rest-frame optical \cite{conselice2003} (Fig.~\ref{fig:clumpiness}). 
Note that this $S$ parameter has been reported to increase in the rest-frame UV, compared to that in the rest-frame optical \cite{mager2018}. However, we confirm that the late-type spirals and irregulars, the best local analog of our target, still show the average $S$ value of $S=0.35\pm0.10$ in the rest-frame UV \cite{mager2018}, which is still much below what we observe in \targ.
While the spatial resolutions are different between our and these local observations, the $S$ parameter increases with better resolution, and the gap could be even larger in the same resolution. On the other hand, remarkably clumpy local dwarf galaxies, similar to \targ, have also been identified in recent studies (see e.g., Mrk178, DDO155\cite{elmegreen2012}; J0921 \cite{rowland2024}), while these galaxies do not perfectly match with the observed properties of \targ\ (e.g., mass, size). 
Albeit less clumpy, dynamical studies with {H\,{\sc i}} gas for local dwarf irregulars show their global-scale rotations\cite{iorio2017}, while their velocity dispersions are even much smaller ($< 10$~km~s$^{-1}$) than \targ. Although it falls outside the scope of this study, investigating whether analogs of \targ\ exist in the local Universe across various aspects (e.g., size, mass, kinematics, chemical enrichment) is crucial, offering valuable insights on whether the physical properties observed in \targ\ result from unique processes exclusive to the early Universe or certain mechanisms observable in the local Universe as well.

\subsection{12. Clump Luminosity Function: }

We also derive the cumulative clump \textit{luminosity function} (LF)\cite{livermore2012,livermore2015} as a function of SFR in each clump to independently evaluate the clumpiness in \targ. 
In Fig.~\ref{fig:clumpiness}, we show our cumulative clump LF measurements for \targ. 
For comparison, we also show measurements for similarly lensed galaxies at $z\sim1$--3 using the H$\alpha$ line in the literature\cite{livermore2015}. Here we convert the H$\alpha$ line luminosity to SFR\cite{murphy2011} and select galaxies with SFR $=1$--10 $M_{\odot}$~yr$^{-1}$, which is comparable to that of \targ, and recalculate the average cumulative LF per galaxy. 
Note that UV-based and H$\alpha$-based SFR measurements, including the dust correction, are consistent in \targ\ within $<$10\% \cite{valentino2024}, and thus the impact from the different SFR tracers is negligible in this comparison.
We find that the cumulative clump LF in \targ\ is systematically higher than that of the lower-redshift measurement, indicating the increasing trend of the clumpiness towards high redshifts.

\subsection{13. Cosmological zoom-in simulations: }
\stepcounter{extfigure} 
\begin{figure*}[t]
\begin{center}
\includegraphics[angle=0,width=1.0\textwidth]{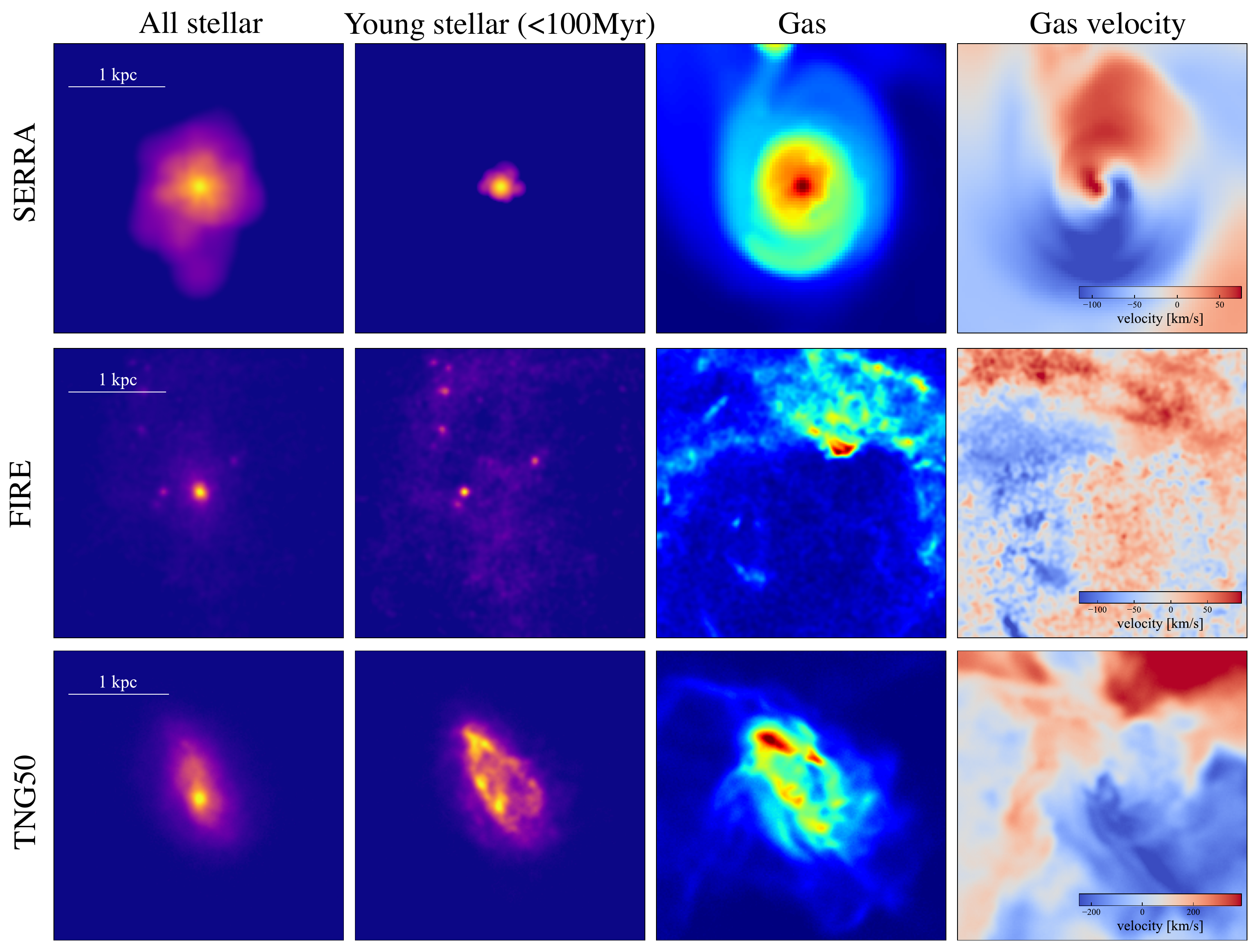}
\end{center}
\vspace{-0.6cm}
\caption{\small 
{\small 
\textbf{\boldmath Examples of simulated early galaxies in the cosmological zoom-in models of SERRA\cite{pallottini2022} (top), FIRE\cite{wetzel2023} (middle), and TNG50\cite{pillepich2019} (bottom)}.
The panels show the noise-free maps of stellar (all), young stellar ($<100$~Myr), gas mass, and gas velocity distributions in the physical scale of 3~kpc $\times$ 3~kpc (see \method). 
The simulated maps are smoothed with Gaussian kernels to match the spatial resolution with the NIRCam/F150W observation of $\simeq0\farcs01$ after the lens correction (Fig.~\ref{fig:SP-morph}). 
For a fair comparison, we use simulated galaxies whose physical properties are close to \targ; SFR $\simeq$ 1--10~$M_{\odot}$~yr$^{-1}$, $M_{\star}\simeq10^{8-9}\,M_{\odot}$, and $z\simeq$ 6--8. We also select rotation-dominated galaxies when available, while no similarly rotation-dominated systems are identified in FIRE due to frequent starbursts and associated feedback effects. Thus, kinematic properties are not considered when selecting the FIRE galaxies for our comparison. 
}
}
\label{fig:model}
\end{figure*}

To investigate whether such a highly clumpy structure observed in \targ\ is reproduced in state-of-the-art cosmological zoom-in simulations for early galaxies, we also quantify the $S$ value and the cumulative clump LF using simulated galaxies in SERRA\cite{zanella2021, pallottini2022, kohandel2023}, FIRE\cite{angles-alcazar2017,wetzel2023}, and TNG50\cite{nelson2019,pillepich2019}.
We refer the reader to the details of the SERRA, FIRE, and TNG50 simulations in Pallottini et al. (2022)\cite{pallottini2022}, Wetzel et al. (2023)\cite{wetzel2023}, Pillepich et al. (2019)\cite{pillepich2019}, respectively, while here we briefly describe each simulation. 

The first set is SERRA, using a customized version of the adaptive mesh refinement (AMR) code \texttt{RAMSES}\cite{teyssier2002} to simulate the dynamics of dark matter, gas, and stars. Gas evolution is modeled using a second-order Godunov scheme, while dark matter and stars are handled via a multigrid particle-mesh solver. Radiative transfer processes are integrated on-the-fly with \texttt{RAMSES-RT}\cite{rosdahl2013}, utilizing a momentum-based framework with M1 closure for the Eddington tensor\cite{aubert2008}. 
Gas-photon interactions, along with the system of equations governing the thermal evolution of the gas, are managed using the \texttt{KROME} package\cite{grassi2014}, which employs a detailed non-equilibrium chemical network. The integration of \texttt{KROME} with \texttt{RAMSES-RT} is achieved by sub-cycling the absorption steps to ensure convergence\cite{decataldo2019, pallottini2019}. 
Emission lines are computed by interpolating grids from the photoionization code \texttt{CLOUDY} c17 \cite{ferland2017}, extending the models from Vallini et al. (2017)\cite{vallini2017} as detailed in Pallottini et al. (2019)\cite{pallottini2019}. Attenuation of UV and continuum FIR emissions is calculated with the Monte Carlo code \texttt{SKIRT} V8\cite{baes2015,camps2015}, informed by the modeling provided by Behrens et al. (2018)\cite{behrens2018}.  The spatial resolution reaches $\simeq$30~pc in the densest regions at $z=6$, closely modeling the characteristics of Galactic Molecular Clouds. 
We select a total of 140 SERRA galaxies at $z=6$--8 for our comparison, whose physical properties are similar to \targ\ with SFR $\simeq$ 1--10 $M_{\odot}$~yr$^{-1}$, $M_{\star}\simeq10^{8-9}\,M_{\odot}$, and the gas kinematics likely to be rotation supported via the visual checks. 

The second set is FIRE, utilizing the N-body+hydrodynamics code \texttt{GIZMO} to re-simulate a set of haloes from the MassiveFIRE galaxy simulations, originally presented without black hole (BH) physics by Feldman et al. (2017)\cite{feldmann2017}. The updated simulations, employing the FIRE-2 code\cite{hopkins2018}, cover diverse halo formation histories with halo masses around \(10^{12.5} M_{\odot}\) at \(z = 2\), and are evolved down to \( z = 1 \) 
with baryonic and dark matter particle masses of $3.3\times10^{4}M_{\odot}$ and $1.7\times10^{5}M_{\odot}$ with force softenings of 0.7~pc, 7~pc, and 57~pc for gas, stellar, and DM particles, respectively. 
The FIRE-2 enhancements include the meshless finite mass (MFM) hydrodynamics solver and refined stellar feedback algorithms. These improvements, alongside the assumed \(\Lambda\)CDM cosmology consistent with \cite{planck2016}, enable detailed modeling of galaxy evolution. 
Among the simulated galaxies publicly available\footnote{https://flathub.flatironinstitute.org/fire}, 
we choose four galaxies of z5m11c, z5m11d, z5m11e, and z5m12a whose $M_{\star}$ values evolve to $(0.5-3.0)\times10^{9}\,M_{\odot}$ at $z=5$\cite{ma2018}, comparable to that of \targ, and use their simulated maps at redshift slices of $z=6.0, 6.5, 7.0, 7.5$, and $8.0$. Note that the systems with the rotating gas disk are rarely found among the simulated galaxies in FIRE due to more busty SFH implemented, where the gas disk is blown away\cite{pallottini2023, sun2023}. Thus, we do not consider the kinematic properties in selecting the FIRE galaxies, while we do not use obvious major merger systems for our comparison. We select a total of 13 FIRE galaxies at $z=6$--8. 

The third set is from TNG50, the highest-resolution configuration of the IllustrisTNG project\cite{nelson2019,pillepich2019}. TNG50 employs the moving-mesh code \texttt{AREPO}\cite{springel2010} to solve the coupled equations of self-gravity and magnetohydrodynamics. The simulation evolves a $(51.7\,\mathrm{cMpc})^3$ volume from $z=127$ to $z=0$ by adopting a dark matter particle mass resolution of $4.5\times10^5\,M_{\odot}$ and an average gas cell mass of $8.5\times10^4\,M_{\odot}$. The spatial resolution reaches 70~pc (comoving) for the star-forming gas. TNG50 includes comprehensive models for galaxy formation physics, including gas cooling, star formation, stellar evolution and feedback, black hole growth and feedback, and magnetic fields\cite{weinberger2017,pillepich2018}.
For our comparison, we initially select at $z=6$--8 a total of 300 TNG50 galaxies with SFR $\simeq$ 0.1--10 $M_{\odot}$~yr$^{-1}$ and $M_{\star}\simeq10^{8-9}\,M_{\odot}$, i.e. with physical properties similar to \targ.  
Note that the typical TNG50 galaxy at  $z=6$--$8$ is not expected to be rotation-dominated, since on average TNG50 galaxies show $V/\sigma<2$ at these redshifts\cite{pillepich2019,kohandel2023}, Thus, we visually inspect the sample and select only those galaxies showing a symmetric velocity gradient. After such a selection, the TNG50 sample is composed of 12 galaxies.

To perform the comparison, we produce four cutout images of the stellar distribution, young ($<100$~Myr) stellar distribution, gas mass, and gas velocity in the physical scale of 3~kpc $\times$ 3~kpc around galaxies at $z=6$--8 from random lines of sight. 
We then inject random noise and smooth the maps using a Gaussian kernel to match the sensitivity and the spatial resolution as the observation of $0\farcs01$ (after the lens correction) to achieve a fair comparison. 
In Extended Data Fig.~\ref{fig:model}, we show four maps of example simulated galaxies in the SERRA, FIRE, and TNG50 models after smoothing.
As highlighted in these examples, we find that SERRA (FIRE) galaxies are generally characterized as rotation- (dispersion-) dominated systems with one or two (a few) stellar clump components. 
The difference between SERRA and FIRE simulations may originate from their different handling of star formation and feedback, which results in a burstier star formation history\cite{pallottini2023, sun2023} — and, accordingly, more intense starbursts — in FIRE compared SERRA. 
Such frequent starbursts may produce more small clumps\cite{ma2020}, while it leads to more dynamically chaotic structures due to subsequent feedback effects. The FIRE galaxy in Extended Data Fig.~\ref{fig:model} actually shows the situation that the gas disk is blown away due to a strong feedback effect. 
Compared to SERRA and FIRE, TNG50 galaxies typically display intermediate characteristics -- exhibiting moderately clumpy structures in systems that are neither strongly chaotic nor strongly rotation-dominated. The number of clumps in TNG50 galaxies is less than FIRE and thus still significantly less than observed in \targ. 
The average $V/\sigma$ value of $<2$ in TNG50 galaxies at these redshifts\cite{pillepich2019} is less than SERRA\cite{kohandel2023} and \targ.  
In Extended Data Fig.~\ref{fig:model}, we highlight the most clumpy TNG50 galaxy within a rotating disk structure. However, we emphasize that, on average, TNG50 galaxies exhibit less clumpiness and less stable disks than what is highlighted here.

Given the NIRCam/F150W map corresponds to the rest-frame UV continuum representing the young stellar distribution of \targ\ at $z=6.072$, we measure the $S$ parameter and the clump LF in the same manner as the procedures done for the observed map. 
For SFR, we adopt the time-averaged values over $\sim$50 or 100~Myr to make it consistent with the rest-frame UV based SFR that is adopted for the measurements of \targ.
The results are summarized in Fig.~\ref{fig:clumpiness}. 
We find that the clumpiness in \targ\ exceeds the values measured in the SERRA, FIRE, and TNG50 galaxies in both quantities of $S$ and the clump LF. These results indicate challenges in the current simulations to reconcile the two distinct properties -- the numerous clumps and the rotating disk; more frequent starbursts may be required to reproduce the numerous clumps observed in \targ, while the rotating disk needs to remain in the galaxy. 
One plausible solution is that the feedback effects are significantly weaker than what is implemented in the current simulations.
When the surface gas density exceeds 1,000 $M_{\odot}$~pc$^{-2}$, the star-formation efficiency is estimated to be larger than 0.6 in radiation hydrodynamical simulations\cite{fukushima2021} for clouds with  $0.1Z_{\odot}$ and $M_{\rm gas}=10^{6}\,M_{\odot}$, which are comparable to the individual clumps identified in \targ. 
In the same context, we might be witnessing the effect of a feedback-free scenario\cite{dekel2023}, as the low $Q$ Toomre parameter promotes disk fragmentation, which leads to clump formation, boosting the already relatively high gas density ($>$100--1000$M_{\odot}$~pc$^{-2}$; see Extended Data Fig.~\ref{fig:clump_prop}c) and making star formation very efficient, i.e. effectively feedback free up in the burst phase.
On the one hand, however, the low dust obscuration observed in \targ\ may imply that radiation pressure is effectively driving out the gas and dust from the system, a mechanism which is invoked to explain the overabundance of high-$z$ galaxies\cite{ferrara2023}. 
In this process, the self-shielding of high-density molecular clouds within star-forming regions and the dense ISM prevents feedback energy from propagating across the galaxy. Instead, the energy may predominantly escape through face-on directions. This suggests that the presence of outflows is consistent with weak feedback mechanisms in high gas-density environments, which maintain high star-formation efficiency within the star-forming regions. Additionally, dust may be ejected through these "caves" along face-on directions.
We note that the galaxies at $z=6-8$ in SERRA simulations show the gas fraction ($\equiv M_{\rm gas}/(M_{\rm gas}+M_{\star})$) of $\simeq40$\% and the surface gas density of $\simeq$10--100 $M_{\odot}$~pc$^{-2}$ at maximum. On the other hand, the gas fraction and its surface density in \targ\ are estimated to be ~80--90\% and $\simeq10^{3}$--$10^{5}M_{\odot}$~pc$^{-2}$, respectively (see Extended Data Table~3). This indicates that a part of the gap between the observation and simulation results may also be ascribed to the physical mechanisms related to the gas fraction such as the fragmentation induced by the gas supply from the circum- and inter-galactic media\cite{dekel2009b}.
We also note that variations in the source-plane PSF shapes due to differential magnification are found to alter the $S$ parameter and clump LF measurements for the simulated galaxies no more than $5\%$ by assuming the lowest and highest magnification cases, indicating a negligible impact on our results.

\stepcounter{extfigure} 
\begin{figure*}[t]
\begin{center}
\includegraphics[angle=0,width=1.0\textwidth]{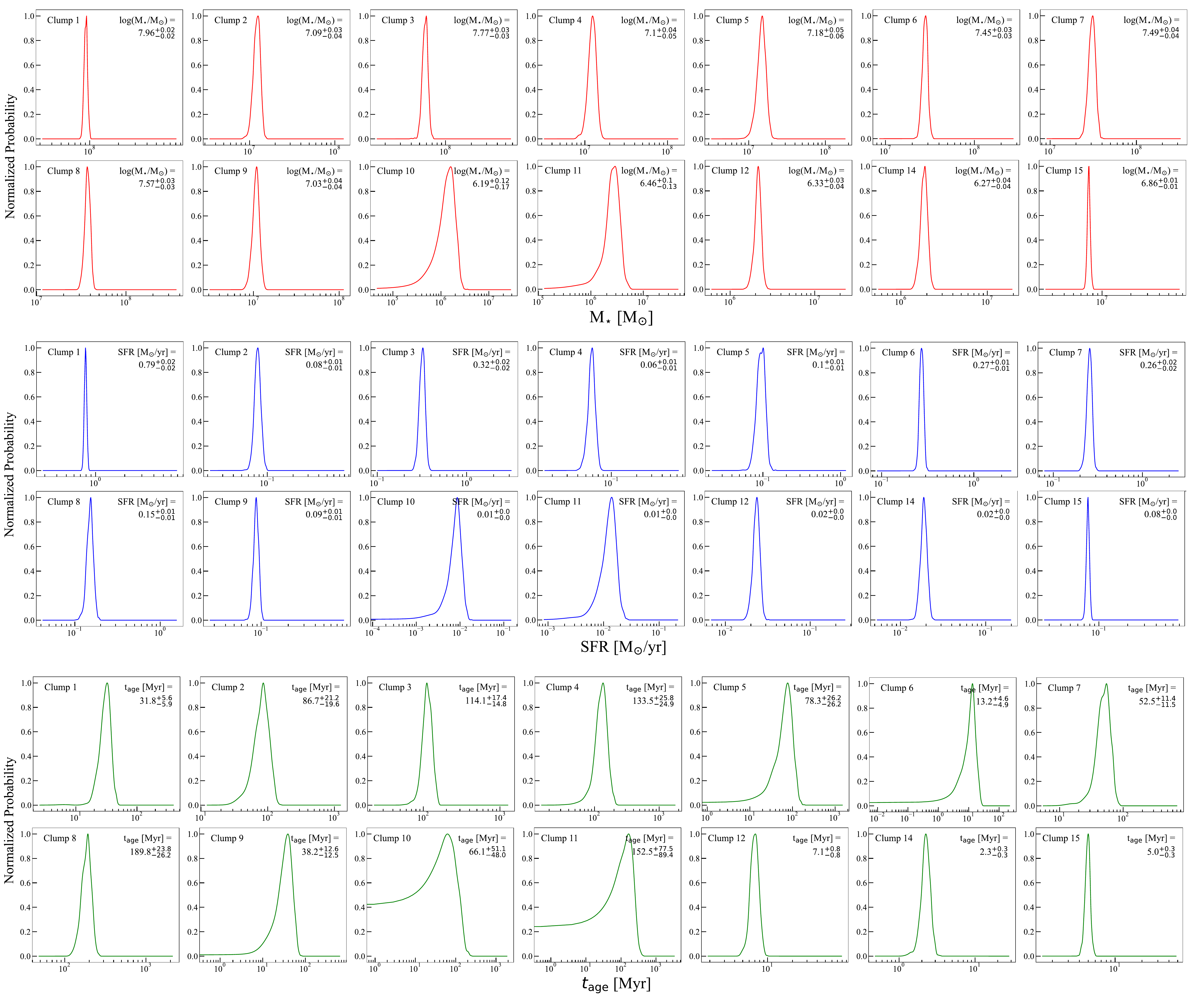}
\end{center}
\vspace{-1cm}
\caption{\small 
{\small 
\textbf{Extended Data Figure 11: Posterior distributions of key parameters for individual clumps.} 
Each row displays the posterior probability distributions for a different parameter: stellar mass ($M_*$, top row), star formation rate (SFR, middle row), and stellar age ($t_\mathrm{age}$, bottom row). The distributions are shown for each of the 15 identified clumps, labeled from 1 to 15. The median value and $1\sigma$ confidence intervals are indicated in the upper right corner of each panel. 
}
}
\label{fig:clump_pdf}
\end{figure*}

\subsection{Extended Data Tables}

\begin{table*}[h]
\label{tab:photometry}
\begin{center}
{\small \textbf{Extended Data Table 1$|$ Optical--NIR observed photometry of \targbf}}
\scalebox{1.}{
\begin{tabular}{cccc}
\hline
Instrument              & Observed $\lambda$ & Flux density & reference \\\ 
                        &  [$\mu$m]          & [$\mu$Jy]             &         \\ \hline
 \textit{HST} ACS/F606W & 0.61     & $< 0.07$&   F21\cite{fujimoto2021}  \\
 \textit{HST} ACS/F814W & 0.81     & $0.32\pm0.04$&   F21\cite{fujimoto2021}  \\
 \textit{HST} ACS/F105W & 1.05     & $1.17\pm0.07$ &   F21\cite{fujimoto2021}  \\
 \textit{HST} ACS/F125W & 1.25     & $1.41\pm0.13$ &   F21\cite{fujimoto2021}  \\
 \textit{HST} ACS/F140W & 1.40     & $1.42\pm0.11$ &   F21\cite{fujimoto2021}  \\
 \textit{HST} ACS/F160W & 1.60     & $1.34\pm0.07$ &   F21\cite{fujimoto2021}  \\
 \textit{JWST} NIRCam/F115W & 1.15 & $1.176\pm0.017$ &    Here    \\
 \textit{JWST} NIRCam/F150W & 1.50  & $1.521\pm0.008$ &   Here    \\
 \textit{JWST} NIRCam/F277W & 2.79  & $2.378\pm0.008$ &   Here    \\
 \textit{JWST} NIRCam/F356W & 3.56  & $4.633\pm0.008$ &   Here    \\
\textit{JWST} NIRCam/F444W  & 4.42   & $3.130\pm0.010$ &   Here   \\ \hline
\end{tabular}
}
\end{center}
\end{table*}

\begin{table*}[h]
\label{tab:lineflux}
\begin{center}
{\small \textbf{Extended Data Table 2$|$ Observed line fluxes of \targbf\ with NIRSpec IFU}}
\scalebox{1.}{
\begin{tabular}{ccc}
\hline
Line                & Line flux$^{\ddag}$                    & FWHM$^{\dagger}$             \\ 
                    &    [10$^{-18}~$erg~s$^{-1}$~cm$^{-2}$]  &  [km~s$^{-1}$]   \\  \hline
H$\delta$           & $11.3\pm0.9$                         & $166\pm12$         \\
H$\gamma$           & $18.3\pm0.7$                         & $180\pm6^{\sharp a}$    \\
\oiii$\lambda$4363  & $4.4\pm0.4$                          & $180\pm6^{\sharp a}$    \\
H$\beta$            & $40.9\pm0.7$                         & $162\pm6 $    \\
\oiii$\lambda$5008  & $269.4\pm2.1$                        & $154\pm1$    \\
HeI$\lambda$5877    & $9.8\pm2.0$                          & $292\pm48$    \\
H$\alpha$           & $136.4\pm1.4$                        & $148\pm1^{\sharp b}$    \\
\nii$\lambda$6585   & $8.1\pm0.8$                         & $148\pm1^{\sharp b}$  \\
\sii$\lambda$6718   & $4.6\pm0.7$                         & $106\pm16^{\sharp c}$  \\
\sii$\lambda$6733   & $3.8\pm0.6$                         & $106\pm16^{\sharp c}$  \\ \hline
\end{tabular}
}
\end{center}
\flushleft{\footnotesize
{$\ddag$ Within the $0\farcs7$-radius aperture. \\
$\dagger$ We show the intrinsic FWHM, FWHM(int) $= \sqrt{\rm FWHM(observed)^{2} - (spectral\,resolution)^2}$, assuming the spectral resolution of 100~km~s$^{-1}$.\\
$\sharp$ We assume the same line width in the double Gaussian fitting for the neighboring emission lines of $a$, $b$, and $c$.   
}}
\end{table*}

\setlength{\tabcolsep}{2pt}
\begin{table*}[h]
\label{tab:phy_prop}
\begin{center}
{\small \textbf{Extended Data Table 3$|$ Intrinsic Physical properties of \targbf\ on the global galaxy scale}}
{\footnotesize
\scalebox{1.}{
\begin{tabular}{llll}
\hline
Parameter          & Integrated Value              & Description                       & Reference  \\ \hline
R.A.              &    06:00:09.5647   & Right Ascension (J2000) in the observed frame & F21\cite{fujimoto2021}, L21\cite{laporte2021}   \\
Decl.             &   $-$20:08:10.993  & Declination (J2000) in the observed frame     & F21\cite{fujimoto2021}, L21\cite{laporte2021}   \\
$z_{\rm spec}$    &    6.072           & Redshift from \cii\ line                 & F21   \\
$\mu$             &    32.5$_{-6.8}^{+0.7}$  & Average magnification              & F24\cite{furtak2024prep} \\
$r_{\rm e}^{\dagger}$  [kpc]     &     $0.68^{+0.09}_{-0.01}$   & Circularized effective radius in rest-frame UV       & Here \\ 
$M_{\rm UV}^{\dagger}$  [mag]        &      $-19.29^{+0.02}_{-0.26}$         & Absolute UV magnitude  & Here \\
$E(B-V)$  [mag]             &  $0.13\pm0.02$   & Dust reddening via H$\alpha$/H$\beta$ & Here \\
$n_{\rm e}$   [cm$^{-3}$]       &   260$^{+400}_{-230}$  & Electron density via \sii$\lambda$6718/\sii$\lambda$6733 & Here \\
SFR [$M_{\odot}$~yr$^{-1}$]   &   $2.6^{+1.7}_{-1.5}$  & Average star-formation rate over 100 Myrs from SED fitting  & G24\cite{clara2024} \\
$M_{\rm dyn}$ [$M_{\odot}$] &  $1.7^{+0.7}_{-0.8}\times10^{9}$ & Dynamical mass within 2$\times r_{\rm e}$ from \cii\ kinematics  & Here \\
$M_{\star}$ [$M_{\odot}$]   &   $4.5^{+2.7}_{-1.1}\times10^{8}$  & Stellar mass from SED fitting                 & G24\cite{clara2024} \\
$M_{\rm dust}$ [$M_{\odot}$]   &   $1.2^{+1.2}_{-0.7}\times10^{6}$  & Dust mass from FIR SED fitting                 & V24\cite{valentino2024} \\
$M_{\rm gas}$ (\cii) [$M_{\odot}$] &  $1.1^{+0.4}_{-0.1}\times10^{9}$ & Gas mass inferred from \cii\ luminosity\cite{zanella2018}  & Here \\
$M_{\rm gas}$ (dyn.) [$M_{\odot}$] &  $(1.3\pm0.8)\times10^{9}$ & Gas mass inferred from $M_{\rm dyn} - M_{\rm star}$                            & Here \\
$M_{\rm gas}$ (dust) [$M_{\odot}$] &  $1.2^{+1.2}_{-0.7}\times10^{9}$    & Gas mass inferred from the gas-to-dust ($\propto$ metallicity)\cite{remy-ruyer2014} & Here \\
$T_{\rm e}$(\oiii)   [K]  & $14100\pm600$  & Electron temperature via \oiii$\lambda$4363/\oiii$\lambda$5008 & Here \\
12+$\log$(O/H) ($T_{\rm e}$) &  $8.11\pm0.03$  & Direct $T_{\rm e}$ method\cite{izotov2006, pilyugin2005} &  Here \\
12+$\log$(O/H) (R3)          &   $8.00\pm0.09$            & Strong line method with the R3 index\cite{curti2020b}  &  Here \\
12+$\log$(O/H) (N2)          &    $8.30\pm0.16$             & Strong line method with the N2 index\cite{curti2020b}  &  Here \\
12+$\log$(O/H) (O3N2)        &    $8.21\pm0.21$           & Strong line method with the O3N2 index\cite{curti2020b}   &  Here \\
12+$\log$(O/H) (S2)          &    $8.03\pm0.11$             & Strong line method with the S2 index\cite{curti2020b}   &  Here \\
12+$\log$(O/H) (O3S2)       &    $7.98\pm0.11$           & Strong line method with the O3S2 index\cite{curti2020b}   &  Here \\
$\log$(N/O)                 &  $-1.46\pm0.04$          & Nitrogen abundance        & Here \\
\hline
\end{tabular}
}
}
\end{center}
$\dagger$ These estimates are taken from Fujimoto et al. (2021)\cite{fujimoto2021}, while the updated magnification estimate and uncertainty are applied to the estimates. 
\end{table*}

\setlength{\tabcolsep}{2.5pt}
\begin{table*}
\label{tab:clump}
\begin{center}
{\small \textbf{Extended Data Table 4$|$ Intrinsic Physical properties of the star-forming clumps in \targbf}}
\scalebox{1.}{
\begin{tabular}{ccccccccc}
\hline
ID & $r_{\rm e}^{\dagger}$ & $r_{\rm e}$ & $M_{\star}$                   &             SFR                       &    $t_{\rm age}$    &  $L_{\rm [CII]}^{\ddag}$     & $M_{\rm gas}^{\natural}$          & $\mu$\\ 
    &             [pix]                 &     [pc]         &   [$10^{6}M_{\odot}$]    &   [$M_{\odot}$~yr$^{-1}$]  &  [10$^{6}$ years]  &  [$10^{6}\, L_{\odot}$]        &  [$10^{8}\, M_{\odot}$]& \\  
\hline
1 & $0.6\pm0.1$ & $10.7\pm2.2$ & $90.8_{-3.8}^{+3.7}$ & $0.79_{-0.02}^{+0.02}$ & $6.9_{-19.3}^{+30.9}$ & $12.8\pm2.34$ & $3.97\pm0.73$ & $29.3_{-2.7}^{+1.7}$ \\
2 & $1.9\pm0.4$ & $33.9\pm6.7$ & $12.4_{-1.1}^{+1.0}$ & $0.08_{-0.01}^{+0.01}$ & $13.8_{-54.0}^{+93.0}$ & -- & -- & $38.1_{-4.8}^{+1.0}$ \\
3 & $2.9\pm0.2$ & $62.1\pm4.0$ & $58.4_{-4.1}^{+4.3}$ & $0.32_{-0.02}^{+0.02}$ & $128.2_{-29.8}^{+2.6}$ & $2.94\pm1.76$ & $0.91\pm0.55$ & $26.0_{-2.0}^{+0.7}$ \\
4 & $3.8\pm0.5$ & $48.4\pm7.1$ & $12.4_{-1.3}^{+1.2}$ & $0.06_{-0.01}^{+0.01}$ & $138.1_{-31.5}^{+22.6}$ & $0.93\pm0.89$ & $0.29\pm0.28$ & $42.5_{-5.5}^{+1.3}$ \\
5 & $2.7\pm0.8$ & $48.9\pm14.2$ & $15.0_{-1.9}^{+1.8}$ & $0.1_{-0.01}^{+0.01}$ & $67.5_{-13.6}^{+40.1}$ & -- & -- & $35.9_{-4.4}^{+1.2}$ \\
6 & $2.1\pm0.2$ & $42.3\pm4.7$ & $28.0_{-1.8}^{+2.0}$ & $0.27_{-0.01}^{+0.01}$ & $4.1_{-4.2}^{+13.4}$ & $4.40\pm1.76$ & $1.36\pm0.55$ & $27.9_{-2.9}^{+0.2}$ \\
7 & $1.1\pm0.1$ & $22.4\pm2.8$ & $31.0_{-2.9}^{+2.7}$ & $0.26_{-0.02}^{+0.02}$ & $18.8_{-22.7}^{+44.9}$ & $2.81\pm1.72$ & $0.87\pm0.53$ & $23.8_{-1.7}^{+0.7}$ \\
8 & $1.1\pm0.2$ & $16.0\pm2.7$ & $37.3_{-2.4}^{+2.7}$ & $0.15_{-0.01}^{+0.01}$ & $209.9_{-45.3}^{+3.1}$ & $2.08\pm1.08$ & $0.64\pm0.33$ & $37.3_{-5.5}^{+0.00}$ \\
9 & $1.4\pm0.3$ & $23.2\pm4.8$ & $10.9_{-0.9}^{+0.9}$ & $0.09_{-0.01}^{+0.01}$ & $3.4_{-23.1}^{+46.1}$ & $1.70\pm0.94$ & $0.53\pm0.29$ & $33.0_{-4.5}^{+0.2}$ \\
10 & $4.0\pm0.5$ & $50.8\pm7.2$ & $1.5_{-0.5}^{+0.5}$ & $0.01_{-0.00}^{+0.00}$ & $56.8_{-32.8}^{+60.6}$ & -- & -- & $36.5_{-4.7}^{+0.8}$ \\
11 & $2.4\pm0.9$ & $46.6\pm17.4$ & $2.9_{-0.8}^{+0.8}$ & $0.01_{-0.00}^{+0.00}$ & $195.1_{-120.8}^{+37.5}$ & -- & -- & $33.4_{-4.1}^{+1.2}$ \\
12 & $1.6\pm1.1$ & $28.3\pm18.3$ & $2.2_{-0.2}^{+0.2}$ & $0.02_{-0.00}^{+0.00}$ & $8.0_{-1.8}^{+-0.1}$ & -- & -- & $26.3_{-2.5}^{+0.8}$ \\
13 & $0.9\pm1.3$ & $7.0\pm10.2$ & -- & -- & -- & -- & -- & $22.1_{-1.4}^{+0.6}$ \\
14 & $1.1\pm0.4$ & $19.8\pm7.9$ & $1.9_{-0.2}^{+0.2}$ & $0.02_{-0.00}^{+0.00}$ & $1.8_{-0.2}^{+0.8}$ & -- & -- & $35.0_{-3.4}^{+1.0}$ \\
15 & $<0.9$ & $<18$ & $7.3_{-0.3}^{+0.2}$ & $0.08_{-0.00}^{+0.00}$ & $3.0_{-1.6}^{+2.2}$ & -- & -- & $32.4_{-3.7}^{+1.3}$ \\
\hline
\end{tabular}
}

\end{center}
\flushleft{\footnotesize
{
$\dagger$ Circularized effective radius. \\
$\sharp$ The clump location is not included in our pixel selection procedure for the SED fitting.  \\
$\ddag$ \cii\ line luminosity extracted with an aperture adjusted to the segmentation area. We only count for the clumps whose segmentation areas are larger than the ALMA beam in the \cii\ line map. \\ 
$\natural$ Using a conversion factor from $L_{\rm [CII]}$\cite{zanella2018}.\\
For the SED fitting outputs ($M_{\star}$, SFR, $t_{\rm age}$), we compute the 1$\sigma$ error by integrating the posterior distribution for each parameter across the pixels that constitute each clump.
}}
\end{table*}

\setlength{\tabcolsep}{1.pt}
\begin{table*}
\label{tab:clump}
\begin{center}
{\small \textbf{Extended Data Table 5$|$ Our $V$ and $\sigma$ measurements of \targbf}}
\begin{tabular}{ccccccccccccccc}
\hline
Line & \multicolumn{4}{c}{\cii} & \multicolumn{5}{c}{H$\alpha$} & \multicolumn{5}{c}{\oiii} \\ \hline
Radius [kpc]   $|$&0.09 &0.26 &0.43 &0.60 $|$ &0.09 &0.26 &0.43 &0.60 &0.77 $|$ &0.10 &0.31 &0.51 &0.71 &0.92 \\
$V$(R)$^{\dagger}$ [km/s]  $|$& 50$^{+4}_{-3}$  &50$^{+4}_{-5}$  &60$^{+5}_{-6}$  &82$^{+11}_{-15}$ $|$  &19$^{+4}_{-4}$  &26$^{+4}_{-4}$  &52$^{+17}_{-17}$  &56$^{+7}_{-7}$  &64$^{+5}_{-5}$ $|$  &44$^{+8}_{-8}$  &60$^{+9}_{-9}$  &54$^{+17}_{-17}$  &59$^{+6}_{-6}$  &59$^{+11}_{-11}$  \\
$\sigma$(R) [km/s]  $|$& 18$^{+7}_{-7}$  &31$^{+4}_{-5}$  &28$^{+5}_{-6}$  &15$^{+7}_{-10}$ $|$  &70$^{+10}_{-9}$  &59$^{+11}_{-10}$  &52$^{+13}_{-13}$  &44$^{+13}_{-13}$  &55$^{+10}_{-10}$ $|$ &70$^{+12}_{-13}$  &66$^{+13}_{-13}$  &59$^{+12}_{-13}$  &42$^{+13}_{-12}$  &22$^{+12}_{-13}$   \\ \hline 
$V_{\rm max}/\sigma_{0}^{\dagger}$ (each)  $|$ & \multicolumn{4}{c}{3.58 $\pm$ 0.74} $|$ & \multicolumn{5}{c}{\tcb{(}2.77 $\pm$ 0.42\tcb{)$^{\natural}$}} $|$ & \multicolumn{5}{c}{\tcb{(}2.60 $\pm$ 0.52\tcb{)$^{\natural}$}} \\ \hline 
\end{tabular}
\label{tab:rotation_curve}

\end{center}
\flushleft{\footnotesize
{
$\sharp$ Inclination is corrected. \\
$\dagger$ Because the systematically high $\sigma(R)$ observed in H$\alpha$ and \oiii\ lines could be attributed to the feedback effects\cite{kohandel2023}, or limited spectral resolution of NIRSpec ($\simeq100$~km/s), we determine the average $\sigma(R)$ value ($\sigma_{0}$) using \cii\ line results. \\
$\natural$ We adopt the \cii-based measurement as our fiducial $V/\sigma$ value throughout the paper, as ionized gas tracers could be affected by feedback effects. 
}}
\end{table*}

\end{methods}
\renewcommand{\figurename}{Figure}
\renewcommand{\theHfigure}{\arabic{figure}}

\subsection{Code availability.}
The NIRCam data were processed with {\sc grizli} available at \url{https://github.com/gbrammer/grizli}. 
The NIRCam F160W image is analyzed with \texttt{GALFIT} 
available at \url{https://users.obs.carnegiescience.edu/peng/work/galfit/galfit.html}.  
The ALMA data were reduced using the CASA software version 6.4.1.12
available at \url{https://casa.nrao.edu/casa\_obtaining.shtml}. 
The NIRSpec data reduction code developed by the TEMPLATE team, which is available at \url{https://github.com/JWST-Templates/Notebooks/blob/main/nirspec_ifu_cookbook.ipynb}. 
The \cii, H$\alpha$, and \oiii$\lambda$5008 data cubes are analyzed with \texttt{3DBarolo} available at \url{https://editeodoro.github.io/Bbarolo/}. 

\subsection{Data availability.}

This paper makes use of the \jwst\ data from \#GO-1567, available at  \url{https://archive.stsci.edu/}.  
The reduced \jwst\ NIRCam images are available at \url{https://dawn-cph.github.io/dja/}.
The ALMA data supporting our finding is from \#2021.1.00055.S, available at \url{http://almascience.nao.ac.jp/}. 
The simulated galaxies in FIRE used in this study are all available at \url{https://flathub.flatironinstitute.org/fire}.  
Other datasets generated and/or analyzed during the current study are available from the corresponding author upon reasonable request.

\bibliographystyle{naturemag2}
\bibliography{reference}

\begin{thebibliography}{100}

\bibitem{hopkins2014}
{Hopkins}, P.~F., {Kere{\v s}}, D., {O{\~n}orbe}, J., {Faucher-Gigu{\`e}re},
  C.-A. et al., ``{Galaxies on FIRE (Feedback In Realistic Environments):
  stellar feedback explains cosmologically inefficient star formation}''.
\newblock {\em \mnras}, { \bf 445}, 581--603,  November  (2014).

\bibitem{rizzo2020}
{Rizzo}, F., {Vegetti}, S., {Powell}, D., {Fraternali}, F. et al., ``{A
  dynamically cold disk galaxy in the early Universe}''.
\newblock {\em \nat}, { \bf 584}(7820), 201--204,  August  (2020).

\bibitem{arrabal-halo2023a}
{Arrabal Haro}, P., {Dickinson}, M., {Finkelstein}, S.~L., {Kartaltepe}, J.~S.
  et al., ``{Confirmation and refutation of very luminous galaxies in the early
  Universe}''.
\newblock {\em \nat}, { \bf 622}(7984), 707--711,  October  (2023).

\bibitem{laporte2021}
{Laporte}, N., {Zitrin}, A., {Ellis}, R.~S., {Fujimoto}, S. et al., ``{ALMA
  Lensing Cluster Survey: a strongly lensed multiply imaged dusty system at z
  {\ensuremath{\geq}} 6}''.
\newblock {\em \mnras}, { \bf 505}(4), 4838--4846,  August  (2021).

\bibitem{fujimoto2021}
{Fujimoto}, S., {Oguri}, M., {Brammer}, G., {Yoshimura}, Y. et al., ``{ALMA
  Lensing Cluster Survey: Bright [C II] 158 {\ensuremath{\mu}}m Lines from a
  Multiply Imaged Sub-L$^{{\ensuremath{\star}}}$ Galaxy at z = 6.0719}''.
\newblock {\em \apj}, { \bf 911}(2), 99,  April  (2021).

\bibitem{clara2024}
{Gim{\'e}nez-Arteaga}, C., {Fujimoto}, S., {Valentino}, F., {Brammer}, G.~B. et
  al., ``{Outshining in the Spatially Resolved Analysis of a Strongly-Lensed
  Galaxy at z=6.072 with JWST NIRCam}''.
\newblock {\em arXiv e-prints}, { \bf }, arXiv:2402.17875,  February  (2024).

\bibitem{shibuya2015}
{Shibuya}, T., {Ouchi}, M. and {Harikane}, Y., ``{Morphologies of $\sim$190,000
  Galaxies at z = 0-10 Revealed with HST Legacy Data. I. Size Evolution}''.
\newblock {\em \apjs}, { \bf 219}, 15,  August  (2015).

\bibitem{iyer2018}
{Iyer}, K., {Gawiser}, E., {Dav{\'e}}, R., {Davis}, P. et al., ``{The SFR-M
  $_{*}$ Correlation Extends to Low Mass at High Redshift}''.
\newblock {\em \apj}, { \bf 866}(2), 120,  October  (2018).

\bibitem{behroozi2018}
{Behroozi}, P. and {Silk}, J., ``{The most massive galaxies and black holes
  allowed by {\ensuremath{\Lambda}}CDM}''.
\newblock {\em \mnras}, { \bf 477}(4), 5382--5387,  July  (2018).

\bibitem{clara2023}
{Gim{\'e}nez-Arteaga}, C., {Oesch}, P.~A., {Brammer}, G.~B., {Valentino}, F. et
  al., ``{Spatially Resolved Properties of Galaxies at $5 < z < 9$ in the SMACS
  0723 JWST ERO Field}''.
\newblock {\em \apj}, { \bf 948}(2), 126,  May  (2023).

\bibitem{livermore2015}
{Livermore}, R.~C., {Jones}, T.~A., {Richard}, J., {Bower}, R.~G. et al.,
  ``{Resolved spectroscopy of gravitationally lensed galaxies: global dynamics
  and star-forming clumps on {\ensuremath{\sim}}100 pc scales at $1 < z <
  4$}''.
\newblock {\em \mnras}, { \bf 450}(2), 1812--1835,  June  (2015).

\bibitem{vanzella2023}
{Vanzella}, E., {Claeyssens}, A., {Welch}, B., {Adamo}, A. et al.,
  ``{JWST/NIRCam Probes Young Star Clusters in the Reionization Era Sunrise
  Arc}''.
\newblock {\em \apj}, { \bf 945}(1), 53,  March  (2023).

\bibitem{adamo2024}
{Adamo}, A., {Bradley}, L.~D., {Vanzella}, E., {Claeyssens}, A. et al., ``{The
  discovery of bound star clusters 460 Myr after the Big Bang}''.
\newblock {\em arXiv e-prints}, { \bf }, arXiv:2401.03224,  January  (2024).

\bibitem{ono2023}
{Ono}, Y., {Harikane}, Y., {Ouchi}, M., {Yajima}, H. et al., ``{Morphologies of
  Galaxies at z {\ensuremath{\gtrsim}} 9 Uncovered by JWST/NIRCam Imaging:
  Cosmic Size Evolution and an Identification of an Extremely Compact Bright
  Galaxy at z 12}''.
\newblock {\em \apj}, { \bf 951}(1), 72,  July  (2023).

\bibitem{zanella2018}
{Zanella}, A., {Daddi}, E., {Magdis}, G., {Diaz Santos}, T. et al., ``{The [C
  II] emission as a molecular gas mass tracer in galaxies at low and high
  redshifts}''.
\newblock {\em Monthly Notices of the Royal Astronomical Society}, { \bf
  481}(2), 1976--1999,  Dec  (2018).

\bibitem{vizgan2022}
{Vizgan}, D., {Greve}, T.~R., {Olsen}, K.~P., {Zanella}, A. et al., ``{Tracing
  Molecular Gas Mass in $z\simeq6$ Galaxies with [C II]}''.
\newblock {\em \apj}, { \bf 929}(1), 92,  April  (2022).

\bibitem{cacciato2012}
{Cacciato}, M., {Dekel}, A. and {Genel}, S., ``{Evolution of violent
  gravitational disc instability in galaxies: late stabilization by transition
  from gas to stellar dominance}''.
\newblock {\em \mnras}, { \bf 421}(1), 818--831,  March  (2012).

\bibitem{genzel2011}
{Genzel}, R., {Newman}, S., {Jones}, T., {F{\"o}rster Schreiber}, N.~M. et al.,
  ``{The Sins Survey of z \raisebox{-0.5ex}\textasciitilde 2 Galaxy Kinematics:
  Properties of the Giant Star-forming Clumps}''.
\newblock {\em \apj}, { \bf 733}(2), 101,  June  (2011).

\bibitem{tadaki2018}
{Tadaki}, K., {Iono}, D., {Yun}, M.~S., {Aretxaga}, I. et al., ``{The
  gravitationally unstable gas disk of a starburst galaxy 12 billion years
  ago}''.
\newblock {\em \nat}, { \bf 560}(7720), 613--616,  August  (2018).

\bibitem{kohandel2023}
{Kohandel}, M., {Pallottini}, A., {Ferrara}, A., {Zanella}, A. et al.,
  ``{Dynamically cold disks in the early Universe: Myth or reality?}''.
\newblock {\em \aap}, { \bf 685}, A72,  May  (2024).

\bibitem{conselice2003}
{Conselice}, C.~J., ``{The Relationship between Stellar Light Distributions of
  Galaxies and Their Formation Histories}''.
\newblock {\em \apjs}, { \bf 147}(1), 1--28,  July  (2003).

\bibitem{pallottini2022}
{Pallottini}, A., {Ferrara}, A., {Gallerani}, S., {Behrens}, C. et al., ``{A
  survey of high-z galaxies: SERRA simulations}''.
\newblock {\em \mnras}, { \bf 513}(4), 5621--5641,  July  (2022).

\bibitem{wetzel2023}
{Wetzel}, A., {Hayward}, C.~C., {Sanderson}, R.~E., {Ma}, X. et al., ``{Public
  Data Release of the FIRE-2 Cosmological Zoom-in Simulations of Galaxy
  Formation}''.
\newblock {\em \apjs}, { \bf 265}(2), 44,  April  (2023).

\bibitem{pillepich2019}
{Pillepich}, A., {Nelson}, D., {Springel}, V., {Pakmor}, R. et al., ``{First
  results from the TNG50 simulation: the evolution of stellar and gaseous discs
  across cosmic time}''.
\newblock {\em \mnras}, { \bf 490}(3), 3196--3233,  December  (2019).

\bibitem{livermore2012}
{Livermore}, R.~C., {Jones}, T., {Richard}, J., {Bower}, R.~G. et al.,
  ``{Hubble Space Telescope H{\ensuremath{\alpha}} imaging of star-forming
  galaxies at $z\simeq1-1.5$: evolution in the size and luminosity of giant H
  II regions}''.
\newblock {\em \mnras}, { \bf 427}(1), 688--702,  November  (2012).

\bibitem{mager2018}
{Mager}, V.~A., {Conselice}, C.~J., {Seibert}, M., {Gusbar}, C. et al.,
  ``{Galaxy Structure in the Ultraviolet: The Dependence of Morphological
  Parameters on Rest-frame Wavelength}''.
\newblock {\em \apj}, { \bf 864}(2), 123,  September  (2018).

\bibitem{ma2020}
{Ma}, X., {Grudi{\'c}}, M.~Y., {Quataert}, E., {Hopkins}, P.~F. et al.,
  ``{Self-consistent proto-globular cluster formation in cosmological
  simulations of high-redshift galaxies}''.
\newblock {\em \mnras}, { \bf 493}(3), 4315--4332,  April  (2020).

\bibitem{finkelstein2022b}
{Finkelstein}, S.~L., {Bagley}, M.~B., {Haro}, P.~A., {Dickinson}, M. et al.,
  ``{A Long Time Ago in a Galaxy Far, Far Away: A Candidate z 12 Galaxy in
  Early JWST CEERS Imaging}''.
\newblock {\em \apjl}, { \bf 940}(2), L55,  December  (2022).

\bibitem{harikane2023}
{Harikane}, Y., {Ouchi}, M., {Oguri}, M., {Ono}, Y. et al., ``{A Comprehensive
  Study of Galaxies at z 9-16 Found in the Early JWST Data: Ultraviolet
  Luminosity Functions and Cosmic Star Formation History at the
  Pre-reionization Epoch}''.
\newblock {\em \apjs}, { \bf 265}(1), 5,  March  (2023).

\bibitem{grudic2021}
{Grudi{\'c}}, M.~Y., {Guszejnov}, D., {Hopkins}, P.~F., {Offner}, S. S.~R. et
  al., ``{STARFORGE: Towards a comprehensive numerical model of star cluster
  formation and feedback}''.
\newblock {\em \mnras}, { \bf 506}(2), 2199--2231,  September  (2021).

\bibitem{fukushima2021}
{Fukushima}, H. and {Yajima}, H., ``{Radiation hydrodynamics simulations of
  massive star cluster formation in giant molecular clouds}''.
\newblock {\em \mnras}, { \bf 506}(4), 5512--5539,  October  (2021).

\bibitem{elzant2001}
{El-Zant}, A., {Shlosman}, I. and {Hoffman}, Y., ``{Dark Halos: The Flattening
  of the Density Cusp by Dynamical Friction}''.
\newblock {\em \apj}, { \bf 560}(2), 636--643,  October  (2001).

\bibitem{planck2014}
{Planck Collaboration}, {Ade}, P.~A.~R., {Aghanim}, N., {Armitage-Caplan}, C.
  et al., ``{Planck 2013 results. XVI. Cosmological parameters}''.
\newblock {\em \aap}, { \bf 571}, A16,  November  (2014).

\bibitem{ebeling2001}
{Ebeling}, H., {Edge}, A.~C. and {Henry}, J.~P., ``{MACS: A Quest for the Most
  Massive Galaxy Clusters in the Universe}''.
\newblock {\em \apj}, { \bf 553}(2), 668--676,  June  (2001).

\bibitem{coe2013}
{Coe}, D., {Zitrin}, A., {Carrasco}, M., {Shu}, X. et al., ``{CLASH: Three
  Strongly Lensed Images of a Candidate z {\ensuremath{\approx}} 11 Galaxy}''.
\newblock {\em \apj}, { \bf 762}(1), 32,  January  (2013).

\bibitem{postman2012}
{Postman}, M., {Coe}, D., {Ben{\'\i}tez}, N., {Bradley}, L. et al., ``{The
  Cluster Lensing and Supernova Survey with Hubble: An Overview}''.
\newblock {\em \apjs}, { \bf 199}(2), 25,  April  (2012).

\bibitem{lotz2017}
{Lotz}, J.~M., {Koekemoer}, A., {Coe}, D., {Grogin}, N. et al., ``{The Frontier
  Fields: Survey Design and Initial Results}''.
\newblock {\em \apj}, { \bf 837}(1), 97,  March  (2017).

\bibitem{oguri2010}
{Oguri}, M., ``{The Mass Distribution of SDSS J1004+4112 Revisited}''.
\newblock {\em \pasj}, { \bf 62}, 1017--,  aug  (2010).

\bibitem{jullo2007}
{Jullo}, E., {Kneib}, J.~P., {Limousin}, M., {El{\'\i}asd{\'o}ttir}, {\'A}. et
  al., ``{A Bayesian approach to strong lensing modelling of galaxy
  clusters}''.
\newblock {\em New Journal of Physics}, { \bf 9}(12), 447,  December  (2007).

\bibitem{zitrin2015}
{Zitrin}, A., {Fabris}, A., {Merten}, J., {Melchior}, P. et al., ``{Hubble
  Space Telescope Combined Strong and Weak Lensing Analysis of the CLASH
  Sample: Mass and Magnification Models and Systematic Uncertainties}''.
\newblock {\em \apj}, { \bf 801}(1), 44,  March  (2015).

\bibitem{fujimoto2024prep}
{Fujimoto, S. et al. in preperation}.
\newblock { \bf },  (2024).

\bibitem{bouwens2017}
{Bouwens}, R.~J., {van Dokkum}, P.~G., {Illingworth}, G.~D., {Oesch}, P.~A. et
  al., ``{Very low-luminosity galaxies in the early universe have observed
  sizes similar to single star cluster complexes}''.
\newblock {\em ArXiv e-prints}, { \bf },  November  (2017).

\bibitem{fujimoto2023b}
{Fujimoto}, S., {Kohno}, K., {Ouchi}, M., {Oguri}, M. et al., ``{ALMA Lensing
  Cluster Survey: Deep 1.2 mm Number Counts and Infrared Luminosity Functions
  at $z\simeq1-8$}''.
\newblock {\em arXiv e-prints}, { \bf }, arXiv:2303.01658,  March  (2023).

\bibitem{valentino2023}
{Valentino}, F., {Brammer}, G., {Gould}, K. M.~L., {Kokorev}, V. et al., ``{An
  Atlas of Color-selected Quiescent Galaxies at z > 3 in Public JWST Fields}''.
\newblock {\em \apj}, { \bf 947}(1), 20,  April  (2023).

\bibitem{brammer2021}
{Brammer}, G. and {Matharu}, J.
\newblock ``{gbrammer/grizli: Release 2021}''.
\newblock Zenodo,  June  (2021).

\bibitem{brammer2023}
Brammer, G.
\newblock ``grizli'',  March  (2023).
\newblock Please cite this software using these metadata.

\bibitem{gaia2021}
{Gaia Collaboration}, {Brown}, A.~G.~A., {Vallenari}, A., {Prusti}, T. et al.,
  ``{Gaia Early Data Release 3. Summary of the contents and survey
  properties}''.
\newblock {\em \aap}, { \bf 649}, A1,  May  (2021).

\bibitem{casa2022}
{CASA Team}, {Bean}, B., {Bhatnagar}, S., {Castro}, S. et al., ``{CASA, the
  Common Astronomy Software Applications for Radio Astronomy}''.
\newblock {\em \pasp}, { \bf 134}(1041), 114501,  November  (2022).

\bibitem{rigby2023}
{Rigby}, J.~R., {Vieira}, J.~D., {Phadke}, K.~A., {Hutchison}, T.~A. et al.,
  ``{JWST Early Release Science Program TEMPLATES: Targeting Extremely
  Magnified Panchromatic Lensed Arcs and their Extended Star formation}''.
\newblock {\em arXiv e-prints}, { \bf }, arXiv:2312.10465,  December  (2023).

\bibitem{welch2023}
{Welch}, B., {Rigby}, J.~R. and {Hutchison}, T.~A., ``{TEMPLATES: Tests of
  NIRSpec Observing Strategy, using SGAS1723}''.
\newblock {\em Research Notes of the American Astronomical Society}, { \bf
  7}(1), 17,  January  (2023).

\bibitem{birkin2023}
{Birkin}, J.~E., {Hutchison}, T.~A., {Welch}, B., {Spilker}, J.~S. et al.,
  ``{JWST's TEMPLATES for Star Formation: The First Resolved Gas-phase
  Metallicity Maps of Dust-obscured Star-forming Galaxies at z 4}''.
\newblock {\em \apj}, { \bf 958}(1), 64,  November  (2023).

\bibitem{rauscher2023}
{Rauscher}, B.~J., ``{NSClean: An Algorithm for Removing Correlated Noise from
  JWST NIRSpec Images}''.
\newblock {\em arXiv e-prints}, { \bf }, arXiv:2306.03250,  June  (2023).

\bibitem{cresci2023}
{Cresci}, G., {Tozzi}, G., {Perna}, M., {Brusa}, M. et al., ``{Bubbles and
  outflows: The novel JWST/NIRSpec view of the z = 1.59 obscured quasar
  XID2028}''.
\newblock {\em \aap}, { \bf 672}, A128,  April  (2023).

\bibitem{marshall2023}
{Marshall}, M.~A., {Perna}, M., {Willott}, C.~J., {Maiolino}, R. et al.,
  ``{GA-NIFS: Black hole and host galaxy properties of two $z\simeq6.8$ quasars
  from the NIRSpec IFU}''.
\newblock {\em \aap}, { \bf 678}, A191,  October  (2023).

\bibitem{perna2023}
{Perna}, M., {Arribas}, S., {Marshall}, M., {D'Eugenio}, F. et al., ``{GA-NIFS:
  The ultra-dense, interacting environment of a dual AGN at z
  {\ensuremath{\sim}} 3.3 revealed by JWST/NIRSpec IFS}''.
\newblock {\em \aap}, { \bf 679}, A89,  November  (2023).

\bibitem{vanzella2023b}
{Vanzella}, E., {Loiacono}, F., {Bergamini}, P., {Mestric}, U. et al., ``{An
  extremely metal poor star complex in the reionization era: Approaching
  Population III stars with JWST}''.
\newblock {\em arXiv e-prints}, { \bf }, arXiv:2305.14413,  May  (2023).

\bibitem{peng2010}
{Peng}, C.~Y., {Ho}, L.~C., {Impey}, C.~D. and {Rix}, H.-W., ``{Detailed
  Decomposition of Galaxy Images. II. Beyond Axisymmetric Models}''.
\newblock {\em \aj}, { \bf 139}, 2097--2129,  June  (2010).

\bibitem{deugenio2024}
{D'Eugenio}, F., {P{\'e}rez-Gonz{\'a}lez}, P.~G., {Maiolino}, R., {Scholtz}, J.
  et al., ``{A fast-rotator post-starburst galaxy quenched by supermassive
  black-hole feedback at z = 3}''.
\newblock {\em Nature Astronomy}, { \bf },  September  (2024).

\bibitem{isobe2023}
{Isobe}, Y., {Ouchi}, M., {Nakajima}, K., {Harikane}, Y. et al., ``{Redshift
  Evolution of Electron Density in the Interstellar Medium at z 0-9 Uncovered
  with JWST/NIRSpec Spectra and Line-spread Function Determinations}''.
\newblock {\em \apj}, { \bf 956}(2), 139,  October  (2023).

\bibitem{coe2019}
{Coe}, D., {Salmon}, B., {Brada{\v{c}}}, M., {Bradley}, L.~D. et al.,
  ``{RELICS: Reionization Lensing Cluster Survey}''.
\newblock {\em \apj}, { \bf 884}(1), 85,  October  (2019).

\bibitem{furtak2024}
{Furtak}, L.~J., {Zitrin}, A., {Richard}, J., {Eckert}, D. et al., ``{A complex
  node of the cosmic web associated with the massive galaxy cluster MACS
  J0600.1-2008}''.
\newblock {\em \mnras}, { \bf 533}(2), 2242--2261,  September  (2024).

\bibitem{pascale2022}
{Pascale}, M., {Frye}, B.~L., {Diego}, J., {Furtak}, L.~J. et al.,
  ``{Unscrambling the Lensed Galaxies in JWST Images behind SMACS 0723}''.
\newblock {\em \apjl}, { \bf 938}(1), L6,  October  (2022).

\bibitem{furtak2023a}
{Furtak}, L.~J., {Zitrin}, A., {Weaver}, J.~R., {Atek}, H. et al.,
  ``{UNCOVERing the extended strong lensing structures of Abell 2744 with the
  deepest JWST imaging}''.
\newblock {\em \mnras}, { \bf 523}(3), 4568--4582,  August  (2023).

\bibitem{jaffe1983}
{Jaffe}, W., ``{A simple model for the distribution of light in spherical
  galaxies.}''.
\newblock {\em \mnras}, { \bf 202}, 995--999,  March  (1983).

\bibitem{eliasdottir2007}
{El{\'\i}asd{\'o}ttir}, {\'A}., {Limousin}, M., {Richard}, J., {Hjorth}, J. et
  al., ``{Where is the matter in the Merging Cluster Abell 2218?}''.
\newblock {\em arXiv e-prints}, { \bf }, arXiv:0710.5636,  October  (2007).

\bibitem{papovich2001}
{Papovich}, C., {Dickinson}, M. and {Ferguson}, H.~C., ``{The Stellar
  Populations and Evolution of Lyman Break Galaxies}''.
\newblock {\em \apj}, { \bf 559}(2), 620--653,  October  (2001).

\bibitem{maraston2010}
{Maraston}, C., {Pforr}, J., {Renzini}, A., {Daddi}, E. et al., ``{Star
  formation rates and masses of z \raisebox{-0.5ex}\textasciitilde 2 galaxies
  from multicolour photometry}''.
\newblock {\em \mnras}, { \bf 407}(2), 830--845,  September  (2010).

\bibitem{perrin2012}
{Perrin}, M.~D., {Soummer}, R., {Elliott}, E.~M., {Lallo}, M.~D. et al.
\newblock ``{Simulating point spread functions for the James Webb Space
  Telescope with WebbPSF}''.
\newblock In {\em Space Telescopes and Instrumentation 2012: Optical, Infrared,
  and Millimeter Wave, }{Clampin}, M.~C., {Fazio}, G.~G., {MacEwen}, H.~A. and
  {Oschmann}, Jacobus~M., J., editors, volume 8442 of {\em Society of
  Photo-Optical Instrumentation Engineers (SPIE) Conference Series},  84423D,
  September  (2012).

\bibitem{perrin2014}
{Perrin}, M.~D., {Sivaramakrishnan}, A., {Lajoie}, C.-P., {Elliott}, E. et al.
\newblock ``{Updated point spread function simulations for JWST with
  WebbPSF}''.
\newblock In {\em Space Telescopes and Instrumentation 2014: Optical, Infrared,
  and Millimeter Wave, }{Oschmann}, Jacobus~M., J., {Clampin}, M., {Fazio},
  G.~G. and {MacEwen}, H.~A., editors, volume 9143 of {\em Society of
  Photo-Optical Instrumentation Engineers (SPIE) Conference Series},  91433X,
  August  (2014).

\bibitem{carnall2018}
{Carnall}, A.~C., {McLure}, R.~J., {Dunlop}, J.~S. and {Dav{\'e}}, R.,
  ``{Inferring the star formation histories of massive quiescent galaxies with
  BAGPIPES: evidence for multiple quenching mechanisms}''.
\newblock {\em \mnras}, { \bf 480}(4), 4379--4401,  November  (2018).

\bibitem{ferland2017}
{Ferland}, G.~J., {Chatzikos}, M., {Guzm{\'a}n}, F., {Lykins}, M.~L. et al.,
  ``{The 2017 Release Cloudy}''.
\newblock {\em \rmxaa}, { \bf 53}, 385--438,  October  (2017).

\bibitem{bruzual2003}
{Bruzual}, G. and {Charlot}, S., ``{Stellar population synthesis at the
  resolution of 2003}''.
\newblock {\em \mnras}, { \bf 344}, 1000--1028,  October  (2003).

\bibitem{calzetti2000}
{Calzetti}, D., {Armus}, L., {Bohlin}, R.~C., {Kinney}, A.~L. et al., ``{The
  Dust Content and Opacity of Actively Star-forming Galaxies}''.
\newblock {\em \apj}, { \bf 533}, 682--695,  April  (2000).

\bibitem{kroupa2001}
{Kroupa}, P., ``{On the variation of the initial mass function}''.
\newblock {\em \mnras}, { \bf 322}(2), 231--246,  April  (2001).

\bibitem{nakajima2023}
{Nakajima}, K., {Ouchi}, M., {Isobe}, Y., {Harikane}, Y. et al., ``{JWST Census
  for the Mass-Metallicity Star Formation Relations at z = 4-10 with
  Self-consistent Flux Calibration and Proper Metallicity Calibrators}''.
\newblock {\em \apjs}, { \bf 269}(2), 33,  December  (2023).

\bibitem{sandles2023}
{Sandles}, L., {D'Eugenio}, F., {Maiolino}, R., {Looser}, T.~J. et al.,
  ``{JADES: Balmer Decrement Measurements at redshifts 4 < z < 7}''.
\newblock {\em arXiv e-prints}, { \bf }, arXiv:2306.03931,  June  (2023).

\bibitem{shapley2022}
{Shapley}, A.~E., {Sanders}, R.~L., {Salim}, S., {Reddy}, N.~A. et al., ``{The
  MOSFIRE Deep Evolution Field Survey: Implications of the Lack of Evolution in
  the Dust Attenuation-Mass Relation to z 2}''.
\newblock {\em \apj}, { \bf 926}(2), 145,  February  (2022).

\bibitem{pilyugin2012}
{Pilyugin}, L.~S., {V{\'\i}lchez}, J.~M., {Mattsson}, L. and {Thuan}, T.~X.,
  ``{Abundance determination from global emission-line SDSS spectra: exploring
  objects with high N/O ratios}''.
\newblock {\em \mnras}, { \bf 421}(2), 1624--1634,  April  (2012).

\bibitem{reddy2023}
{Reddy}, N.~A., {Topping}, M.~W., {Sanders}, R.~L., {Shapley}, A.~E. et al.,
  ``{Paschen-line Constraints on Dust Attenuation and Star Formation at z 1-3
  with JWST/NIRSpec}''.
\newblock {\em \apj}, { \bf 948}(2), 83,  May  (2023).

\bibitem{sanders2023}
{Sanders}, R.~L., {Shapley}, A.~E., {Topping}, M.~W., {Reddy}, N.~A. et al.,
  ``{Excitation and Ionization Properties of Star-forming Galaxies at z =
  2.0-9.3 with JWST/NIRSpec}''.
\newblock {\em \apj}, { \bf 955}(1), 54,  September  (2023).

\bibitem{kauffmann2003}
{Kauffmann}, G., {Heckman}, T.~M., {Tremonti}, C., {Brinchmann}, J. et al.,
  ``{The host galaxies of active galactic nuclei}''.
\newblock {\em \mnras}, { \bf 346}(4), 1055--1077,  December  (2003).

\bibitem{osterbrock1989}
{Osterbrock}, D.~E.
\newblock {\em {Astrophysics of gaseous nebulae and active galactic nuclei}}.
\newblock  (1989).

\bibitem{izotov2006}
{Izotov}, Y.~I., {Stasi{\'n}ska}, G., {Meynet}, G., {Guseva}, N.~G. et al.,
  ``{The chemical composition of metal-poor emission-line galaxies in the Data
  Release 3 of the Sloan Digital Sky Survey}''.
\newblock {\em \aap}, { \bf 448}(3), 955--970,  March  (2006).

\bibitem{pilyugin2005}
{Pilyugin}, L.~S. and {Thuan}, T.~X., ``{Oxygen Abundance Determination in H II
  Regions: The Strong Line Intensities-Abundance Calibration Revisited}''.
\newblock {\em \apj}, { \bf 631}(1), 231--243,  September  (2005).

\bibitem{campbell1986}
{Campbell}, A., {Terlevich}, R. and {Melnick}, J., ``{The stellar populations
  and evolution of H II galaxies - I. High signal-to-noise optical
  spectroscopy.}''.
\newblock {\em \mnras}, { \bf 223}, 811--825,  December  (1986).

\bibitem{garnett1992}
{Garnett}, D.~R., ``{Electron Temperature Variations and the Measurement of
  Nebular Abundances}''.
\newblock {\em \aj}, { \bf 103}, 1330,  April  (1992).

\bibitem{curti2020b}
{Curti}, M., {Mannucci}, F., {Cresci}, G. and {Maiolino}, R., ``{The
  mass-metallicity and the fundamental metallicity relation revisited on a
  fully T$_{e}$-based abundance scale for galaxies}''.
\newblock {\em \mnras}, { \bf 491}(1), 944--964,  January  (2020).

\bibitem{nakajima2022}
{Nakajima}, K., {Ouchi}, M., {Xu}, Y., {Rauch}, M. et al., ``{EMPRESS. V.
  Metallicity Diagnostics of Galaxies over
  12+log(O/H)=\raisebox{-0.5ex}\textasciitilde6.9-8.9 Established by a Local
  Galaxy Census: Preparing for JWST Spectroscopy}''.
\newblock {\em arXiv e-prints}, { \bf }, arXiv:2206.02824,  June  (2022).

\bibitem{bunker2023}
{Bunker}, A.~J., {Saxena}, A., {Cameron}, A.~J., {Willott}, C.~J. et al.,
  ``{JADES NIRSpec Spectroscopy of GN-z11: Lyman-{\ensuremath{\alpha}} emission
  and possible enhanced nitrogen abundance in a z = 10.60 luminous galaxy}''.
\newblock {\em \aap}, { \bf 677}, A88,  September  (2023).

\bibitem{marques-chaves2024}
{Marques-Chaves}, R., {Schaerer}, D., {Kuruvanthodi}, A., {Korber}, D. et al.,
  ``{Extreme N-emitters at high redshift: Possible signatures of supermassive
  stars and globular cluster or black hole formation in action}''.
\newblock {\em \aap}, { \bf 681}, A30,  January  (2024).

\bibitem{isobe2023b}
{Isobe}, Y., {Ouchi}, M., {Tominaga}, N., {Watanabe}, K. et al., ``{JWST
  Identification of Extremely Low C/N Galaxies with [N/O]
  {\ensuremath{\gtrsim}} 0.5 at z 6-10 Evidencing the Early CNO-cycle
  Enrichment and a Connection with Globular Cluster Formation}''.
\newblock {\em \apj}, { \bf 959}(2), 100,  December  (2023).

\bibitem{topping2024}
{Topping}, M.~W., {Stark}, D.~P., {Senchyna}, P., {Plat}, A. et al.,
  ``{Metal-poor star formation at z > 6 with JWST: new insight into hard
  radiation fields and nitrogen enrichment on 20 pc scales}''.
\newblock {\em \mnras}, { \bf 529}(4), 3301--3322,  April  (2024).

\bibitem{ubler2023}
{{\"U}bler}, H., {Maiolino}, R., {Curtis-Lake}, E., {P{\'e}rez-Gonz{\'a}lez},
  P.~G. et al., ``{GA-NIFS: A massive black hole in a low-metallicity AGN at z
  {\ensuremath{\sim}} 5.55 revealed by JWST/NIRSpec IFS}''.
\newblock {\em \aap}, { \bf 677}, A145,  September  (2023).

\bibitem{norris2014}
{Norris}, M.~A., {Kannappan}, S.~J., {Forbes}, D.~A., {Romanowsky}, A.~J. et
  al., ``{The AIMSS Project - I. Bridging the star cluster-galaxy
  divide$^{\star}${\textdagger}{\textdaggerdbl}{\textsection}{\textparagraph}}''.
\newblock {\em \mnras}, { \bf 443}(2), 1151--1172,  Sep  (2014).

\bibitem{kennicutt2012}
{Kennicutt}, R.~C. and {Evans}, N.~J., ``{Star Formation in the Milky Way and
  Nearby Galaxies}''.
\newblock {\em \araa}, { \bf 50}, 531--608,  September  (2012).

\bibitem{morishita2023}
{Morishita}, T. and {Stiavelli}, M., ``{Physical Characterization of Early
  Galaxies in the Webb's First Deep Field SMACS J0723.3-7323}''.
\newblock {\em \apjl}, { \bf 946}(2), L35,  April  (2023).

\bibitem{bertin1996}
{Bertin}, E. and {Arnouts}, S., ``{SExtractor: Software for source
  extraction}''.
\newblock {\em \aap}, { \bf 117}, 393--404,  jun  (1996).

\bibitem{kalita2024}
{Kalita}, B.~S., {Silverman}, J.~D., {Daddi}, E., {Mercier}, W. et al.,
  ``{Near-IR clumps and their properties in high-z galaxies with
  JWST/NIRCam}''.
\newblock {\em arXiv e-prints}, { \bf }, arXiv:2402.02679,  February  (2024).

\bibitem{shibuya2016}
{Shibuya}, T., {Ouchi}, M., {Kubo}, M. and {Harikane}, Y., ``{Morphologies of
  \raisebox{-0.5ex}\textasciitilde190,000 Galaxies at z = 0-10 Revealed with
  HST Legacy Data. II. Evolution of Clumpy Galaxies}''.
\newblock {\em \apj}, { \bf 821}(2), 72,  April  (2016).

\bibitem{mowla2024}
{Mowla}, L., {Iyer}, K., {Asada}, Y., {Desprez}, G. et al., ``{The Firefly
  Sparkle: The Earliest Stages of the Assembly of A Milky Way-type Galaxy in a
  600 Myr Old Universe}''.
\newblock {\em arXiv e-prints}, { \bf }, arXiv:2402.08696,  February  (2024).

\bibitem{king1966}
{King}, I.~R., ``{The structure of star clusters. III. Some simple dynamical
  models}''.
\newblock {\em \aj}, { \bf 71}, 64,  February  (1966).

\bibitem{fujimoto2020b}
{Fujimoto}, S., {Silverman}, J.~D., {Bethermin}, M., {Ginolfi}, M. et al.,
  ``{The ALPINE-ALMA [C II] Survey: Size of Individual Star-forming Galaxies at
  z = 4-6 and Their Extended Halo Structure}''.
\newblock {\em \apj}, { \bf 900}(1), 1,  September  (2020).

\bibitem{welch2023a}
{Welch}, B., {Coe}, D., {Zitrin}, A., {Diego}, J.~M. et al., ``{RELICS:
  Small-scale Star Formation in Lensed Galaxies at z = 6-10}''.
\newblock {\em \apj}, { \bf 943}(1), 2,  January  (2023).

\bibitem{valentino2024}
{Valentino}, F., {Fujimoto}, S., {Gim{\'e}nez-Arteaga}, C., {Brammer}, G. et
  al., ``{The cold interstellar medium of a normal sub-$L^\star$ galaxy at the
  end of reionization}''.
\newblock {\em arXiv e-prints}, { \bf }, arXiv:2402.17845,  February  (2024).

\bibitem{ono2018}
{Ono}, Y., {Ouchi}, M., {Harikane}, Y., {Toshikawa}, J. et al., ``{Great
  Optically Luminous Dropout Research Using Subaru HSC (GOLDRUSH). I. UV
  luminosity functions at z {\ensuremath{\sim}} 4-7 derived with the
  half-million dropouts on the 100 deg$^{2}$ sky}''.
\newblock {\em \pasj}, { \bf 70}, S10,  January  (2018).

\bibitem{rizzo2022}
{Rizzo}, F., {Kohandel}, M., {Pallottini}, A., {Zanella}, A. et al.,
  ``{Dynamical characterization of galaxies up to z {\ensuremath{\sim}} 7}''.
\newblock {\em \aap}, { \bf 667}, A5,  November  (2022).

\bibitem{diteodoro2015}
{Di Teodoro}, E.~M. and {Fraternali}, F., ``{$^{\rm 3D}$BAROLO: a new 3D
  algorithm to derive rotation curves of galaxies}''.
\newblock {\em \mnras}, { \bf 451}(3), 3021--3033,  Aug  (2015).

\bibitem{spilker2022}
{Spilker}, J.~S., {Hayward}, C.~C., {Marrone}, D.~P., {Aravena}, M. et al.,
  ``{Chaotic and Clumpy Galaxy Formation in an Extremely Massive
  Reionization-era Halo}''.
\newblock {\em \apjl}, { \bf 929}(1), L3,  April  (2022).

\bibitem{nakazato2024}
{Nakazato}, Y., {Ceverino}, D. and {Yoshida}, N., ``{A merger-driven scenario
  for clumpy galaxy formation in the epoch of reionization: Physical properties
  of clumps in the FirstLight simulation}''.
\newblock {\em arXiv e-prints}, { \bf }, arXiv:2402.08911,  February  (2024).

\bibitem{bacchini2024}
{Bacchini}, C., {Nipoti}, C., {Iorio}, G., {Roman-Oliveira}, F. et al., ``{A 3D
  view on the local gravitational instability of cold gas discs in star-forming
  galaxies at 0 {\ensuremath{\lesssim}} z {\ensuremath{\lesssim}} 5}''.
\newblock {\em \aap}, { \bf 687}, A115,  July  (2024).

\bibitem{sommovigo2021}
{Sommovigo}, L., {Ferrara}, A., {Carniani}, S., {Zanella}, A. et al., ``{Dust
  temperature in ALMA [C II]-detected high-z galaxies}''.
\newblock {\em \mnras}, { \bf 503}(4), 4878--4891,  May  (2021).

\bibitem{remy-ruyer2014}
{R{\'e}my-Ruyer}, A., {Madden}, S.~C., {Galliano}, F., {Galametz}, M. et al.,
  ``{Gas-to-dust mass ratios in local galaxies over a 2 dex metallicity
  range}''.
\newblock {\em \aap}, { \bf 563}, A31,  March  (2014).

\bibitem{elmegreen2012}
{Elmegreen}, B.~G., {Zhang}, H.-X. and {Hunter}, D.~A., ``{In-spiraling Clumps
  in Blue Compact Dwarf Galaxies}''.
\newblock {\em \apj}, { \bf 747}(2), 105,  March  (2012).

\bibitem{rowland2024}
{Rowland}, L.~E., {McLeod}, A.~F., {Fattahi}, A., {Belfiore}, F. et al.,
  ``{Pre-supernova stellar feedback in nearby starburst dwarf galaxies}''.
\newblock {\em \aap}, { \bf 685}, A46,  May  (2024).

\bibitem{iorio2017}
{Iorio}, G., {Fraternali}, F., {Nipoti}, C., {Di Teodoro}, E. et al., ``{LITTLE
  THINGS in 3D: robust determination of the circular velocity of dwarf
  irregular galaxies}''.
\newblock {\em \mnras}, { \bf 466}(4), 4159--4192,  April  (2017).

\bibitem{murphy2011}
{Murphy}, E.~J., {Condon}, J.~J., {Schinnerer}, E., {Kennicutt}, R.~C. et al.,
  ``{Calibrating Extinction-free Star Formation Rate Diagnostics with 33 GHz
  Free-free Emission in NGC 6946}''.
\newblock {\em \apj}, { \bf 737}, 67,  August  (2011).

\bibitem{zanella2021}
{Zanella}, A., {Pallottini}, A., {Ferrara}, A., {Gallerani}, S. et al.,
  ``{Early galaxy growth: mergers or gravitational instability?}''.
\newblock {\em \mnras}, { \bf 500}(1), 118--137,  January  (2021).

\bibitem{angles-alcazar2017}
{Angl{\'e}s-Alc{\'a}zar}, D., {Faucher-Gigu{\`e}re}, C.-A., {Quataert}, E.,
  {Hopkins}, P.~F. et al., ``{Black holes on FIRE: stellar feedback limits
  early feeding of galactic nuclei}''.
\newblock {\em \mnras}, { \bf 472}(1), L109--L114,  November  (2017).

\bibitem{nelson2019}
{Nelson}, D., {Pillepich}, A., {Springel}, V., {Pakmor}, R. et al., ``{First
  results from the TNG50 simulation: galactic outflows driven by supernovae and
  black hole feedback}''.
\newblock {\em \mnras}, { \bf 490}(3), 3234--3261,  December  (2019).

\bibitem{teyssier2002}
{Teyssier}, R., ``{Cosmological hydrodynamics with adaptive mesh refinement. A
  new high resolution code called RAMSES}''.
\newblock {\em \aap}, { \bf 385}, 337--364,  April  (2002).

\bibitem{rosdahl2013}
{Rosdahl}, J., {Blaizot}, J., {Aubert}, D., {Stranex}, T. et al., ``{RAMSES-RT:
  radiation hydrodynamics in the cosmological context}''.
\newblock {\em \mnras}, { \bf 436}(3), 2188--2231,  December  (2013).

\bibitem{aubert2008}
{Aubert}, D. and {Teyssier}, R., ``{A radiative transfer scheme for
  cosmological reionization based on a local Eddington tensor}''.
\newblock {\em \mnras}, { \bf 387}(1), 295--307,  June  (2008).

\bibitem{grassi2014}
{Grassi}, T., {Bovino}, S., {Schleicher}, D.~R.~G., {Prieto}, J. et al.,
  ``{KROME - a package to embed chemistry in astrophysical simulations}''.
\newblock {\em \mnras}, { \bf 439}(3), 2386--2419,  April  (2014).

\bibitem{decataldo2019}
{Decataldo}, D., {Pallottini}, A., {Ferrara}, A., {Vallini}, L. et al.,
  ``{Photoevaporation of Jeans-unstable molecular clumps}''.
\newblock {\em \mnras}, { \bf 487}(3), 3377--3391,  August  (2019).

\bibitem{pallottini2019}
{Pallottini}, A., {Ferrara}, A., {Decataldo}, D., {Gallerani}, S. et al.,
  ``{Deep into the structure of the first galaxies: SERRA views}''.
\newblock {\em \mnras}, { \bf 487}(2), 1689--1708,  Aug  (2019).

\bibitem{vallini2017}
{Vallini}, L., {Ferrara}, A., {Pallottini}, A. and {Gallerani}, S.,
  ``{Molecular cloud photoevaporation and far-infrared line emission}''.
\newblock {\em \mnras}, { \bf 467}, 1300--1312,  May  (2017).

\bibitem{baes2015}
{Baes}, M. and {Camps}, P., ``{SKIRT: The design of a suite of input models for
  Monte Carlo radiative transfer simulations}''.
\newblock {\em Astronomy and Computing}, { \bf 12}, 33--44,  September  (2015).

\bibitem{camps2015}
{Camps}, P. and {Baes}, M., ``{SKIRT: An advanced dust radiative transfer code
  with a user-friendly architecture}''.
\newblock {\em Astronomy and Computing}, { \bf 9}, 20--33,  March  (2015).

\bibitem{behrens2018}
{Behrens}, C., {Pallottini}, A., {Ferrara}, A., {Gallerani}, S. et al.,
  ``{Dusty galaxies in the Epoch of Reionization: simulations}''.
\newblock {\em \mnras}, { \bf 477}, 552--565,  June  (2018).

\bibitem{feldmann2017}
{Feldmann}, R., {Quataert}, E., {Hopkins}, P.~F., {Faucher-Gigu{\`e}re}, C.-A.
  et al., ``{Colours, star formation rates and environments of star-forming and
  quiescent galaxies at the cosmic noon}''.
\newblock {\em \mnras}, { \bf 470}(1), 1050--1072,  September  (2017).

\bibitem{hopkins2018}
{Hopkins}, P.~F., {Wetzel}, A., {Kere{\v{s}}}, D., {Faucher-Gigu{\`e}re}, C.-A.
  et al., ``{FIRE-2 simulations: physics versus numerics in galaxy
  formation}''.
\newblock {\em \mnras}, { \bf 480}(1), 800--863,  October  (2018).

\bibitem{planck2016}
{Planck Collaboration}, {Ade}, P.~A.~R., {Aghanim}, N., {Arnaud}, M. et al.,
  ``{Planck 2015 results. XIII. Cosmological parameters}''.
\newblock {\em \aap}, { \bf 594}, A13,  September  (2016).

\bibitem{ma2018}
{Ma}, X., {Hopkins}, P.~F., {Garrison-Kimmel}, S., {Faucher-Gigu{\`e}re}, C.-A.
  et al., ``{Simulating galaxies in the reionization era with FIRE-2: galaxy
  scaling relations, stellar mass functions, and luminosity functions}''.
\newblock {\em \mnras}, { \bf 478}(2), 1694--1715,  August  (2018).

\bibitem{pallottini2023}
{Pallottini}, A. and {Ferrara}, A., ``{Stochastic star formation in early
  galaxies: Implications for the James Webb Space Telescope}''.
\newblock {\em \aap}, { \bf 677}, L4,  September  (2023).

\bibitem{sun2023}
{Sun}, G., {Faucher-Gigu{\`e}re}, C.-A., {Hayward}, C.~C., {Shen}, X. et al.,
  ``{Bursty Star Formation Naturally Explains the Abundance of Bright Galaxies
  at Cosmic Dawn}''.
\newblock {\em \apjl}, { \bf 955}(2), L35,  October  (2023).

\bibitem{springel2010}
{Springel}, V., ``{E pur si muove: Galilean-invariant cosmological
  hydrodynamical simulations on a moving mesh}''.
\newblock {\em \mnras}, { \bf 401}(2), 791--851,  January  (2010).

\bibitem{weinberger2017}
{Weinberger}, R., {Springel}, V., {Hernquist}, L., {Pillepich}, A. et al.,
  ``{Simulating galaxy formation with black hole driven thermal and kinetic
  feedback}''.
\newblock {\em \mnras}, { \bf 465}(3), 3291--3308,  March  (2017).

\bibitem{pillepich2018}
{Pillepich}, A., {Springel}, V., {Nelson}, D., {Genel}, S. et al.,
  ``{Simulating galaxy formation with the IllustrisTNG model}''.
\newblock {\em \mnras}, { \bf 473}(3), 4077--4106,  January  (2018).

\bibitem{dekel2023}
{Dekel}, A., {Sarkar}, K.~C., {Birnboim}, Y., {Mandelker}, N. et al.,
  ``{Efficient formation of massive galaxies at cosmic dawn by feedback-free
  starbursts}''.
\newblock {\em \mnras}, { \bf 523}(3), 3201--3218,  August  (2023).

\bibitem{ferrara2023}
{Ferrara}, A., {Pallottini}, A. and {Dayal}, P., ``{On the stunning abundance
  of super-early, luminous galaxies revealed by JWST}''.
\newblock {\em \mnras}, { \bf 522}(3), 3986--3991,  July  (2023).

\bibitem{dekel2009b}
{Dekel}, A., {Sari}, R. and {Ceverino}, D., ``{Formation of Massive Galaxies at
  High Redshift: Cold Streams, Clumpy Disks, and Compact Spheroids}''.
\newblock {\em \apj}, { \bf 703}(1), 785--801,  September  (2009).

\bibitem{furtak2024prep}
{Furtak, L.~et~al.~in~preperation}, ``{The Anglerfish cluster --- Discovery of
  large galaxy cluster structures associated with MACS~J0600.1-2008}''.
\newblock { \bf },  (2024).

\end{thebibliography}

\end{document}